\documentclass[journal]{IEEEtran}

\usepackage{cite}
\usepackage{algorithm,algorithmic,amsbsy,amsmath,amssymb,epsfig,bbm,mathrsfs,multirow,amsthm}
\usepackage{amsmath,amssymb,amsfonts}
\usepackage{algorithmic}
\usepackage{graphicx}
\usepackage{textcomp}
\usepackage{url}

\usepackage[table]{xcolor}

\def\BibTeX{{\rm B\kern-.05em{\sc i\kern-.025em b}\kern-.08em
    T\kern-.1667em\lower.7ex\hbox{E}\kern-.125emX}}

\usepackage[T1]{fontenc}
\usepackage{makecell}
\usepackage{amstext} 
\usepackage{subcaption} 
\definecolor{mygray}{gray}{0.6}
\definecolor{myblue}{rgb}{0.8,0.85,1}

\usepackage{array, tabularx, boldline}
\newcolumntype{L}[1]{>{\raggedright\let\newline\\\arraybackslash\hspace{0pt}}m{#1}}
\newcolumntype{C}[1]{>{\centering\let\newline\\\arraybackslash\hspace{0pt}}m{#1}}
\newcolumntype{R}[1]{>{\raggedleft\let\newline\\\arraybackslash\hspace{0pt}}m{#1}}

\usepackage{cellspace}
\begin{document}

\title{A Survey of Coded Distributed Computing}

\author{
Jer Shyuan Ng\thanks{JS.~Ng and WYB.~Lim are with Alibaba Group and Alibaba-NTU Joint Research Institute, Nanyang Technological University, Singapore. }, 
Wei Yang Bryan Lim, 
Nguyen Cong Luong\thanks{N.~C.~Luong is with Faculty of Information Technology, PHENIKAA University, Hanoi 12116, Vietnam, and is with PHENIKAA Research and Technology Institute (PRATI), A\&A Green Phoenix Group JSC, No.167 Hoang Ngan, Trung Hoa, Cau Giay, Hanoi 11313, Vietnam.},
Zehui Xiong\thanks{Z.~Xiong is with Alibaba-NTU Joint Research Institute, and also with School of Computer Science and Engineering, Nanyang Technological University, Singapore. }, 
Alia Asheralieva,\thanks{A.~Asheralieva is with Department of Computer Science and Engineering, Southern University of Science and Technology, Shenzhen, China.}
Dusit~Niyato,\thanks{D.~Niyato is with School of Computer Science and Engineering, Nanyang Technological University, Singapore. }~\textit{IEEE~Fellow}, 
Cyril Leung,\thanks{C. Leung is with The University of British Columbia and Joint NTU-UBC Research Centre of Excellence in Active Living for the Elderly (LILY).}
Chunyan Miao \thanks{C.~Miao is with Joint NTU-UBC Research Centre of Excellence in Active Living for the Elderly (LILY) and School of Computer Science and Engineering, Nanyang Technological University, Singapore.}
}

\maketitle

\begin{abstract}
Distributed computing has become a common approach for large-scale computation of tasks due to benefits such as high reliability, scalability, computation speed, and cost-effectiveness. However, distributed computing faces critical issues related to communication load and straggler effects. In particular, computing nodes need to exchange intermediate results with each other in order to calculate the final result, and this significantly increases communication overheads. Furthermore, a distributed computing network may include straggling nodes that run intermittently slower. This results in a longer overall time needed to execute the computation tasks, thereby limiting the performance of distributed computing. To address these issues, coded distributed computing (CDC), i.e., a combination of coding theoretic techniques and distributed computing, has been recently proposed as a promising solution. Coding theoretic techniques have proved effective in WiFi and cellular systems to deal with channel noise. Therefore, CDC may significantly reduce communication load, alleviate the effects of stragglers, provide fault-tolerance, privacy and security. In this survey, we first introduce the fundamentals of CDC, followed by basic CDC schemes. Then, we review and analyze a number of CDC approaches proposed to reduce the communication costs, mitigate the straggler effects, and guarantee privacy and security. Furthermore, we present and discuss applications of CDC in modern computer networks. Finally, we highlight important challenges and promising research directions related to CDC.  
\end{abstract}


\section{Introduction}
\label{sec:intro}

In recent years, distributed computing has used for large-scale computation~\cite{cristea2010large} since it offers several advantages over centralized computing. First, distributed computing is able to provide computing services with high reliability and fault-tolerance. In particular, distributed computing systems can efficiently and reliably work even if some of the computing nodes, i.e., computers or workers, fail. Second, distributed computing has high computation speed as the computation load is shared among various computing nodes. Third, distributed computing systems are scalable since computing nodes can easily be added. Fourth, distributed computing is economical by using computing nodes with low-cost hardware. Likewise, distributed computing is adopted in cloud computing and other emerging services. Given the aforementioned advantages, distributed computing has been applied in numerous real-life applications such as telecommunication networks~\cite{kumar2005distributed} (e.g., telephone networks and wireless sensor networks), network applications~\cite{corbett2013spanner} (e.g., world-wide web networks, massively multiplayer online games and virtual reality communities, distributed database management systems, and network file systems), real-time process control~\cite{gonzaga2009ann} (e.g., aircraft control systems), and parallel computation~\cite{de2015structured} (e.g., cluster computing, grid computing, and computer graphics).


However, distributed computing faces serious challenges. Let us consider one of the most common distributed computation frameworks, i.e., MapReduce~\cite{dean2008mapreduce}. MapReduce is a software framework and programming
model for processing a computation task across large datasets using a large number of computing nodes, i.e., workers. A set of the computing nodes is referred to as a cluster or a grid. In general, the overall computation task is decomposed into three phases, i.e., the ``Map'' phase, ``Shuffle'' phase, and ``Reduce'' phase. In the Map phase, a master node splits the computation task into multiple subtasks and assigns the subtasks to the computing nodes. The computing nodes compute the subtasks according to the allocated Map functions to generate intermediate results. Then, the intermediate results are exchanged among the computing nodes, namely ``data shuffling'', during the Shuffle phase. In the Reduce phase, the computing nodes use these results to compute the final result in a distributed manner by using their allocated Reduce functions. 

Distributed computing has two major challenges. First, the computing nodes need to exchange a number of intermediate results over the network with each other in order to calculate the final result; this significantly increases communication overheads and limits the performance of distributed computing applications such as Self-Join\cite{ahmad2012tarazu}, Terasort \cite{guo2017ishuffle}, and machine learning \cite{chowdhury2011managing}. For example, for the Hadoop cluster at Facebook, it is observed that on average, the data shuffling phase accounts for 33\% of the overall job execution time~\cite{chowdhury2011managing}. 65\% and 70\% of the overall job execution time is spent on the Shuffle phase when running a TeraSort and Self-Join application respectively on a heterogeneous Amazon EC2 cluster~\cite{zhang2013performance}. In fact, the communication bottleneck is worse in the trainings of convolutional neural networks (CNNs), e.g., Resnet-50 \cite{he2016deep} and AlexNet \cite{alex2012imagenet}, which include updates of millions of model parameters. Second, distributed computing is executed by a large number of computing nodes which may have very different computing and networking resources. As a result, there are straggling nodes or stragglers, i.e., computing nodes which run unintentionally slower than others, thereby increasing the overall time needed to complete the computing tasks. To address the straggler effects, traditional approaches such as work exchange~\cite{attia2017combating} and naive replication \cite{wang2015replication} have been adopted for the distributed computing. However, such approaches either introduce redundancy or require coordination among the nodes that significantly increases communication costs and computation loads. This motivates the need for a novel technique that is able to more effectively and completely address the straggler effects and communication load of distributed computing. 


Coding theoretic techniques, e.g., channel coding such as low-density parity-check (LDPC)~\cite{maity2019ldpc}, have been widely used in WiFi and cellular systems to combat the impact of channel noise and impairments. They have also been applied in distributed storage systems and cache networks~\cite{maddah2014fundamental} to reduce storage cost and network traffic. The basic principle of the coding theoretic techniques is that redundant information, i.e., redundancy, is introduced in messages/signals before they are transmitted to a receiver. The redundancy is included in the messages in a controller manner such that it can be utilized by the receiver to correct errors caused by the channel noise. Coding theoretic techniques have been recently regarded as promising solutions to cope with the challenges in distributed computing~\cite{li2018tradeoff}, \cite{li2015mapreduce}. For example, coding theoretic techniques can be used to encode the Map tasks of the computing nodes such that the master node is able to recover the final result from partially finished nodes, thus alleviating the straggler effects~\cite{lee2018speeding}. Another example is that coding theoretic techniques enable coding opportunities across intermediate results of the distributed computation tasks which significantly reduces the communication load by reducing the number and the size of data transmissions among the processing nodes~\cite{li2018tradeoff}. The combination of coding techniques and distributed computing is called coded distributed computing (CDC)~\cite{li2018tradeoff}. Apart from reducing communication load and alleviating the effects of stragglers, CDC can provide fault-tolerance, preserve privacy~\cite{nodehi2018limited}, and improve security~\cite{yu2018lagrange} in distributed computing. As a result, CDC approaches have recently received a lot of attention.

CDC schemes can be applied in modern networks such as Network Function Virtualization (NFV) and edge computing. With the data mainly generated by end devices, e.g., Internet of Things (IoT), that have significant sensing as well as computational and storage capabilities, it is natural to perform some computations at the end devices, instead of the cloud which may not be able to handle the large amounts of data generated. As such, edge computing \cite{shi2016edge} has been proposed as a solution to perform distributed computation tasks. In order to perform complex computations, e.g., the training of deep neural networks that involve a large number of training layers, resource-constrained devices may need to pool their resources to perform their computations collaboratively \cite{lim2020incentive}. This results in high communication costs and computation latency. CDC schemes can be implemented to overcome these challenges. Furthermore, CDC schemes can be implemented on edge computing networks that involve constantly-moving end devices \cite{mao2017mec}, e.g., vehicles and smartphones, which imposes additional communication constraints.

To the best of our knowledge, although there are several surveys and books related to distributed computing, there is no survey paper on CDC. In particular, large-scale distributed computing and applications are discussed in \cite{cristea2010large}. Surveys related to distributed computing include grid resource management systems for distributed computing~\cite{krauter2002taxonomy}, resource allocation in high performance distributed computing~\cite{hussain2013survey}, wireless distributed computing~\cite{datla2012wireless}, and wireless grid computing~\cite{ahuja2006survey}. This motivates the need for this survey on CDC. In summary, our survey has the following contributions:
\begin{itemize}
\item We describe the fundamentals of CDC. In particular, we introduce the commonly used distributed computation frameworks for the implementation of coding techniques and algorithms. We then discuss basic CDC schemes.
\item We review and discuss a number of CDC schemes to reduce the communication costs for distributed computing. The approaches include file allocation, coded shuffling design, and function allocation. We further analyze and compare the advantages and disadvantages of the CDC schemes.  
\item We review, discuss, and analyze CDC schemes which mitigate the straggler effects of distributed computing. The approaches include computation load allocations, approximate coding, and exploitation of stragglers.  
\item We review and present CDC schemes that can improve the privacy and security in distributed computing.
\item We analyze and provide insights into the existing approaches and solutions in the CDC literature.
\item We present and discuss applications of CDC in modern networks such as NFV and edge computing. 
\item We highlight challenges and discuss promising research directions related to CDC.
\end{itemize}

\begin{figure}
\centering
\includegraphics[width=\linewidth]{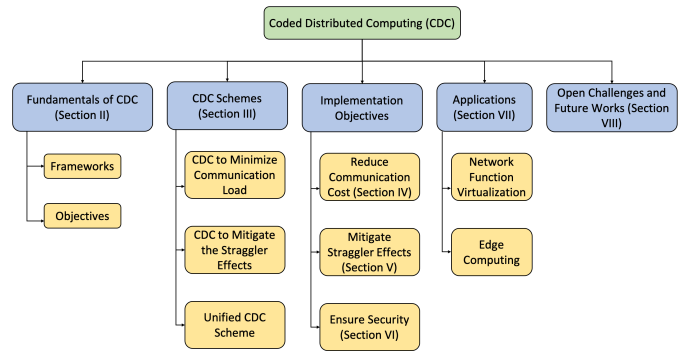}
\caption{\small Structure of survey.}
\label{fig:structure}
\end{figure}

For the reader's convenience, we classify the related CDC studies according to the challenges that need to be handled. In particular, the issues are communication costs, straggler effects, and security. As such, readers who are interested in or working on related issues will benefit greatly from our insightful reviews and in-depth discussions of existing approaches, remaining/open problems, and potential solutions. The rest of this paper is organized as follows. Section~\ref{sec:fundamentals} introduces the fundamentals of CDC. Section~\ref{sec:schemes} presents basic CDC schemes. Section~\ref{sec:comms} reviews CDC approaches that have been proposed to reduce communication costs. Section~\ref{sec:stragglers} discusses CDC approaches that have been proposed to mitigate straggler effects. Section~\ref{sec:security} presents CDC approaches that have been proposed to enhance privacy and security in distributed computing.~Section~\ref{sec:applications} discusses applications of CDC. Section~\ref{sec:challenges} highlights important challenges and promising research directions. The structure of the survey is presented in Figure~\ref{fig:structure}. Section~\ref{sec:conclusion} concludes the paper. A list of abbreviations commonly used in this paper is given in Table~\ref{tab:table_abb}.

\begin{table}[h!]
\scriptsize
  \caption{\small List of common abbreviations used in this paper.}
  \label{tab:table_abb}
  \centering
  \begin{tabularx}{8.7cm}{|Sl|X|}
  \hline
  \rowcolor{mygray}
 \textbf{Abbreviation} & \textbf{Description} \\   \hline
ARIMA & Auto Regressive Integrated Moving Average \\ \hline
BCC & Batch Coupon's Collector\\ \hline
BGC & Bernoulli Gradient Code\\ \hline
BGW & Ben-Or, Goldwasser, and Wigderson \\ \hline
BPCC & Batch-Processing Based Coded Computing\\ \hline
C3P & Coded Cooperative Computation Protocol\\ \hline
CDC & Coded Distributed Computing\\ \hline
CNN & Convolutional Neural Network\\ \hline
CPGC & Coded Partial Gradient Computation\\ \hline
DAG & Directed Acyclic Graphs\\ \hline
DNN & Deep Neural Network \\ \hline
FRC & Fractional Repetition Coding\\ \hline
HCMM & Heterogeneous Coded Matrix Multiplication\\ \hline
IoT & Internet of Things\\ \hline
LCC & Lagrange Coded Computing\\ \hline
LDPC & Low-Density Parity-Check\\ \hline
LT & Luby Transform\\ \hline
MDS & Maximum Distance Separable\\ \hline
MMC & Multi-Message Communication\\ \hline
MPC & Multi-Party Computation \\ \hline
NFV & Network Function Virtualization \\ \hline
PCR & Polynomially Coded Regression\\ \hline
PDA & Placement Delivery Array \\ \hline
SDMM & Secure Distributed Matrix Multiplication\\ \hline
SGC & Stochastic Gradient Coding \\ \hline
SGD & Stochastic Gradient Descent\\ \hline
SVD & Singular Vector Decomposition\\ \hline
UAV & Unmanned Aerial Vehicle\\ \hline
 
\end{tabularx}
\end{table}

\section{Fundamentals of Coded Distributed Computing}
\label{sec:fundamentals}

Distributed computing has been an important solution to large-scale, complex computation problems, which involves massive amounts of data. Various distributed computing models, e.g., cluster computing \cite{valentini2013overview}, grid computing \cite{sadashiv2011comparison} and cloud computing \cite{hussain2013survey, idrissi2013cloud}, have been developed to perform the distributed computation tasks while providing high quality of services (QoS) to the users. Among the distributed computing models, cloud computing is gaining much popularity recently as it eliminates the need for users to purchase expensive hardware and software resources since the users only need to pay for the cloud services based on their usage needs in an on-demand basis. A comparison between cluster, grid and cloud computing models is summarized in Table~\ref{tab:compare}.

Distributed computing has been widely implemented in a variety of applications, e.g., sensor networks\cite{xu2004distributed}, healthcare applications\cite{khan2017handbook}, the development of smart cities\cite{tang107incorporating}, automated manufacturing processes\cite{raghavan2002dpac} and vehicular applications\cite{juan2013vehicle}. In order to improve the performance of the distributed computing systems, various aspects such as resource allocation strategies \cite{hussain2013survey}, task allocation strategies \cite{ahmad1995task, kafil1998optimal}, scheduling algorithms \cite{woo1997task,  lopes2016taxonomy}, incentive mechanisms \cite{ranjan2008incentive, duan2012incentive}, energy efficiency \cite{valentini2013overview}, network security \cite{xiao2007security} and the performance modelling \cite{pllana2007performance} of the distributed computing systems have been extensively studied in the literature. However, the performance of the distributed computing systems is still limited by the high communication costs and straggler effects which lead to a longer time needed to execute the computation tasks. As a result, recent research has focused on coding techniques to overcome these implementation challenges of the distributed computing systems, the aims of which are to minimize the communication load as well as to mitigate the straggler effects.

In this section, we discuss commonly used distributed computation frameworks for the implementation of coding techniques and algorithms. Note that while the different computation frameworks are useful for different computing applications, we focus specifically on the MapReduce framework \cite{dean2008mapreduce} as the majority of the research works on CDC schemes are based on the MapReduce computation framework. We also introduce the two main lines of works in CDC, i.e., to reduce communication load and to mitigate the straggler effects, which aim to solve the challenges in distributed computing.


\begin{table*}[t]
\caption{\small Comparison between cluster, grid and cloud computing models \cite{hussain2013survey}. }
\label{tab:compare}
\centering
\begin{tabular}{|l |c| c| c|}

\hline
\rowcolor{mygray}
\multicolumn{1}{c|}{\textbf{Feature}} & \textbf{Cluster} & \textbf{Grid} & \textbf{Cloud} \\ \hline
Size & Small to medium & Large & Small to large\\ \hline
Network type & Private, LAN & Private, WAN & Public, WAN\\ \hline
Job management and scheduling & Centralized & Decentralized & Both\\ \hline
Coupling & Tight & Loose/tight & Loose \\ \hline
Resource reservation & Pre-reserved & Pre-reserved & On-demand\\ \hline
Service-level agreement (SLA) constraint & Strict & High & High\\ \hline
Resource support & Homogeneous and heterogeneous (GPU) & Heterogeneous & Heterogenous\\ \hline
Virtualization & Semi-virtualized & Semi-virtualized & Completely virtualized\\ \hline
Security type & Medium & High & Low \\ \hline
Service-oriented architecture and heterogeneity support & Not supported & Supported & Supported\\ \hline
User interface & Single system image & Diverse and dynamic & Single system image \\ \hline
Initial infrastructure cost & Very high & High & Low \\ \hline
Self service and elasticisty & No & No & Yes\\ \hline
Administrative domain & Single & Multi & Both\\ \hline

\end{tabular}
\end{table*}

\subsection{Coded Distributed Computation Frameworks}
\label{subsec:mapreduce}

While the distributed computation frameworks have moved beyond a simple MapReduce framework, the majority of the studies on CDC have focused on the MapReduce framework. 

\begin{figure}
\centering
\includegraphics[width=\linewidth]{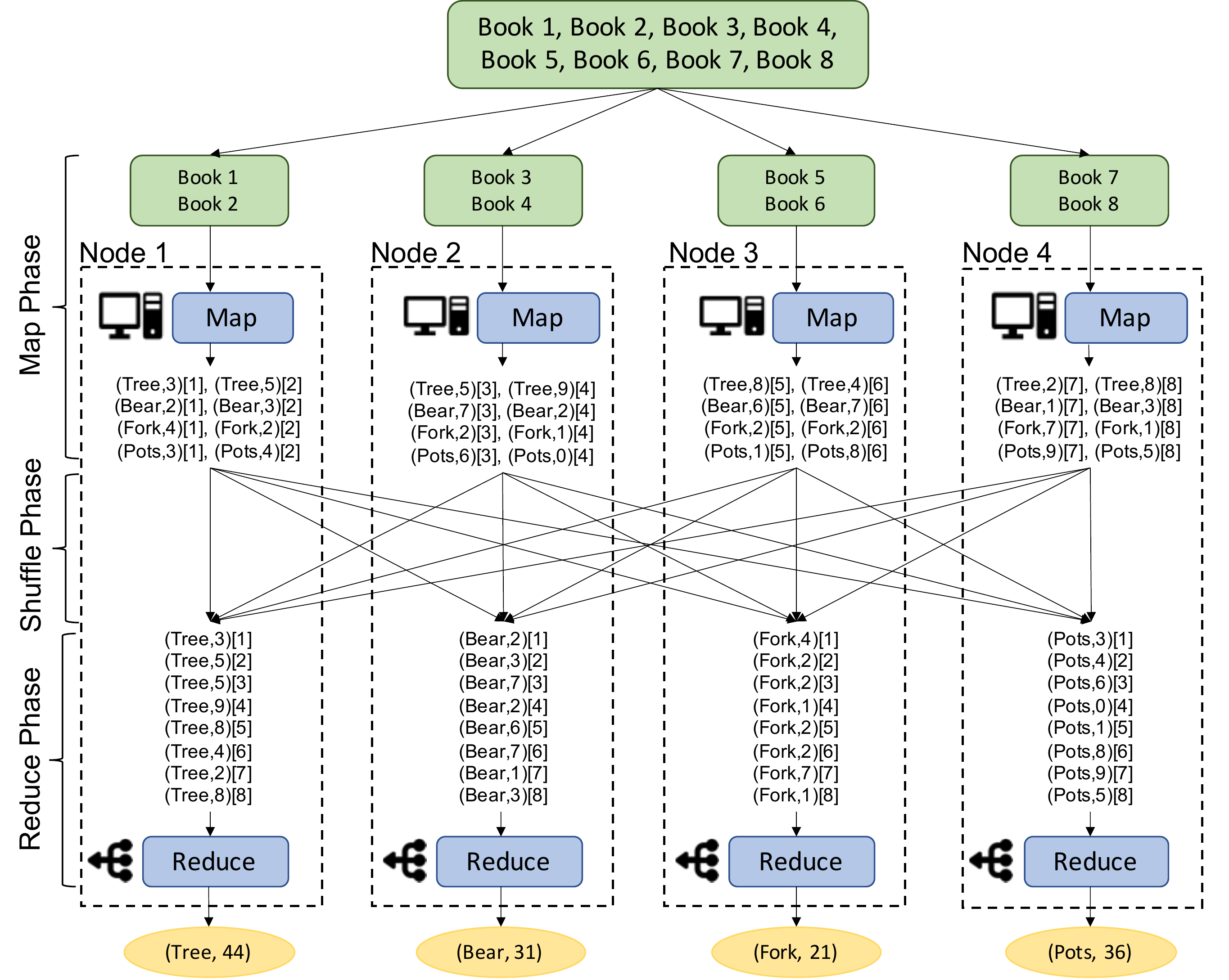}
\caption{\small Illustration of conventional MapReduce framework. The intermediate output pairs are represented by (key,frequency)[book number] and the output pairs are represented by (key,frequency).}
\label{fig:conventional}
\end{figure}

MapReduce \cite{dean2008mapreduce} is a software framework and programming model that runs on a large cluster of commodity machines for the processing of large-scale datasets in a distributed computing environment. The cluster of computers is modelled as a master-worker system which consists of a single master node and multiple workers to store and analyzes massive amount of unstructured data. Due to its scalability and its ability to tolerate machines' failure \cite{jiang2010performance}, the MapReduce framework is commonly used in a wide range of applications \cite{dean2008mapreduce}, e.g., the analysis of web access logs, the clustering of documents, the construction of web-link graph that matches the all source URLs to a target URL and the development of machine learning algorithms. Generally, the MapReduce computation framework involves the processing of a large input file to generate multiple output pairs of which each pair consists of a key and a corresponding value. Figure~\ref{fig:conventional} demonstrates the implementation of the conventional MapReduce framework to determine the frequency of occurrence of 4 specific words in the books, where the 4 processing nodes, i.e., the workers are to compute the 4 output pairs. There are three important phases in the MapReduce computation framework:

\begin{enumerate}
\item In the \textbf{\emph{Map}} phase, there are two stages, namely the allocation of Map tasks and the execution of Map tasks. Generally, a Map task is a function that generates a key-value output pair based on the allocated subfiles. Firstly, as seen in Fig.~\ref{fig:conventional}, the master node splits the input file into 8 subfiles of smaller sizes and allocates the subfiles to the 4 workers. Secondly, each worker produces 4 intermediate key-value pairs for each allocated subfile using the map functions. Since the workers are allocated 2 subfiles each, each worker generates 8 computed intermediate results.

\item In the \textbf{\emph{Shuffle}} phase, the workers exchange their computed intermediate results to obtain the required intermediate results for the computation of the Reduce functions. In particular, in each time slot, one of the workers creates a message that contains information of the intermediate output pairs from the Map phase and transmits the message to other workers. The shuffling process continues until all workers have received the required intermediate output pairs for the Reduce phase. 

\item In the \textbf{\emph{Reduce}} phase, the workers aggregate the 8 intermediate key-value pairs obtained from the Shuffle phase and compute the final result which is a smaller set of key-value pairs using the reduce functions. In particular, the reduction tasks are evenly distributed among the workers. Each reduce function is responsible for the evaluation of a key. For example, node 1 in Fig.~\ref{fig:conventional} is responsible for the evaluation of ``Tree''. Therefore, the total number of reduce functions needed equals the total number of keys of the output, i.e., 4 reduce functions are needed to compute 4 output pairs.
\end{enumerate}


Apart from the MapReduce framework, there are other distributed computation frameworks that provide support for the processing of large-scale datasets such as:
\begin{itemize}
\item \emph{Spark \cite{zaharia2010spark}: }It supports applications that need to reuse a working dataset across the multiple parallel processes. These applications cannot be expressed as efficiently as acyclic data flows which are required in popular computation frameworks such as MapReduce. There are two use cases for the implementation of Spark computation framework: (i) iterative machine learning algorithms which operate on the same dataset repeatedly, and (ii) iterative data analysis tools, where different users query for a subset of data from the same dataset.
\item \emph{Dryad \cite{isard2007dryad}: }By allowing the developers to construct their own communication graphs and the subroutines at the vertices through simple, high-level programming language, Dryad executes large-scale data-intensive computations over clusters consisting multiple computers. It does not require the developers to express their code in Map, Shuffle and Reduce phases in order to adopt the MapReduce framework for computations. Besides, the Dryad execution engine, which is based on the constructed data flow graph, takes care of the implementation issues of the distributed computation tasks such as the scheduling of tasks, allocation of resources and the recovery from communication and computation failures.
\item \emph{CIEL \cite{murray2011ciel}: }The main characteristic of the CIEL computation framework is that it allows data-dependent data flows where the directed acyclic graphs (DAG) are built dynamically based on the execution of previous computations, rather than a statistically predetermined DAG. Instead of maximizing throughput, CIEL aims to minimize latency of individual tasks, which is very useful for the implementation of iterative algorithms, where latency grows significantly as the number of iteration increases.
\end{itemize}

A comparison between the distributed computation frameworks is presented in Table~\ref{tab:frameworks}.

\begin{table}[t]
\caption{\small Comparison between distributed computation frameworks \cite{murray2011ciel}. }
\label{tab:frameworks}
\centering
\begin{tabular}{|m{2.8cm} | >{\centering\arraybackslash}m{1.5cm} |>{\centering\arraybackslash}m{1.5cm} |>{\centering\arraybackslash}m{1.3cm} |}
\hline
\rowcolor{mygray}
\multicolumn{1}{c|}{\textbf{Feature}} & \textbf{MapReduce \cite{dean2008mapreduce}} & \textbf{Dyrad \cite{isard2007dryad}} & \textbf{CIEL \cite{murray2011ciel}} \\ \hline
Dynamic Control Flow & No & No & Yes \\ \hline
Task Dependencies & Fixed (2-stage) & Fixed (DAG) & Dynamic \\ \hline
Fault Tolerant & Transparent & Transparent & Transparent \\ \hline
Data Locality & Yes & Yes & Yes \\ \hline
Transparent Scaling & Yes & Yes & Yes \\ \hline

\end{tabular}
\end{table}

\subsection{Objectives of CDC Schemes}

There are two main lines of works in CDC. Firstly, the CDC schemes are implemented to minimize the \emph{communication load} in distributed computing systems. Secondly, the CDC schemes aim to mitigate the \emph{straggler effects} which cause a delay in the computation of the distributed tasks. For each of the objectives, we discuss the importance of solving these issues to improve the performance of the distributed computing systems. Then, we briefly discuss the existing solutions that have been proposed in the literature to meet these objectives. However, the current existing solutions do not adopt coding approaches. Different from the existing solutions, the CDC schemes are able to meet these objectives by introducing coded redundancy. In fact, the CDC schemes outperform the replication methods, e.g., naive replication \cite{li2015mapreduce} and fork-join model \cite{joshi2017redundancy, wang2015replication}, which introduce redundancy without coding techniques, in terms of time taken to execute the tasks. 

\subsubsection{\textit{Communication Load}} \label{subsub:comms} Among the three phases of the MapReduce computation framework, the Shuffle phase dominates the time required to complete the computation tasks  \cite{weets2015limitations, guo2017ishuffle, ahmad2012tarazu} since multiple communications between the processing nodes are needed to exchange their intermediate results. For the Hadoop cluster at Facebook, it is observed that on average, the data shuffling phase accounts for 33\% of the overall job execution time \cite{chowdhury2011managing}. In fact, the data shuffling phase is more time consuming when running on heterogeneous clusters with diverse computational, communication and storage capabilities. When running TeraSort \cite{hadoop_terasort}, which is a conventional distributed sorting algorithm for large amount of data, and Self-Join applications on heterogeneous Amazon EC2 clusters, 65\% and 70\% of the overall job execution time is spent on the Shuffle phase respectively \cite{zhang2013performance}. The data shuffling process is also an important step in implementing distributed learning algorithms. In particular, to train machine learning models with distributed algorithms, it is common to shuffle the data randomly and run the algorithms iteratively for a number of times such that the processing nodes compute a different subset of the data at each iteration until there is a convergence \cite{retch2013parallel, bottou2012stochastic, ioffe2015batch}. \emph{For the logistic regression application which requires at least 100 iteration to converge, 42\% of the iteration time is spent on communication \cite{chowdhury2011managing}}. For each time the data shuffling process is performed, the entire training dataset is communicated over the network, resulting in high communication costs which limit the performance of the distributed computing systems.  

Since the performance of the data shuffling process has a significant impact on the overall performance of the distributed computing systems, it has been extensively studied in the literature, e.g., \cite{ahmad2013marco, nicolae2017leveraging, yu2015virtual, isard2009quincy, suresh2014scheduling, xie2013research, hadoop_2020, zaharia2010delay}. Various data shuffling strategies are proposed to achieve different objectives such as minimizing the job execution time, maximizing the utilization of resources and accommodating interactive workloads. While the overlap between the map computations and the shuffle communication helps to reduce the latency of the distributed computation tasks \cite{ahmad2013marco}, the computing nodes require large storage capacities for buffering. An efficient and adaptive data shuffling strategy is proposed in the study of \cite{nicolae2017leveraging} to manage the tradeoff between the accumulation of the shuffle blocks and the minimization of the utilization of memory space to reduce the overall job execution time and improve the scalability of the distributed computing systems. In \cite{yu2015virtual}, the authors propose a virtual data shuffling strategy which reduces storage space and traffic load in the network by delaying the actual movement of the data until it is needed to complete the computations in the Reduce phase. 

To improve the performance of the data shuffling process, task scheduling algorithms such as Quincy scheduler \cite{isard2009quincy}, Hadoop Fair Scheduler \cite{hadoop_2020} and delay scheduling algorithm \cite{zaharia2010delay} are also designed to allocate tasks to the workers. In the design of optimal task scheduling and task selection algorithms, the communication load can be minimized through various approaches such as by optimizing the placement of computation tasks, distributing the computing resources fairly to the nodes and maximizing the resource utilization of the systems.  
Since the task scheduling schemes are not the focus of this survey, we refer interested readers to the study of \cite{rao2012survey} and the references therein for more detailed information on the scheduling techniques.

\begin{figure}
\centering
\includegraphics[width=\linewidth]{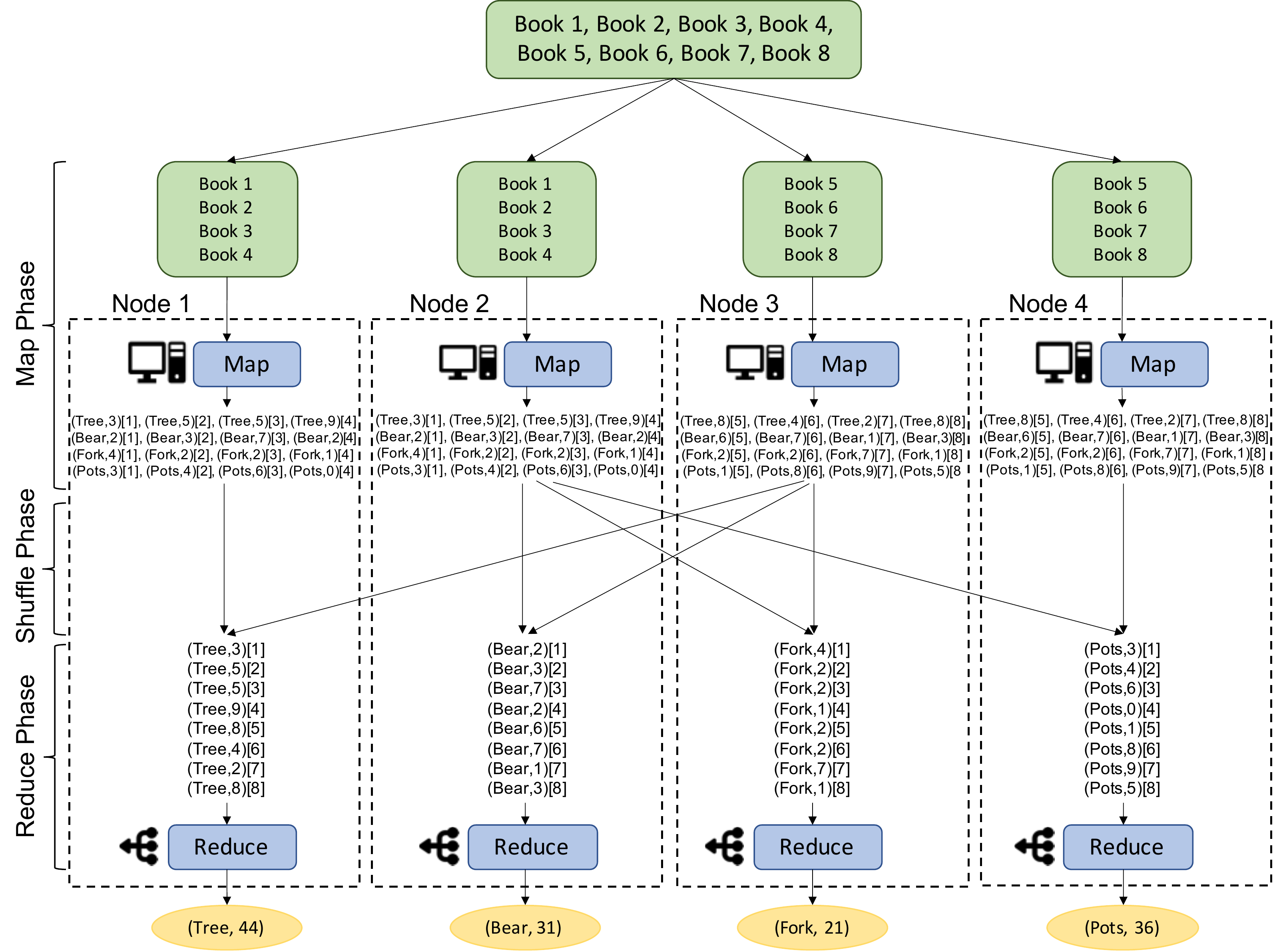}
\caption{\small Illustration of naive replication MapReduce framework.}
\label{fig:uncoded}
\end{figure}

One of the ways to reduce communication load in the shuffling phase is by repeating the computation tasks \cite{li2015mapreduce}. Figure~\ref{fig:uncoded} illustrates the implementation of the MapReduce framework by using the naive replication method in which 4 processing nodes are required to compute 4 output pairs, which is the same computation task illustrated in Fig.~\ref{fig:conventional}. By simply replicating the Map tasks where each worker is required to compute more intermediate output pairs, i.e., 16 in Fig.~\ref{fig:conventional} instead of 8 intermediate output pairs in Fig.~\ref{fig:uncoded}, the communication load is reduced as fewer intermediate output pairs are communicated. For example, in the conventional MapReduce framework in Fig.~\ref{fig:conventional}, node 1 needs to obtain 6 intermediate output pairs from other workers whereas in the naive replication scheme in Fig.~\ref{fig:uncoded}, node 1 only needs to obtain 4 intermediate output pairs from other workers. 

However, the aforementioned non-coding methods have limits to which the communication load in the data shuffling phase can be minimized. Given that the naive replication method reduces communication load in the data shuffling phase by introducing redundancy to the systems (which is also discussed in Section~\ref{subsec:cdccommunication}), coding techniques can be used to introduce redundancy to further minimize communication load, which will be discussed in-depth in Section~\ref{sec:comms}.

\subsubsection{\textit{Straggler Effects}}In distributed computing systems, the processing nodes may have heterogeneous computing capabilities, i.e., different processing speeds. As such, another line of work in the CDC literature is to solve the bottleneck that results from a variation in time taken to complete the allocated tasks. In distributed computing systems, there are \emph{stragglers}, which are the processing nodes that run unexpectedly slower than the average or nodes that may be disconnected from the network due to several factors such as insufficient power, contention of shared resources, imbalance work allocation and network congestion \cite{dean2013tail, ananthanarayanan2010reining}. As a result, the overall time needed to execute the tasks is determined by the slowest processing node. We briefly discuss the existing approaches (which are summarized in Table~\ref{tab:priorstragglers}) to handle the straggler effects as follows:

\begin{table*}[t]
\caption{\small Approaches to mitigate the straggler effects.} 
\label{tab:priorstragglers}
\centering
\begin{tabular}{|>{\centering\arraybackslash}m{2.6cm}  |m{6.5cm} |m{6.5cm}|}
\hline

\rowcolor{mygray}
\textbf{Approach} & \multicolumn{1}{c|}{\textbf{Key Ideas}} & \multicolumn{1}{c|}{\textbf{Shortcomings}} \\ \hline
Stragglers Detection & Detects the straggling nodes, determines the cause of delay and implements targeted solutions & Difficult to identify the cause of delays \\ \hline
Work Stealing & Reallocates the remaining computation tasks from the slower workers to the faster workers & The capabilities of the slower workers are not maximized\\ \hline
Work Exchange & Reallocates the computation tasks every time a worker completes its tasks & Incurs high communication costs due to the feedback from the workers to the master node as well as the reallocation of data \\ \hline
Naive Replication & Introduces redundancy where each computation subtask is performed by more than one processing node & Incurs high communication costs and computation load\\  \hline
Coded Redundancy & Uses coding techniques to introduce redundancy such that the master node can recover the final result from any decodable set of workers & Still incurs high communication costs and computation load, but lower than that of naive replication \\ \hline

\end{tabular}
\end{table*}

\begin{itemize}
\item \emph{Stragglers Detection: }The most direct approach to mitigate the straggler effects is to detect the stragglers and act on them early in their lifetime. For example, Mantri \cite{ananthanarayanan2010reining} detects the stragglers by identifying the tasks that are processed at a rate slower than the average. The system determines the cause of the delay and implements targeted solutions to mitigate the stragglers. The solutions include restarting the tasks allocated to the stragglers at other processing nodes, optimally allocating the tasks based on network resources as well as protecting against interim data loss by replicating the outputs of valuable tasks. 

\item \emph{Work Stealing \cite{blumofe1999stealing}: }The basic idea of the work stealing algorithm is to allow the faster processing nodes to take over the remaining computation tasks from the slower processing nodes so that the overall job execution time is minimized. By adopting this approach, the faster processing nodes operate continuously while leaving the slower processing nodes idle after their jobs are taken over by the faster processing nodes till the end of the computation session. 

\item \emph{Work Exchange \cite{attia2017combating}: } By leveraging on the information of computational heterogeneity in the system, the master node first allocates the tasks to the workers based on their computational capabilities. Upon receiving the first computed result from any of the workers, the master node pauses the computation process and redistribute the remaining incomplete work to be computed by the workers. The process is performed for a number of iterations until all work is done. Since the workers need to inform the master node of the amount of work done at each time that the computation process is paused, additional communication costs are incurred. The higher communication costs are also a result of the reallocation of data to the workers. 

\item \emph{Naive Replication: }One of the solutions to handle stragglers in the distributed computing systems is by introducing redundancy to minimize computation latency. The computation task is replicated and executed over multiple processing nodes. Since all processing nodes are working on the same computation task, the time required to complete the computation task is determined by the fastest processing node. The partial computations of the remaining processing nodes are discarded. Experiments on Google Trace data \cite{wang2015replication} have shown the effectiveness of the use of redundancy in minimizing computation latency by eliminating the need for the computed results by the stragglers. However, the introduction of redundancy comes at the expense of higher cost such as high communication bandwidth and high computation load \cite{wang2015replication, joshi2017redundancy, gardner2015exact, shah2016redundant, aktas2018relaunch, aktas2017clones}. Various redundancy strategies have been analyzed to derive the limiting distribution of the state of the systems \cite{joshi2017redundancy, gardner2015exact}. Although the introduction of redundancy helps to reduce latency, the performance varies under different settings. In fact, in some settings, it is optimal to not use any redundancy strategy. Looking into this, the work in \cite{shah2016redundant} presents the optimal redundant-requesting policies under diverse settings.

\end{itemize}

Similar to the existing methods to reduce communication costs as discussed in the previous section (Section~\ref{subsub:comms}), the existing methods to mitigate the straggler effects do not adopt coding approaches. Coding techniques can also be used to introduce redundancy into the systems to mitigate the straggler effects. The authors in \cite{aktas2018relaunch, aktas2017clones} investigate the tradeoff between latency and cost for both replication-redundancy systems and coded-redundancy systems. Coded-redundancy systems outperform the replication-redundancy systems in both latency and cost. In other words, by using coding techniques, the latency and cost incurred are lower than that of naive replication. The use of coding techniques to mitigate the straggler effects is discussed in more detail in Section~\ref{subsec:codingtech} and Section~\ref{sec:stragglers}.
To better understand the proposed CDC schemes, some of the commonly used performance metrics of the distributed computing systems are defined as follows:
\begin{enumerate}

\item \emph{Storage Space }is defined as the total number of files stored across $K$ processing nodes, normalized by the total number of subfiles $N$ \cite{yan2018storage}.

\item \emph{Computation load }is represented by $r$, where $1\leq r \leq K$, is defined as the total number of Map functions computed across $K$ processing nodes, normalized by the total number of subfiles $N$ \cite{li2018tradeoff}. In particular, when $r=1$, it means that each Map function is only computed by a single processing node. When $r=2$, it means that each Map function is computed by two processing nodes on average. 

\item \emph{Communication load }is represented by $L$, where $0\leq L \leq 1$, is defined as the total number of bits communicated by the $K$ processing nodes in the Shuffle phase, normalized by the total number of subfiles $N$ \cite{li2018tradeoff}.
\end{enumerate}

Given that coding techniques are able to solve the aforementioned implementation challenges of the distributed computing systems, we review various proposed CDC schemes, which is the main focus of this paper. In the following section, we present a tutorial of the simple CDC schemes along these two lines of works, i.e., minimizing communication load and mitigating the straggler effects, of which is useful to better understand the related works discussed in Section~\ref{sec:comms}, \ref{sec:stragglers} and \ref{sec:security}.

\section{Coded Distributed Computing (CDC) Schemes}
\label{sec:schemes}

Recently, coding techniques have become a popular approach to solve the challenges of the distributed computing systems. As mentioned previously, there are two main lines of work in CDC: (i) to reduce the communication costs and (ii) to mitigate the straggler effects. In this section, we introduce the two basic CDC schemes which are the first works that show the effectiveness of using coding techniques to solve these two challenges separately. Then, we discuss a unified CDC scheme that characterizes the tradeoff between computation latency and communication load.

\subsection{CDC to Minimize Communication Load}
\label{subsec:cdccommunication}

In the conventional MapReduce computation framework as shown in Fig.~\ref{fig:conventional}, after the split of the input file into multiple subfiles, each subfile is mapped to only one of the processing nodes, i.e., the workers. The naive replication scheme, i.e., uncoded data shuffling scheme which relaxes this restriction, can reduce the communication costs of the system by allowing each subfile to be replicated and mapped to more than one processing nodes. In the example illustrated in Fig.~\ref{fig:uncoded}, each subfile is repeated twice. Hence, as compared to the conventional MapReduce framework, each processing node has more Map tasks to perform in the naive replication scheme. However, by simple replication, the communication load in the Shuffle phase decreases and this gain is known as the \emph{repetition gain}. Specifically, the communication load for uncoded schemes which include both the conventional MapReduce framework and the naive replication scheme, is denoted as follows \cite{li2018tradeoff}:

\begin{equation}
L_{uncoded}(r)=1-\frac{r}{K},
\label{eqn:uncoded}
\end{equation}
where $K$ is the number of processing nodes in the network. Based on Equation (\ref{eqn:uncoded}), the communication loads achieved by the conventional MapReduce framework in Fig.~\ref{fig:conventional} and the naive replication scheme in Fig.~\ref{fig:uncoded} are $\frac{3}{4}$ where $r=1$ and $\frac{1}{2}$ where $r=2$ respectively.

\begin{figure}
\centering
\includegraphics[width=\linewidth]{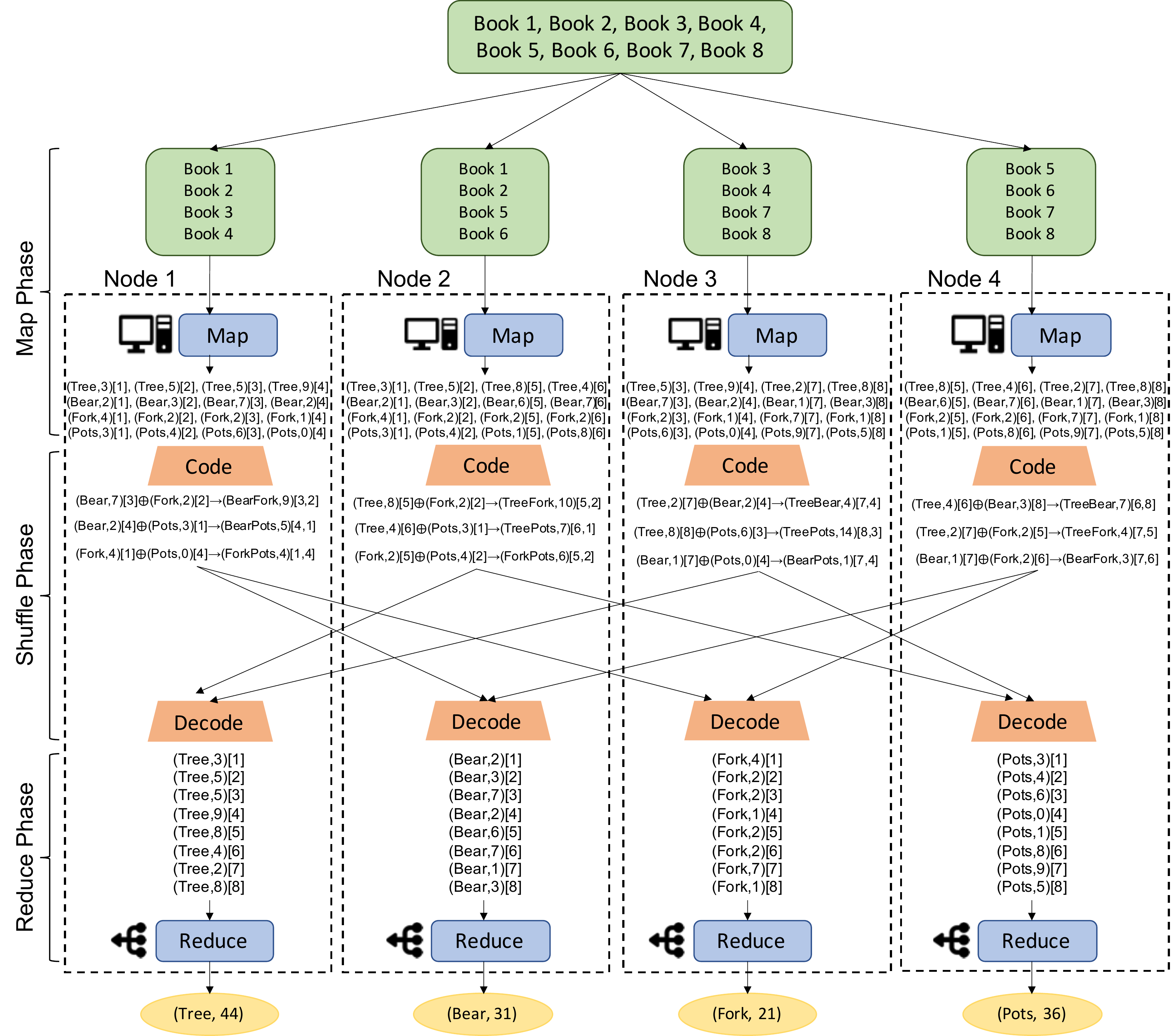}
\caption{\small Illustration of Coded MapReduce framework.}
\label{fig:coded}
\end{figure}

To further reduce the communication load, i.e., to increase the repetition gain, the Coded MapReduce computation framework is proposed in \cite{li2015mapreduce} where the Map tasks are carefully distributed among the processing nodes and the messages are encoded for transmission in the Shuffle phase by using coding theory. Figure~\ref{fig:coded} illustrates the Coded MapReduce framework with 4 processing nodes to determine the 4 output pairs. After the Map phase, the processing node multicasts a bit-wise XOR of two computed intermediate pairs, denoted by $\oplus$, satisfying the requirements of two other processing nodes simultaneously. For example, node 1 multicasts a bit-wise XOR of ``Bear'' and ``Fork'' to both nodes 2 and 3, which involves the transmission of only one packet of information, instead of two packets if the information is sent separately to the nodes in a unicast manner. Since the intermediate output pairs are now coded, there is an additional step of decoding before the reduce functions are applied. Given the coded ``BearFork'' information, node 2 is able to decode and recover the required ``Bear'' information by cancelling the ``Fork'' information since node 2 has also computed the same ``Fork'' information. Similarly, node 3 can recover ``Fork'' information by cancelling the ``Bear'' information. The simulation results in \cite{li2015mapreduce}show that the Coded MapReduce reduces the communication load by 66\% and 50\% as compared to the conventional MapReduce framework and the naive replication scheme respectively.

\begin{figure}
\centering
\includegraphics[width=\linewidth]{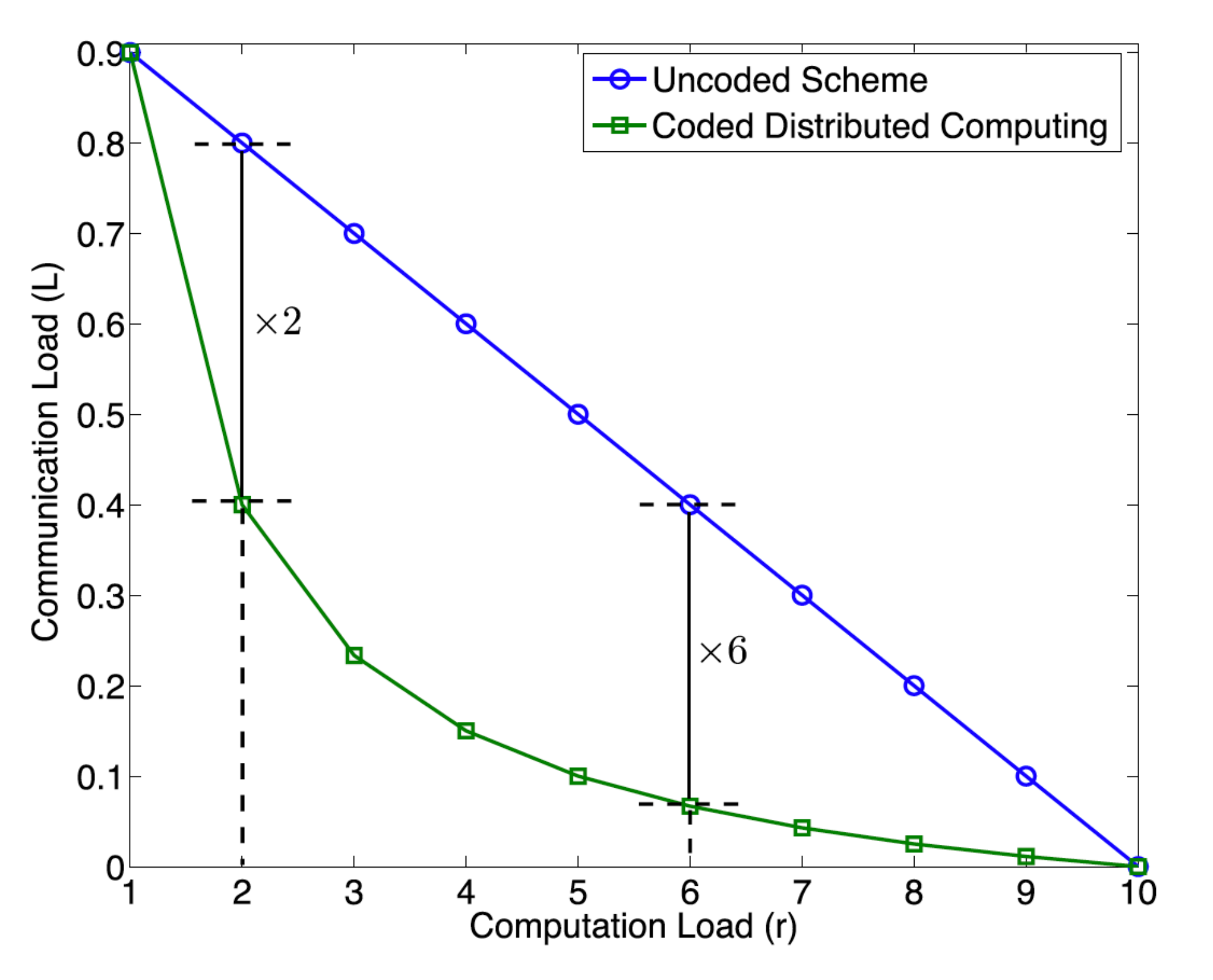}
\caption{\small Comparison of communication load between the CDC scheme and the uncoded scheme \cite{li2018tradeoff}.}
\label{fig:tradeoff}
\end{figure}

Since the use of coding techniques reduces both the latency and cost of the distributed computing systems \cite{aktas2017clones}, a more generalized framework known as the Coded Distributed Computing (CDC) scheme is introduced in \cite{li2018tradeoff}. 
The study of \cite{li2018tradeoff} presents the fundamental inverse relationship between computation load and communication load. Specifically, the communication load in the Shuffle phase can be reduced by a factor $r$ by increasing the computation load in the Map phase by the same factor $r$, as shown in Fig.~\ref{fig:tradeoff}. The communication load achieved by the CDC framework, $L_{coded}$, is given as follows:
\begin{equation}
L_{coded}(r)=\frac{1}{r}\left(1-\frac{r}{K}\right).
\label{eqn:coded}
\end{equation}

Note that the information-theoretic lower bound derived on the minimum communication load $L^*(r)$ equals $L_{coded}(r)$ of the CDC framework. As such, the optimal tradeoff between the computation load and communication load is characterized as follows \cite{li2018tradeoff}:
\begin{equation}
L^*(r)=L_{coded}(r)=\frac{1}{r}\left(1-\frac{r}{K}\right), r \in\mathcal{K}.
\end{equation}

From Equation (\ref{eqn:uncoded}) which applies to the uncoded computation schemes, the communication load $L$ decreases linearly as the computation load $r$ increases. However, when the number of processing nodes $K$ becomes large, there is no significant impact of increasing computation load on the communication load. On the other hand, for the proposed CDC framework, the communication load is inversely proportional to the computation load (Equation (\ref{eqn:coded})). Even when $K$ becomes large, the increase in computation load still significantly reduces the communication load. 

Since the proposed CDC framework can be applied to any distributed computation framework with an underlying MapReduce structure, the performance of the CDC framework on TeraSort \cite{hadoop_terasort} is evaluated. Experimental results on Amazon EC2 clusters show that the Coded TeraSort scheme \cite{li2017terasort} which is a coded distributed sorting algorithm, achieves a reduction in the overall job execution time by factors of 2.16 and 3.39 with 16 processing nodes and computation loads of $r=3$ and $r=5$, respectively as compared to the uncoded TeraSort scheme.

Previously in \cite{li2018tradeoff}, computation load is linearly dependent on the number of replicated Map tasks, i.e., load redundancy, as each processing node is assumed to compute all intermediate values for all subfiles allocated in its memory. However, the processing nodes can be selective in choosing the intermediate values to compute. As such, the load redundancy is no longer a direct measure of computation load. In other words, the storage constraints do not necessarily imply computation constraints. Building on the work in \cite{li2018tradeoff}, the authors in \cite{ezzeldin2017alternative} propose an alternative tradeoff between computation load and communication load under a predefined storage constraint. In fact, the computation load is quadratic in terms of load redundancy. By taking load redundancy into consideration, an alternative computation-communication tradeoff curve is derived. In particular, this alternative tradeoff curve is especially relevant to the processing nodes that do have the sufficient resources or time to perform computations for all the allocated subfiles. Given that the processing nodes can only perform a limited amount of computations below the computation load threshold, the alternative tradeoff curve proposed in \cite{ezzeldin2017alternative} accurately defines the amount of communication load needed for the distributed computation tasks.

Since the processing nodes are not required to compute all intermediate results that can be obtained from their locally stored data, the storage capabilities of the processing nodes should be considered in the CDC computation framework in \cite{li2018tradeoff}. The study of \cite{yan2018storage} characterizes the tradeoff between storage, computation and communication where the minimum communication load is determined given the storage and computational capabilities of the processing nodes. In particular, the optimal computation curve is obtained by characterizing the optimal storage-communication tradeoff given the minimum computation load. As a result, the triangles between the optimal communication curve and the optimal computation curve reflect the pareto-optimal surface of all achievable storage-computation-communication triples. However, as the number of processing nodes in the system increases, the number of input files required increases exponentially, resulting in an increase in the number of transmissions needed and hence high communication costs. As such, it is important to reduce the number of input files, which is discussed in Section~\ref{subsubsec:packet} later.

\begin{table*}[!ht]
\caption{\small Coding techniques to mitigate the straggler effects.} 
\label{tab:stragglers}
\centering
\begin{tabular}{|m{2.0cm}  |>{\centering\arraybackslash}m{1.5cm} |>{\centering\arraybackslash}m{3.3cm} |m{8cm}|}
\hline

\rowcolor{mygray}
\multicolumn{1}{|c|}{\textbf{Problems}}  & \textbf{Ref.} & \textbf{Coding Schemes} & \multicolumn{1}{c|}{\textbf{Key Ideas}}  \\ \hline

\multirow{8}{=}{Matrix-Vector}  & \cite{lee2018speeding} & MDS Codes & Reduce the computation latency as the master node is able to recover the final result without waiting for the slowest processing node \\
\cline{2-4}
 & \cite{mallick2019rateless} & LT Codes & Exploit the rateless property to generate unlimited number of encoded symbol from a finite set of source symbols\\
\cline{2-4}
 & \cite{dutta2016shortdot} & Short-Dot Codes & Reduce the length of dot-products computed at the processing nodes by introducing sparsity to the encoded matrices\\
\cline{2-4}
 & \cite{wang2018fundamental} & s-diagonal codes & Exploit the diagonal structure of the matrices to achieve both optimal recovery threshold and optimal computation load\\
\hline

\multirow{16}{=}{Matrix-Matrix} & \cite{lee2017high} & Product Codes & Instead of encoding the matrices along one dimension as in the MDS-coded schemes, the matrices are encoded with MDS codes along both dimensions, i.e. row and column  \\
\cline{2-4}
  & \cite{yu2017polynomial} & Polynomial Codes & - Design the algebraic structure of the encoded matrices such that the MDS structure is found in both the encoded matrices and the intermediate computations\\
 & & & - Reconstruct the final results by solving the polynomial interpolation problem\\
\cline{2-4}
  & \cite{fahim2017optimal} & MatDot Codes & Achieve lower recovery threshold than Polynomial Codes \cite{yu2017polynomial} at the expense of higher communication costs by computing only the relevant cross-products\\
\cline{2-4}
  & \cite{fahim2017optimal} & PolyDot Codes & Characterize the tradeoff between recovery threshold and communication costs where Polynomial Codes \cite{yu2017polynomial} and MatDot Codes are the two extreme ends on this tradeoff curve\\
\cline{2-4}
  & \cite{wang2018coded} & Sparse Codes & Exploit to sparsity in both input and output matrices to reduce computation load, while achieving near optimal threshold\\
\hline

\multirow{15}{=}{Gradient Descent}  & \cite{tandon2017gradient} & Fractional Repetition Coding & - Workers are divided into multiple groups and data is divided among the workers in each group \\
 & & & - Each partition of data is performed by more than one worker\\
\cline{2-4}
  & \cite{tandon2017gradient} & Cyclic Repetition Coding & Data is allocated based on a cyclic assignment strategy\\
\cline{2-4}
  & \cite{raviv2017gradient}& Cyclic MDS Codes & The entries to the columns of the encoding matrix are cyclic shifts of the entries to the first column \\
\cline{2-4}
  & \cite{halbawi2018solomon} & Reed-Solomon Codes & Use balanced mask matrix and chooses appropriate codewords from the RS codes to construct the encoding matrix \\
\cline{2-4}
  & \cite{li2018near} & Batch Coupons Collector & - Divide the data into multiple batches which are allocated randomly to the workers for computations\\
 & & & - communication costs are significantly reduced as there is no need for communication between the workers and for feedback from the master node to the workers\\
\cline{2-4}
  & \cite{li2018polynomially} & Polynomially Coded Regression & Encode the data batches directly instead of the computed intermediate results\\
\hline

Convolution  & \cite{dutta2017convolution} & Coded Convolution & Split both vectors into multiple parts of specified length and encodes one of the vectors with MDS codes \\
\hline

Fourier Transform  & \cite{yu2017fourier} & Coded Fourier Transform & Leverage on recursive structure and the linearity of the discrete Fourier Transform operations\\
\hline

\end{tabular}
\end{table*}

\begin{figure}
\centering
\includegraphics[width=\linewidth]{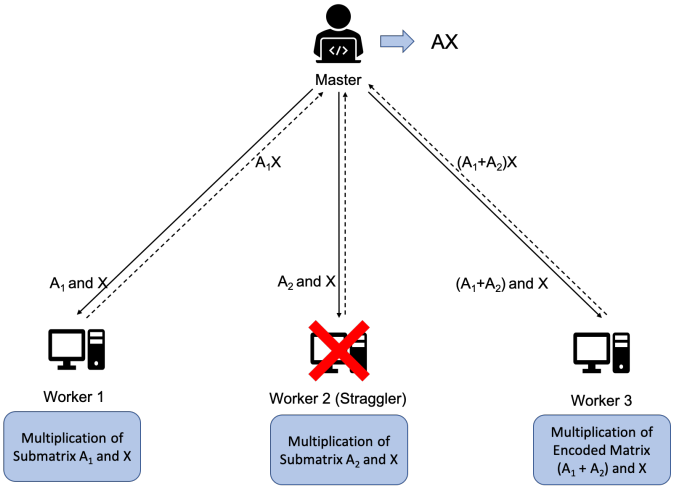}
\caption{\small Illustration of coded computation with 3 workers. The master node is able to recover the final result upon receiving the computed results from any 2 workers.}
\label{fig:codedcomputation}
\end{figure}

\subsection{CDC to Mitigate the Straggler Effects}
\label{subsec:codingtech}

Apart from reducing communication load in the Shuffle phase of the MapReduce framework, coding techniques can also be used to alleviate the straggler effects. Since matrix multiplication is one of the most basic linear operations used in distributed computing systems, a coded computation framework is proposed in \cite{lee2018speeding} to minimize computation latency of distributed matrix multiplication tasks. The coded computation framework uses erasure codes to generate redundant intermediate computations. In particular, 
the master node encodes the equal-sized data blocks, i.e., submatrices and distributes them to the workers to compute the local functions. Upon completion, the workers transmit the computed results to the master node. The master node can recover the final result by using the decoding functions once the local computations from any of the decodable sets are completed. As seen in Fig.~\ref{fig:codedcomputation}, the master node can recover the final result upon receiving the computed results from any 2 workers, instead of all 3 workers. As such, the total computation is not determined by the slowest straggler, but by the time when the master node receives computed results from some decodable set of indices. In this work \cite{lee2018speeding}, the authors explore the effectiveness of encoding the submatrices by using maximum distance separable (MDS) codes \cite{blaum1999lowest} to mitigate the effects of stragglers.

Considering $K$ workers and a shifted-exponential distribution for the job execution time of the distributed algorithm, the simulation results show that the optimal repetition-coded distributed algorithm achieves a lower average job execution time when the straggling parameter is smaller than one, i.e., $\mu<1$ but is still slower than the optimal MDS-coded distributed algorithm by a factor of $\Theta (\log K)$. However, the storage cost of the coded distributed algorithm is higher than that of the uncoded distributed algorithm as more data is required to be stored at the workers' sites for the coded distributed algorithm. The proposed algorithm is tested on an Amazon EC2 cluster and is compared against various parallel matrix multiplication algorithms, e.g., block matrix multiplication, column-partition matrix multiplication and row-partition matrix multiplication. The simulation results show that the proposed algorithm in \cite{lee2018speeding} performs better where the coded matrix multiplication achieves $40\%$ and $39.5\%$ reduction in average job execution time on clusters of m1-small and c1-medium instances with 10 workers each respectively as compared to the best of the three uncoded distributed algorithms. 

Although the MDS codes proposed in \cite{lee2018speeding} is able to mitigate the straggler effects, it cannot be generalized to all types of computation tasks. In order to mitigate the straggler effects of different distributed computation tasks, the coding techniques can be designed by exploiting the algebraic structures of the specific operations. An important performance metric that is introduced in the proposed CDC schemes is the recovery threshold, which refers to the worst-case required number of workers the master needs to wait to recover the final result for job completion \cite{yu2017polynomial}. The smaller the recovery threshold, the shorter the computation latency. The objective is to reduce the recovery threshold so that the final result can be recovered by waiting for a smaller number of workers, thus contributing to a reduction in computation latency. Here, we discuss the coding techniques for various types of computation tasks, namely (i) matrix-vector multiplications, (ii) matrix-matrix multiplications, (iii) gradient descent, (iv) convolution and Fourier transform. Table~\ref{tab:stragglers} summarizes the coding techniques designed for different distributed computation tasks.

\subsubsection{Matrix-vector multiplications}Distributed matrix-vector multiplications are the building blocks of linear transformation computations which are an important step in machine learning and signal processing applications. In particular, the computation of linear transformation on high-dimensional vectors is required for popular dimensionality reduction techniques such as Linear Discriminant Analysis (LDA) \cite{balakrishnama1998linear} and Principal Component Analysis (PCA) \cite{abdi2010principal}.

Instead of using MDS codes that is proposed in \cite{lee2018speeding}, the authors in \cite{mallick2019rateless} propose the use of Luby Transform (LT) codes to mitigate the straggler effects in distributed matrix-vector multiplications problems. Different from the works in \cite{severinson2019block} and \cite{wang2018coded} which use LT codes in fixed-rate settings, the rateless property of the LT codes can be exploited to generate unlimited number of encoded symbol from a finite set of source symbols. There are several advantages of using rateless codes: (i) near-ideal load balancing, (ii) negligible redundant computation, (iii) maximum straggler tolerance, and (iv) low decoding complexity. To further reduce the latency for practical implementations, blockwise communication can be used to transmit the submatrix-vector products. Instead of transmitting each encoded row-vector product separately to the master node, the workers are allowed to transmit the computed results in blocks where each block comprises a few row-vector products, reducing the number of communication rounds needed and hence minimizing the time needed to complete the computation tasks. 

 The authors in \cite{dutta2016shortdot} propose Short-Dot codes to perform the computation of linear transforms reliably and efficiently under the presence of straggling nodes. Specifically, the processing nodes compute shorter dot products by imposing sparsity on the encoded submatrices. However, there is tradeoff between the optimal threshold recovery and the length of the dot-products where the master node needs to wait for computed results from more processing nodes when the length of the dot-products is shorter. The experimental results on the classification of hand-written digits of MNIST show that the Short-Dot codes achieve 32\% faster expected computation time than the MDS codes \cite{lee2018speeding}.

Although the Short-Dot codes \cite{dutta2016shortdot} can offer lower recovery threshold, the greater length of the dot-products means greater computation load for the processing nodes. With this concern, s-diagonal codes \cite{wang2018fundamental} are proposed to achieve both optimal recovery threshold and optimal computation load by exploiting the diagonal structure of the encoding matrix. The computation time can be further reduced by using a low-complexity hybrid decoding algorithm which combines the peeling decoding algorithm and Gaussian elimination techniques.

\subsubsection{Matrix-matrix multiplications}For large-scale distributed matrix-matrix multiplications, the coded computation schemes based on MDS codes are no longer suitable as the encoding and decoding processes scale with system size. Besides, the size of one of the matrices is assumed to be small enough in order to allow individual workers to perform the computations \cite{lee2017high}, restricting the implementation of MDS codes in large-scale multiplications. Hence, for large-scale problems, coded schemes not only need to achieve low computation time, but also require efficient encoding and decoding algorithms in order to minimize the overall job execution time. To deal with the straggler effects in high-dimensional distributed matrix multiplications, four types of coded computation schemes are proposed:

\begin{itemize}
\item \emph{Product codes \cite{lee2017high}: }Product codes are implemented by building a larger code upon smaller MDS codes. Instead of encoding computations along only one dimension in MDS-coded schemes, the product codes encode computations along both dimensions, i.e., rows and columns of the matrices. When the number of backup workers increases sub-linearly with the number of subtasks, the product-coded schemes outperform the MDS-coded schemes in terms of average computation time and decoding time. In the linear regime, the one-dimensional decoding of the MDS-coded schemes is sufficient to recover the missing entries of the computation results. By allowing each row and column of the MDS constituent codes to have different code rates \cite{park2019irregular}, the average computation time can be further reduced, contributing to a decrease in the overall job execution time. Product codes can also be used to solve higher-dimensional linear operations such as tensor operations by exploiting the tensor-structured encoding matrix \cite{baharav2018proofing}. To reduce the decoding time of the product codes, efficient decoding algorithms such as Reed-Solomon codes and LDPC codes can be explored.

\item \emph{Polynomial codes \cite{yu2017polynomial}: }The key advantage of the polynomial codes in solving large-scale matrix multiplication problems is that they provide a lower bound to the optimal recovery threshold. For polynomial codes, the recovery threshold does not scale with the number of workers involved where as for MDS codes and the product codes, the recovery thresholds scale proportionally with the number of workers. By taking advantage of the algebraic structure of the polynomial codes, the master node can recover the final result by using polynomial interpolation algorithms, e.g., the Reed-Solomon codes, to decode the computation results from the workers. In addition to the optimal recovery threshold, the polynomial-coded schemes achieve minimum possible computation latency and communication load for distributed matrix multiplications. However, as the number of workers increases, the encoding and decoding costs are much higher than that of the product codes. Furthermore, by implementing Reed-Solomon codes, there is a limit to the number of workers that can be handled, which is not useful for practical implementations where the systems may involve up to thousands of nodes. As an extension to the polynomial codes proposed in \cite{yu2017polynomial}, the entangled polynomial code that is proposed in \cite{yu2020fundamental} achieves a lower recovery threshold which is only half of that achieved by the PolyDot codes \cite{fahim2017optimal}, to be discussed later. Different from the polynomial codes which only allow column-wise partitioning of the matrices, the entangled polynomial codes allow arbitrary partitioning of the input matrices and evaluate only a subspace of bilinear functions such that unnecessary multiplications are avoided. The issue of numerical stability has also received attention to ensure the scalability of the polynomial-coded schemes \cite{fahim2019numerically}. 

\item \emph{PolyDot codes \cite{fahim2017optimal}: }PolyDot codes characterize the tradeoff between the recovery threshold and communication costs where the polynomial codes and the MatDot codes are special instances of this coding framework by considering two extreme ends of this tradeoff: minimizing either recovery threshold or communication costs. In particular, the MatDot codes achieve lower recovery threshold than the polynomial codes at the expense of much higher communication costs. This is achieved by only computing the relevant cross-products of the submatrices. Building on the work of PolyDot codes \cite{fahim2017optimal}, the Generalized PolyDot codes \cite{dutta2018polydot} are used to compute matrix-vector multiplications and achieve the same recovery threshold as the entangled polynomial codes \cite{yu2020fundamental}. More importantly, the Generalized PolyDot codes can be  extended for the training of large deep neural networks (DNNs), which consists of multiple non-linear layers. 

\item \emph{Sparse codes \cite{wang2018coded}: }Although the polynomial codes \cite{yu2017polynomial} achieve optimal recovery threshold, the computation loads of the workers increase due to the increased density of the input matrix, resulting in an increase in the overall job execution time which is not desirable. By exploiting sparsity, i.e., the number of zero entries of the encoded matrix, not only the recovery threshold is kept low, but the computation loads of the workers also decrease while maintaining a nearly linear decoding time \cite{suh2017sparse}, jointly contributing to the shorter overall job execution time. The basic idea of the algorithm proposed in \cite{suh2017sparse} is to allow the master node to find the linear combination of row vectors such that only the particular relevant sub-blocks are recovered. Then, the entire block of matrix can be recovered by aggregating the partial recovery of sub-blocks. Simulation results show that the sparse codes require an overall shortest time to complete the job as compared to other computation schemes, e.g., uncoded scheme, product codes \cite{lee2017high}, polynomial codes \cite{yu2017polynomial}, sparse MDS codes \cite{dutta2016shortdot} and LT codes \cite{mallick2019rateless}. A further analysis of the different components of subtasks, i.e., communication time, computation time and decoding time, shows that the sparse codes require much shorter time to decode, thus contributing significantly to the shorter overall job execution time.

\end{itemize}

\begin{figure}
\centering
\includegraphics[width=\linewidth]{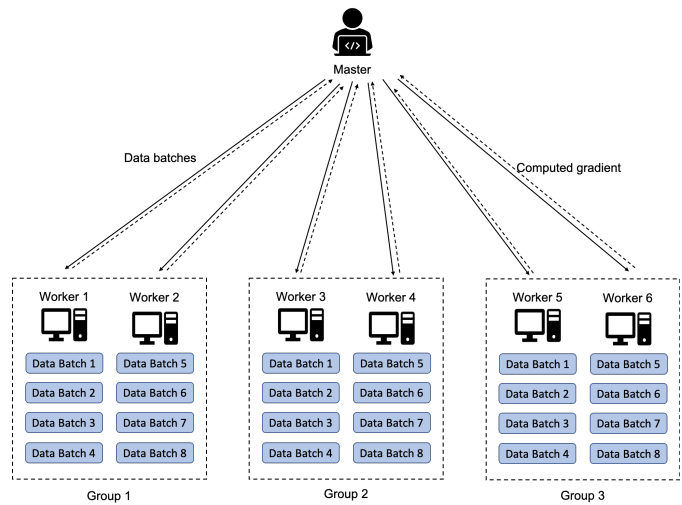}
\caption{\small Fractional repetition coding with 6 workers and 2 stragglers.}
\label{fig:frc}
\end{figure}

\subsubsection{Gradient Descent}Apart from matrices, coding techniques can be applied to recover batch gradients of any loss function of the distributed gradient descent tasks. In \cite{tandon2017gradient}, the authors have introduced the idea of gradient coding which is useful to mitigate stragglers that may slow down the computation tasks. Two gradient coding schemes are proposed, namely (i) fractional repetition coding (FRC) and (ii) cyclic repetition coding. In the FRC scheme, the workers are first divided into several groups. In each group, the data is equally divided and allocated to the workers. As a result, all groups of workers are replica of each other as shown in Fig.~\ref{fig:frc}. Upon completing their subtasks, the workers in each group transmit the sum of partial gradients to the master node. In the cyclic repetition coding scheme, the data partitions are allocated to the workers based on a cyclic assignment strategy. The partial gradients computed by each worker are encoded by linearly combining them, of which is transmitted as a single coded message to the master node. By applying the gradient coding schemes, the distributed computation tasks do not suffer from delays incurred by the straggling nodes as the master node is able to recover the final result with the results from the non-straggling nodes. Other coding theories such as the cyclic MDS codes \cite{raviv2017gradient} and the Reed-Solomon codes \cite{halbawi2018solomon} can be used to compute exact gradients of the distributed gradient descent problems. 

To efficiently mitigate the straggler effects in distributed gradient descent algorithms, Batch Coupon's Collector (BCC) scheme is proposed in \cite{li2018near}. In BCC, there are two important steps, namely (i) batching and (ii) coupon collecting. In batching, the training set is partitioned into batches which are distributed to the workers randomly whereas in coupon collecting, the master node collects the computed results from the workers until the results from all batches of data are received. This decentralized BCC scheme does not require any communication between the workers nodes and each worker is allocated data batches independently of other workers. As a result, it is easy to implement the BCC scheme in practical scenarios. Another important advantage of the BCC scheme is its universality. Different from other coding schemes that are designed to guarantee their robustness to a fixed number of straggling nodes, the BCC scheme does not require any prior knowledge about the straggling nodes which is more practical as it is difficult to estimate the number of straggling nodes present in the clusters. Furthermore, the BCC scheme can be easily extended to solve gradient descent problems over heterogeneous clusters where the workers have different computational and communication capabilities. The simulation results show that the BCC scheme speeds up the overall job execution time by up to 85.4\% and 69.9\% over the uncoded scheme and the cyclic repetition coding scheme \cite{tandon2017gradient} respectively.

The gradient coding schemes proposed in \cite{tandon2017gradient} illustrate the tradeoff between computation load and straggler tolerance. However, in non-linear learning tasks, communication costs dominate the overall job execution time as the number of iterations increases. As such, to generalize the coding schemes in \cite{tandon2017gradient}, the authors in \cite{ye2018communicationcomputation} incorporate communication costs into their framework and present a fundamental tradeoff between the three parameters, namely computation load, straggler tolerance and communication costs. In particular, for a fixed computation load, the communication costs can be reduced by waiting for more workers.

Instead of encoding the partial gradients computed based on uncoded data as seen in the studies of \cite{tandon2017gradient,raviv2017gradient,halbawi2018solomon}, coding techniques can be applied directly to the data batches to reduce the straggler effects and the overall job execution time. Considering the gradient computations for least-square regression problems, the polynomially coded regression (PCR) scheme \cite{li2018polynomially} exploits the underlying algebraic property to generate coded submatrices such that they are linear combinations of the uncoded input matrices. The master node can evaluate the final gradient by interpolating the polynomials from the computed partial gradients by the workers. Compared to the gradient coding schemes proposed in \cite{tandon2017gradient}, the simulation results show that the PCR scheme achieves much lower recovery threshold and hence shorter computation and communication time, resulting in shorter time needed for overall job execution. 

\subsubsection{Convolution and Fourier transform}The polynomial codes proposed in both the studies of \cite{yu2017polynomial} and \cite{yu2020fundamental} can be extended to the applications of distributed coded convolution based on the coded convolution scheme proposed in \cite{dutta2017convolution}. The work in \cite{dutta2017convolution} explores the use of MDS codes to encode the pre-specified vectors such that fast convolution is performed under deadline constraint. In addition to that, MDS codes can be used to mitigate the straggler effects in widely-implemented distributed discrete Fourier Transform operations \cite{yu2017fourier} which are used in many applications such as machine learning algorithms and signal processing frameworks.

\subsection{Unified CDC Scheme}
\label{subsec:unified} 

Given the aforementioned coding schemes of \cite{li2018tradeoff} and \cite{lee2018speeding}, we can observe that coding techniques are used to speed up distributed computing applications with two different approaches. On one hand, the authors in \cite{li2018tradeoff} propose the ``Minimum Bandwidth Code" that minimizes the communication load by repeating the computation tasks in the Map phase to introduce multicasting opportunities in the Shuffle phase. On the other hand, the authors in \cite{lee2018speeding} propose the ``Minimum Latency Code" which minimizes the computation latency by encoding the Map tasks such that the master node is able to recover the final result without waiting for the straggling processing nodes.

Inspired by the aforementioned approaches, a unified coded scheme that characterizes the tradeoff between computation latency and communication load is proposed in \cite{li2016unified} given the computation load. The coding schemes in \cite{li2018tradeoff} and \cite{lee2018speeding} are considered to be the extreme cases of this unified scheme. The unified coded scheme exploits the advantages of the two coding approaches by applying the MDS codes to the Map tasks and replicating the encoded Map tasks. Specifically, the unified coded scheme first encodes the rows of the matrix, following which the coded rows of the matrix are replicated and stored at the processing nodes in a specific pattern. Then, the processing nodes perform the computation until a certain number of the fastest processing nodes complete their tasks. To reduce communication load in the Shuffle phase, coded multitasking is used to exchange the intermediate results that are needed to recover the final results in the Reduce phase. An improvement to the latency-communication tradeoff presented in the unified coded scheme \cite{li2016unified} is proposed in \cite{zhang2019improved} by leveraging on the redundancy created by the repetition code. By increasing the redundancy rate of the repetition code, both the communication load in the Shuffle phase and the computation latency in the Map phase can be simultaneously improved, thus contributing to a improved latency-communication tradeoff.


The aforementioned initial works of coding schemes have shown their effectiveness in minimizing communication costs and alleviating the straggler effects. In the following sections, we review related works that leverage on coding techniques to address the implementation challenges of the distributed computing systems.

\section{Minimization of Communication Load}
\label{sec:comms}

With more computing nodes that are equipped with greater capabilities to collect and process data, massive amounts of data are generated for the computations of user-defined computation tasks. Since the computations are scaled out across a large number of distributed computing nodes, large number of intermediate results need to be exchanged between the computing nodes in the Shuffle phase of the MapReduce framework to complete the computation tasks, resulting in significant data movement. Oftentimes, for the training of a model with distributed learning algorithms, data is shuffled at each iteration, contributing to high communication costs, which is a bottleneck of the distributed computing systems. As a result, there is a need to reduce the communication costs in order to speed up the distributed computation tasks. In this section, we present four approaches to reduce communication costs: 

\begin{itemize}

\item \emph{File Allocation: }In this approach, the studies aim to design an optimal file allocation strategy that considers the heterogeneities of the processing nodes' capabilities in the systems, maximizing data locality or reducing subpacketization level, which refers to the number of subfiles generated \cite{parrinello2018}, \cite{konstantinidis2018leveraging}. These different approaches work towards reducing the communication load in the distributed computing systems.

\item \emph{Coded Shuffling Design: }Since data shuffling phase incurs a large proportion of the communication costs, data is encoded before it is transmitted so that the communication load can be minimized. Apart from combining coding with different techniques, e.g., compression and randomization techniques to improve the performance of the shuffling phase, the coding techniques are also designed to solve different computation problems, e.g., distributed graph computation problems \cite{prakash2018graph}, \cite{srini2018random} and multistage MapReduce computations \cite{li2016multistage}.

\item \emph{Consideration of Underlying Network Architecture: }Generally, the communications between the workers as well as between the workers and the master node are affected by the way that they are connected to each other. For example, server-rack architecture (Fig.~\ref{fig:server}) is one of the most commonly used methods to connect the various servers. By taking the underlying architecture into consideration, the effectiveness of the coding implementation in reducing communication costs can be greatly improved.

\item \emph{Function Allocation: }Similar to the allocation of files, the studies apply this approach on heterogeneous systems. In addition, some studies consider a cascaded system \cite{woolsey2019cascaded} where each Reduce function is allowed to be computed at multiple processing nodes. In some cases where the data is randomly stored at the processing nodes, e.g., when the processing nodes are constantly moving, an optimal function allocation strategy is useful in reducing the number of broadcast transmissions and thus minimizing the communication load. 

\end{itemize}

\subsection{File Allocation}
\label{subsec:files}

The design of file allocation at each processing node is one of the major steps for the implementation of CDC scheme. There are a few approaches to an optimal file allocation strategy: (i) considering heterogeneous systems, (ii) maximizing data locality, and (iii) reducing subpacketization level.

\subsubsection{\textit{Considering Heterogeneous Systems}}As discussed in Section~\ref{subsec:cdccommunication}, although the CDC scheme proposed in \cite{li2018tradeoff} carefully allocates the subfiles to the processing nodes in order to introduce coded multitasking opportunities, it considers a homogeneous system which may not be useful for practical implementation. In order to appropriately allocate the files to the distributed computing nodes, heterogeneous systems where the processing nodes have diverse storage, computational and communication capabilities, should be considered in determining the optimal file allocation strategy and coding scheme that minimize the communication load \cite{kiamari2017heterogeneous}. 

By leveraging on the extra storage capacity of the workers, the communication costs between the master node and the workers in the process of data shuffling are minimized. The reason is that if more data can be stored at the workers, fewer communication rounds are needed for the workers to receive shuffled data from the master node. In the extreme case, if the worker can store the entire dataset, there is no communication needed for the worker to receive shuffled data in any iteration. As a result, there is a tradeoff between the storage capacity of the workers and the communication overhead in the data shuffling process. In the data shuffling process, there are two phases, namely data delivery and storage update. Instead of a random storage placement \cite{lee2018speeding}, a deterministic and systematic storage update strategy \cite{attia2016theoretic} creates more coding opportunities in transmitting data to the workers at each iteration, reducing the communication load.

\subsubsection{\textit{Maximizing Data Locality}}One of the important factors in determining the optimal file allocation strategy is data locality. Data locality is defined as the percentage of local tasks over the total number of Map tasks, i.e., the fraction of Map tasks that are allocated to the processing nodes having the required data for computations such that no communication is needed to obtain the data. High data locality means that less communication bandwidth is needed for the transmission of subfiles, which is required if the processing node does not have the needed subfiles for the execution of the Map tasks. In order to maximize data locality, the problem of allocation of Map tasks to different processing nodes can be tackled by solving a constrained integer optimization problem \cite{gupta2017locality}.

\subsubsection{\textit{Reducing Subpacketization Level}} \label{subsubsec:packet}As the number of processing nodes in the network increases, the input file needs to be split into a large number of subfiles. Specifically, the number of subfiles generated increases exponentially in the number of processing nodes \cite{parrinello2018}. However, there is a maximum allowable subpacketization level, i.e., number of subfiles, where the dataset can only be partitioned into a limited number of packets, beyond which the communication load increases due to more transmissions required and the unevenly-sized intermediate results. Hence, there are several reasons for the reduction of subpacketization level: (i) to reduce the communication load in the Shuffle phase even when there is a large number of processing nodes, (ii) to reduce the packet overheads which increases with the number of broadcast transmissions, and (iii) to reduce the number of unevenly-mapped outputs which require zero padding. To keep the subpacketization level below the maximum allowable level, Group-based Coded MapReduce \cite{parrinello2018} allocates the dataset based on the random groupings of the processing nodes and allows the processing nodes to cooperate in the transmission of messages. 

To avoid splitting the input file too finely, the authors in \cite{konstantinidis2018leveraging} use an appropriate resolvable design \cite{stinson2007combinatorial}, which is based on linear error correcting codes, to determine the number of subfiles, the allocation of the subfiles to the processing nodes and the construction of the coded messages in the Shuffle phase. Building on this initial work, the authors in \cite{konstantinidis2019resolvable} use the resolvable design based scheme to solve the limitation of the compressed CDC scheme \cite{li2018compressed} that uses both compression and coding techniques. Although the compressed CDC scheme helps to reduce communication load, it requires large number of jobs to be processed simultaneously. Hence, the resolvable design based scheme is used to reduce the number of subfiles generated. Specifically, for each job in the compressed CDC scheme, the single-parity code is used to split the input file and the resolvable design based scheme is used to allocate the subfiles to the processing nodes. By aggregating the underlying functions and applying the resolvable design based scheme, multiple jobs can be processed in parallel while minimizing the execution time in the Shuffle phase, contributing to the reduction of the overall job execution time. Although the number of subfiles or number of jobs generated still increases exponentially with some of the system parameters, e.g., number of computing nodes and number of output functions, the exponent is much smaller when the resolvable design based scheme is implemented.


In addition to the exponential increase in the number of subfiles required, the number of output functions required also increases exponentially as the number of processing nodes in the network increases. There are other methods to reduce the number of subfiles and the number of output functions such as the hypercube computing scheme \cite{woolsey2018combinatorial} and the placement delivery array (PDA) \cite{jiang2020coded, yan2018pda, ramkumar2019pda}. However, most of the CDC schemes consider non-cascaded systems, i.e., each Reduce function is computed at exactly one processing node \cite{konstantinidis2019resolvable}. In \cite{woolsey2018combinatorial}, a cascaded system is considered but only two values for the number of processing nodes that perform each Reduce function are considered. By applying the concept of PDA to the distributed computation framework, the performance of the proposed computation scheme is evaluated for different number of processing nodes that compute the Reduce functions \cite{jiang2020coded}. Although the implementation of these various methods reduces the number of subfiles generated, it may come at the expense of higher communication load \cite{konstantinidis2018leveraging}, \cite{ramkumar2019pda}.

\subsection{Coded Shuffling Design}

In the design of coded shuffling algorithms, we have classified the approaches into three different categories: (i) compression and randomization, (ii) coding across multiple iterations and (iii) problem-specific coding approaches.

\subsubsection{Compression and randomization}To further reduce the communication costs of the distributed computation tasks, the design of the coded data shuffling scheme can incorporate different techniques to create more coded multicasting opportunities. Besides, the coded shuffling schemes are designed to minimize communication costs for different distributed computation problems such as iterative algorithms, graph computations and multistage dataflow problems. 

The work in \cite{li2018tradeoff} generates replications of the computation tasks in the Map phase in order to reduce communication load in the Shuffle phase by coding and multicasting the intermediate results. To further reduce the communication load, compression and randomization techniques can be applied to the design of the coded shuffling algorithms. 

\begin{itemize}
\item \emph{Compression Techniques: }The compressed CDC computation scheme is proposed in \cite{li2018compressed} by jointly using two techniques, i.e., compression and coding techniques. Each processing node first computes the allocated Map tasks and generates the intermediate results. By using the compression techniques, several intermediate results of a single computation task are compressed into a single pre-combined value. The communication bandwidth needed to transmit a single pre-combined value is much smaller than that of transmitting several uncombined intermediate values since the size of the pre-combined value equals the size of only one intermediate value. With the pre-combined values from different computation tasks, the processing node codes them for multicasting to other processing nodes simultaneously. There are two advantages to this compressed CDC scheme: (i) the communication load is reduced proportional to the storage capacity of each processing node, and (ii) the communication load does not scale linearly, i.e., slower than linear, with the size of the dataset.

In some cases, e.g., parallel stochastic gradient descent (SGD) algorithms, instead of transmitting intermediate results, computed gradient updates are exchanged among the workers. In such cases, Quantized SGD \cite{alistarh2017qsgd}, a compression technique, can be used to reduce communication bandwidth used during the gradient updates between the processing nodes. In each iteration, the processing nodes are allowed to adjust the number of transmitted bits by quantizing each component to a discrete set of values and encoding these quantized gradients.


\item \emph{Randomization Techniques: }Instead of introducing coded multicasting opportunities to reduce the communication load in the Shuffle phase, there are other coding techniques that can be applied to increase efficiency of data shuffling. One of the ways is to perform a semi-random data shuffling and coding scheme based on pliable index coding which introduces randomization in the data shuffling process \cite{song2020pliable}. There are two important modifications made to the conventional pliable index coding scheme \cite{brahma2015pliable}, which is used to minimize the number of broadcast transmissions while satisfying users' demands. Firstly, the correlation of messages between workers is reduced. In order to do so, a message should only be transmitted to a fraction of the workers so that the same message is not held by all workers. As such, the pliable index coding problem is formulated with an objective where the goal is to minimize the number of broadcast transmissions under the constraint of a maximum number of workers that can receive the same message. Secondly, the correlation of messages between iterations is reduced. The reduction of correlation of messages prevents the workers from performing computations on the same dataset after shuffling, which may be redundant. A two-layer hierarchical structure is proposed for data shuffling. In the upper layer, the messages are partitioned into multiple groups of which each group of messages is transmitted to a fraction of workers. In the lower layer, each group of messages and the corresponding allocated workers are formulated as a constrained pliable index coding problem. Randomization occurs in two stages: (i) when the master node selects the messages in each group and transmits them to the workers, and (ii) when the workers discard old messages from their cache. Experimental results 
show that the proposed pliable index coding requires only 12\% of broadcast transmissions needed by an uncoded scheme, i.e., random shuffling with replacement scheme.

\end{itemize}

\subsubsection{Coding across multiple iterations}Most works on coded iterative algorithms focus on the optimization of a single computation iteration or the minimization of communication load in a single communication round \cite{tandon2017gradient,raviv2017gradient,halbawi2018solomon,charles2017approximate}. However, multiple rounds of communications are generally required to solve distributed iterative problems. In the studies of \cite{haddadpour2018cross} and \cite{haddadpour2018stragglerresilient}, the results of multiple iterations are transmitted in a single round of communication by jointly coding across several iterations of the distributed computation task. By leveraging on the computation and storage redundancy of the workers, the number of communication rounds between the master node and the workers is greatly reduced, resulting in a reduction in the communication costs. However, the computation and storage costs may not be optimal as compared to uncoded computing schemes, e.g., \cite{zhang2013statistical} and \cite{martin2010parallelized}, which achieve near-optimal computation and storage costs. 

\subsubsection{Problem-specific coding approaches}Apart from the typical distributed computation problems, the MapReduce framework can also be used to solve distributed graph computation problems\cite{prakash2018graph}, \cite{srini2018random}. However, for graph computing systems, the computation at each vertex is a function of the graph structure which means that each computation only needs data from its neighbouring vertices. More specifically, the communication load in the Shuffle phase depends on the connectivity probabilities of vertices in the graph, in which each vertex is only allowed to communicate reliably with a subset of random vertices \cite{srini2018random}. As a result, the CDC scheme proposed in \cite{li2018tradeoff} (which was previously discussed in Section~\ref{subsec:cdccommunication}) is not applicable to solving the graph computation problems. Looking into this, the authors in \cite{prakash2018graph} propose a coded scheme to solve the problem of random Erdös-Rényi graph while minimizing the communication load in the Shuffle phase. A similar inverse tradeoff curve between the computation load and average communication load is obtained by using coding techniques in solving distributed graph computations. 

Moreover, many distributed computing applications consists of multiple stages of MapReduce computations. The multistage data flow can be represented by a layered DAG \cite{li2016multistage} in which the processing nodes, i.e., vertices in a particular computation stage are grouped into a single layer. Each vertex computes a user-defined function, i.e., Map or Reduce computation that transforms the given input files into intermediate results whereas the edges represent data flow between the processing nodes. By exploiting the redundancy of the computing nodes, coding techniques are applied to the processing nodes individually to minimize the communication costs. The proposed work in \cite{li2016multistage} considers a uniform resource allocation strategy where the computation of each vertex is distributed across all processing nodes. However, the communication load can be further reduced by reducing the number of processing nodes that are used to compute each vertex. Given that fewer processing nodes are used to compute each vertex, the computation load performed by each processing node increases and the processing nodes have more local information, thus reducing the need for communication to obtain the required information. Therefore, a dynamic resource allocation strategy is needed to further minimize the communication load in the multistage MapReduce problems.


\begin{figure}
\centering
\includegraphics[width=\linewidth]{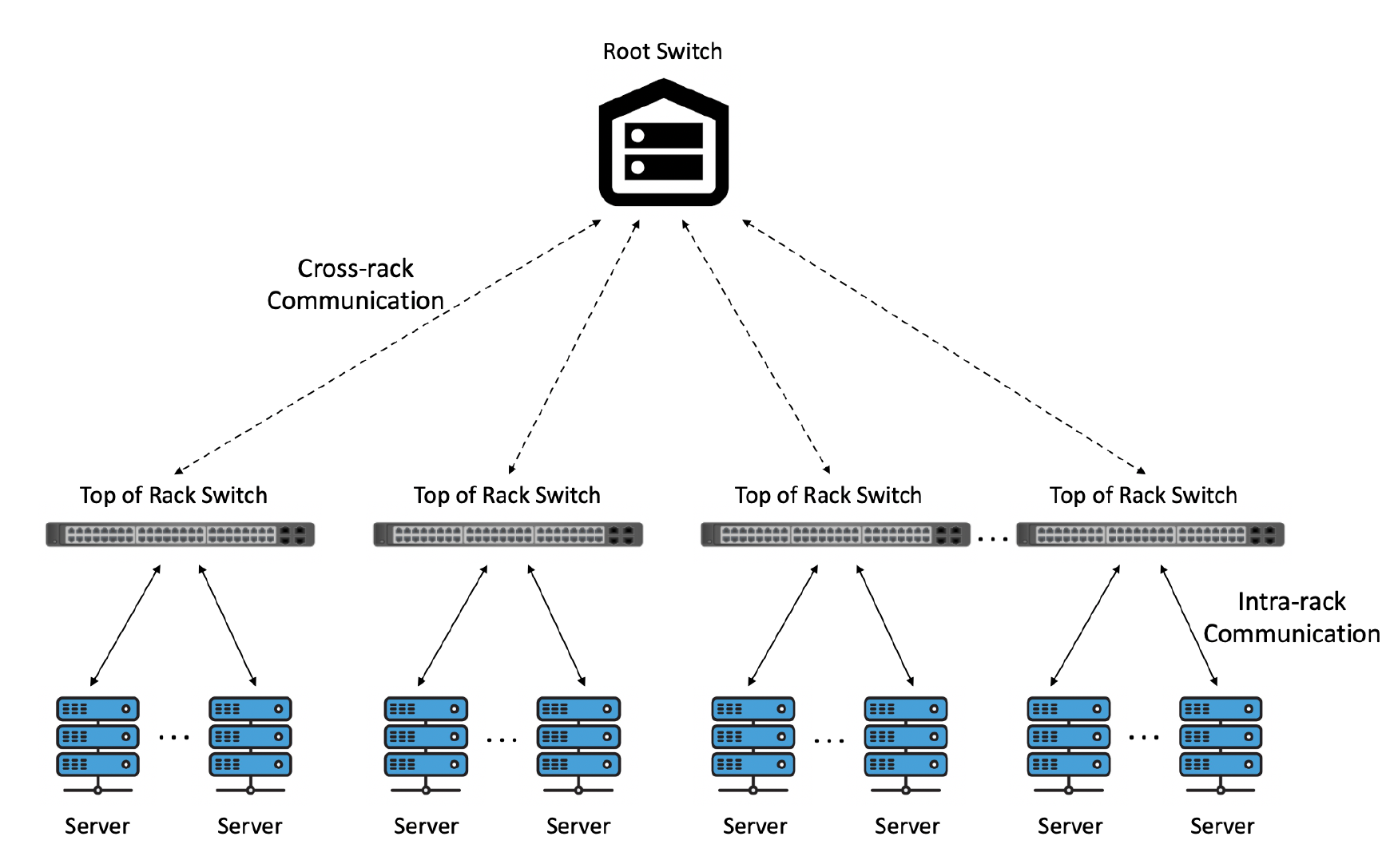}
\caption{\small Server-rack architecture where multiple servers in each rack are connected via a Top of Rack switch and the Root switch connects multiple Top of the Rack switches.}
\label{fig:server}
\end{figure}

\subsection{Consideration of Underlying Network Architecture}

Although various techniques can be used jointly with the coding techniques to increase the efficiency of data shuffling phase, it is important to consider the underlying network architecture, i.e., how the servers are connected to each other, in designing the coded shuffling algorithms. 

In \cite{gupta2017locality}, Hybrid Coded MapReduce is proposed by considering the server-rack architecture (Fig.~\ref{fig:server}) in the distributed computing systems. There are two types of communications in the Shuffle phase: (i) the cross-rack communication where the data is shuffled across different rack layers, and (ii) the intra-rack communication where the data is shuffled within the rack. In the first stage where cross-rack communication takes place, the Coded MapReduce algorithm \cite{li2015mapreduce} (Fig.~\ref{fig:coded}) is used to create multicasting opportunities for the transmission of messages. In the second stage where the intra-rack communication is performed, data is shuffled in a unicast manner where no coding technique is applied. The simulation results show that the cross-rack communication costs incurred by the hybrid scheme are lower than those of both Coded MapReduce \cite{li2015mapreduce} and uncoded scheme \cite{dean2008mapreduce} at the expense of a higher intra-rack communication costs. Although the Coded MapReduce scheme still achieves the lowest total communication costs, the overall communication costs for the hybrid scheme can be further reduced by parallelizing the intra-rack operations to provide a more accurate comparison between the different computation schemes.

The CDC computation scheme proposed in \cite{li2018tradeoff} is useful for networks with processing nodes that are closely located with each other and connected via a common communication bus. However, in practical distributed computing networks, it is hard to implement a common-bus topology for the physically separated processing nodes. Hence, to reduce communication load for the distributed computation tasks, it is important to consider a practical data center network topology to reap the coding benefits of the CDC schemes. As such, in \cite{wan2020topological}, the authors propose a CDC scheme based on a widely-used low-cost network topology which is the t-ary fat-tree topology \cite{xia2017data, fares2008scalable}. It has the characteristics of network symmetry and connections between any two processing nodes in the network. Given a practical network topology design, the proposed topological CDC scheme achieves optimal max-link communication load over all links in the topology of which the optimal tradeoff between the max-link communication load and the computation load is characterized. 

Although the coded shuffling algorithm proposed in \cite{lee2018speeding} reduces communication load of the data shuffling process, there are two main limitations: (i) it assumes a perfect broadcast channel between the master node and the workers, and (ii) the theoretical guarantee of number of broadcast transmissions only holds when the number of data points approaches infinity. To overcome the limitations of the coded shuffling algorithm proposed in \cite{lee2018speeding}, Ubershuffle \cite{chung2017ubershuffle} fills the missing entries in the encoding tables by reallocating the data points between the workers. This reduces the number of transmitted encoded packets, resulting in a reduction of communication load. The performance of the UberShuffle algorithm is evaluated when applied to different problem settings. The experimental results show that in comparison to the coded shuffling algorithm in \cite{lee2018speeding}, the UberShuffle algorithm reduces the shuffling time by up to 47.2\% and 35.7\% when implemented with the distributed SGD algorithm for a low-rank matrix completion problem and the parallel SGD algorithm for a linear regression problem respectively.

\subsection{Function Allocation}

In most of the studies of CDC computation framework, non-cascaded systems are considered. In other words, each Reduce function can only be computed at exactly one processing node. Considering that the processing nodes have different storage, computational and communication capabilities, a non-cascaded heterogeneous system \cite{woolsey2019coded} where each Reduce function is computed exactly once is considered. In this proposed scheme, the processing nodes are allocated with different number of Reduce functions of which the processing nodes with greater storage and computational capabilities are allocated more Reduce functions. This heterogenous Reduce function allocation creates a symmetry among the multicast groups such that each processing node in a group requests the same number of intermediate outputs from the other processing nodes in the same group. The heterogeneous CDC system achieves a lower communication load than an equivalent homogeneous CDC system.

However, it is desirable for the Reduce functions to be computed at multiple processing nodes in some applications, e.g., iterative algorithms in the Spark model  \cite{zaharia2010spark} where the output of a MapReduce procedure acts as the input to the MapReduce procedure in the next iteration. Although the work in \cite{li2018tradeoff} generalizes the CDC framework by allowing each Reduce function to be computed at more than one processing node, it only applies to homogeneous systems. Similar to \cite{woolsey2019coded}, heterogeneities of the systems are also considered in \cite{woolsey2019cascaded}. However, instead of a non-cascaded system, the authors propose a more general framework of cascaded CDC \cite{woolsey2019cascaded} on heterogeneous systems. In other words, each Reduce function is allowed to be computed at multiple processing nodes. Since the processing nodes have different storage capacities, the number of subfiles stored at each processing node differs. The processing nodes with larger storage capacities are allocated more subfiles and thus, compute more intermediate results. Instead of allocating the files and functions randomly as in the work in \cite{li2018tradeoff}, this cascaded CDC scheme uses a hypercube approach \cite{woolsey2018combinatorial} to allocate the files and functions to the processing nodes. The simulation results show that for the same number of processing nodes in the network, the proposed cascaded CDC scheme achieves smaller communication load than the state-of-art schemes that consider homogeneous systems \cite{li2018tradeoff}, \cite{woolsey2018combinatorial}. 

In the study of \cite{song2017benefit}, the allocation of the functions to the processing nodes does not depend on their capabilities but on the data stored at the nodes, of which the data placement is assumed to be random. This is very useful for applications that the processing nodes are mobile and collect data on-the-move. If the probability that the processing nodes contain data is higher than the pre-defined threshold, it is possible to allocate Reduce functions such that the processing nodes do not need to exchange their intermediate results for the computations of Reduce functions, i.e., each processing node can compute the Reduce functions based on its locally stored data. This reduces the number of broadcast transmissions in the Shuffle phase, thus minimizing the communication load.

Although heterogeneities of the systems are taken into consideration in some of the works \cite{kiamari2017heterogeneous}, \cite{woolsey2019coded}, \cite{woolsey2019cascaded}, they merely focus on either file allocation or function allocation. On one hand, the work in \cite{kiamari2017heterogeneous} proposes an optimal file allocation strategy in consideration of the heterogeneous storage capabilities of the processing nodes. However, the Reduce functions are distributed uniformly among the processing nodes. On the other hand, the works in \cite{woolsey2019coded} and \cite{woolsey2019cascaded} propose the allocation of Reduce functions based on the computational capabilities of the processing nodes. However, the input file is split equally and distributed among the processing nodes. Considering a more generalized heterogeneous system, a joint file and function allocation strategy is proposed in \cite{xu2019heterogeneous} to reduce the communication load in the Shuffle phase. The file allocation and function assignment strategies allocate more subfiles and Reduce functions respectively to the processing nodes with higher computational and storage capabilities. Generally, the Reduce function assignment is related to the input file allocation as the processing nodes with more input files have greater storage and computational capabilities and hence are more capable of computing more output functions. In particular, there are two proposed schemes of function assignment, i.e., computation-aware function assignment and shuffle-aware function assignment. For computation-aware function assignment strategy, the number of functions allocated is proportional to the computational capabilities of the processing nodes in order to reduce computation latency. For the shuffle-aware function assignment strategy, the functions are mostly allocated to the processing nodes with high computational capabilities so that the communication load in the Shuffle phase is minimized. The simulation results show that the communication loads achieved by both computation-aware and shuffle-aware function assignment strategies are lower than the uniform function allocation strategy. Besides, the computation-aware function assignment strategy requires fewer number of output functions as compared to the proposed schemes in \cite{woolsey2019cascaded} and \cite{woolsey2019coded}. However, the number of input files required is much larger, especially when the number of processing nodes in the system increases.

While there has been great attention on the design of file allocation, coded shuffling algorithms and function allocation, all these works assume a fixed number of processing nodes in the distributed computing systems. For a given computation task that specifies the number of subfiles and the number of outputs, the allocation of functions and the data shuffling schemes can be implemented with minimum number of processing nodes. In the study of \cite{yu2017allocate}, the resource allocation problem is formulated as an optimization problem that minimizes the overall job execution time with optimal number of processing nodes used. For more practical implementation, the resource allocation strategy should consider the heterogeneity in the processing speed of the nodes, since the straggler effects cause an increase in computation latency which increases the overall job execution time.

By exploiting the fact that the processing nodes have time-varying computing resources, e.g., the processing nodes may have different central processing unit (CPU) power over time, an optimal computation task scheduling scheme helps to reduce the communication load. In the scheduling of tasks under dynamic computing resources, the total communication load is minimized by optimizing the number of allocated tasks and load redundancy at each time slot \cite{zhao2019load}.

\begin{table*}[t]
\caption{\small CDC schemes to reduce communication costs.} 
\label{tab:communication}
\centering
\begin{tabular}{|>{\centering\arraybackslash}m{2cm} |>{\centering\arraybackslash}m{1.4cm} |>{\centering\arraybackslash}m{3.5cm} |m{6.3cm} |>{\centering\arraybackslash}m{2.3cm}|}
\hline

\rowcolor{mygray}
\textbf{Approach} & \textbf{Ref.} & \textbf{Coding Schemes} & \multicolumn{1}{c|}{\textbf{Key Ideas}} & \textbf{Platform Support} \\ \hline

\multirow{12}{=}{File Allocation} &  \cite{attia2016theoretic} & - & Deterministic and systematic storage update strategy  & Heterogeneous\\
\cline{2-5}
& \cite{gupta2017locality} & Hybrid Coded MapReduce & Allocates Map tasks such that data locality is maximized & Homogeneous\\
\cline{2-5}
& \cite{parrinello2018} & Group-based Coded MapReduce & Allocates dataset by using a group-based method in order to avoid high subpacketization level and allow processing nodes to cooperate in the transmission of messages& Homogeneous\\
\cline{2-5}
& \cite{konstantinidis2018leveraging} & Resolvable Design & Uses single-parity code to determine the number of subfiles and allocate the subfiles based on the corresponding resolvable design  & Homogeneous\\
\cline{2-5}
& \cite{jiang2020coded, yan2018pda, ramkumar2019pda} & Placement Delivery Arrays & Construction of CDC schemes based on PDA which has the property of illustrating the placement and delivery phase in a single array & Homogeneous\\ 
\hline

\multirow{18}{=}{Coded Shuffling Design} & \cite{li2018compressed} & Compressed CDC & Pre-combines computed intermediate values of the same function, followed by coding the pre-combined packets for communication between different processing nodes & Homogeneous\\
\cline{2-5}
& \cite{alistarh2017qsgd} & Quantized SGD & Quantizes the component of the gradient vector to a discrete set of values and encodes the quantized gradients given their statistical properties & Homogeneous\\
\cline{2-5}
& \cite{song2020pliable} & Pliable Index Coding & Semi-random data shuffling scheme based on modified pliable index coding to reduce the number of communication rounds & Homogeneous\\
\cline{2-5}
& \cite{haddadpour2018cross} & Cross-iteration Coded Computing & Jointly codes across multiple iterations for a single communication round & Homogeneous\\
\cline{2-5}
& \cite{prakash2018graph} & CDC for distributed graph computing systems & Instead of communicating with all other processing nodes, each processing nodes only needs to communicate with a subset of processing nodes to obtain the required data to complete its computation tasks  & Homogeneous\\
\cline{2-5}
& \cite{li2016multistage} & CDC for multistage dataflow & A more generalized CDC scheme is proposed to handle multistage dataflow computation tasks which are represented by layered DAGs & Homogeneous\\
\hline

\multirow{12}{=}{Consideration of Underlying Network Architecture} & \cite{gupta2017locality}  & Hybrid Coded MapReduce & Reduces cross-rack communication at the expense of higher intra-rack communication based on the server-rack architecture & Homogeneous\\
\cline{2-5}
  & \cite{wan2020topological} & Topological CDC & Considers t-ary fat-tree topology which is a more practical topology to connect the physically separated processing nodes in data center networks & Homogeneous\\
\cline{2-5}
  & \cite{chung2017ubershuffle} & UberShuffle & Considers imperfect communication channel between the workers and the master node & Homogeneous\\
\cline{2-5}
& \cite{lee2017multicore} & - & Considers multicore setup where each computing node can have multiple cores, e.g., the CPU instances of publicly available cloud infrastructure can deliver up to 128 cores & Homogeneous\\
\hline

\multirow{12}{=}{Function Allocation} & \cite{woolsey2019coded}  & - & Considers a non-cascaded system and allocates Reduce functions over a simplified heterogeneous network which comprises multiple homogeneous networks  & Heterogenous \\
\cline{2-5}
& \cite{woolsey2019cascaded} & Cascaded CDC & Reduce functions are computed at more than one processing node and they are allocated based on the combinatorial design in \cite{woolsey2018combinatorial} & Heterogeneous\\
\cline{2-5}
  & \cite{song2017benefit} & - & Allocates functions to maximize data locality such that the number of communication rounds required is reduced & Homogeneous\\
\cline{2-5}
  & \cite{xu2019heterogeneous} & - & Joint file and function allocation strategy & Heterogeneous\\
\cline{2-5}
  & \cite{zhao2019load} & - & Considers the availability of time-varying computing resources & Homogeneous\\
  
\hline

\end{tabular}
\end{table*}

\subsection{Summary and Lessons Learned}
In this section, we have reviewed four approaches to minimize communication costs in distributed computing systems. For each approach, we discuss the solutions that are proposed in different works. We summarize the coding schemes to minimize communication costs in Table~\ref{tab:communication}. The lessons learned are as follows:

\begin{itemize}

\item To handle the increasing amounts of data generated, more processing nodes are needed for the completion of distributed computations given the limited capabilities of the processing nodes. With more processing nodes connected to the network, more communication rounds of the computed intermediate results are required, resulting in higher communication costs which lead to a longer job execution time. Besides, the high communication costs in the data shuffling phase impede the implementation of efficient distributed iterative algorithms which are useful for the training of machine learning models. As such, the minimization of communication costs is a key step to achieve the objective of reducing the overall job execution time of the distributed computation tasks.

\item While having more processing nodes to perform the computation tasks in parallel helps to reduce the computation load of each processing node, the communication costs may increase, slowing down the entire computation process. Instead of generating infinitely large number of subfiles and distributing them among large number of processing nodes, several studies focus on determining the optimal number of subfiles to avoid the input file from splitting too finely. For example, the resolvable design based schemes \cite{konstantinidis2018leveraging, konstantinidis2019resolvable} and PDA approaches \cite{jiang2020coded, yan2018pda, ramkumar2019pda} are adopted to split the input files. Hence, it may not be optimal to use all the processing nodes that are connected to the network. In fact, the authors in \cite{yu2017allocate} propose an optimal resource allocation scheme that determines the minimum number of processing nodes needed to achieve the minimum overall job execution time. 

\item Coding techniques have shown to be effective in reducing communication load in the data shuffling phase at the expense of higher computation load \cite{li2018tradeoff, lee2018speeding}. However, the two-dimensional tradeoff is insufficient to fully evaluate the performance of the CDC schemes. Apart from leveraging on the computational capabilities of the processing nodes, their storage capabilities can be exploited. For example, more data can be stored at processing nodes with higher storage capacities such that the number of communication rounds is reduced \cite{attia2016theoretic}. Moreover, by considering the data stored at the processing nodes, the allocation of functions that maximizes data locality helps to reduce the need for communication bandwidth \cite{gupta2017locality}. Hence, instead of the two-dimensional tradeoff between computation load and communication load, the three-dimensional tradeoff between computation, communication and storage cost \cite{yan2018storage} has to be carefully managed for the implementation of efficient CDC schemes.

\item For effective implementation of CDC schemes in practical distributed computing systems, the underlying architecture has to be considered. Generally, the distributed computing systems operate under the server architecture which consists of multiple racks where each rack has several servers. The Hybrid Coded MapReduce \cite{gupta2017locality} scheme reduces cross-rack communication at the expense of higher intra-rack communication. Besides, the communication channels between the master node and workers are imperfect. As a result, the theoretical gains of the coded data shuffling schemes are not achievable under practical setups. In order to design and analyze the performance of the CDC schemes, the limitations of the distributed computing systems should be taken into consideration. 

\item Most of the CDC schemes focus on the minimization of communication costs in the data shuffling phase at the expense of reasonable increase in computation load. However, the computational overhead of the algorithm is not negligible under some settings. For example, under much faster broadcast environment, the UberShuffle algorithm \cite{chung2017ubershuffle} incurs significant encoding and decoding costs such that the shuffling gain is offset by the high computational overhead. For future works, more practical CDC schemes can be proposed such that the communication costs are minimized while maintaining low computational cost to improve the performance of the distributed computing systems. Given the uncoded computing schemes, e.g., \cite{zhang2013statistical} and \cite{martin2010parallelized}, achieve near-optimal computation and storage costs, one possible research direction is to merge the communication-efficient proposed schemes with the uncoded computing schemes to reduce the computation and storage costs.

\item Although some of the works consider heterogeneities in the capabilities of the processing nodes to allocate files and functions, the presence of stragglers which have slower processing speeds still hinder the performance of the distributed computing systems. Therefore, we further discuss the approaches to mitigate the straggler effects in the next section.

\end{itemize}


\section{Mitigation of Stragglers}
\label{sec:stragglers}

In distributed computing systems, processing nodes have different processing speeds and thus the time taken to complete their allocated subtasks differs from each other. Since the computation task is distributed among the processing nodes, the master node needs to wait for all processing nodes to complete their subtasks and return the computed results before recovering the final result. As such, the time taken to execute a computation task is determined by the slowest processing node. This is also known as the \emph{straggler effects}. 

The problem of straggler effects has been widely observed in the distributed computing systems. Previously, various methods such as straggler detection \cite{ananthanarayanan2010reining}, \cite{zaharia2008improving}, asynchronous execution and naive replication of jobs \cite{wang2015replication}, \cite{shah2016redundant} have been proposed to reduce the overall time taken to execute the computation tasks. Recently, coding approaches have been shown to outperform the aforementioned methods in reducing computation latency of the distributed computing systems. In this section, we discuss three approaches to mitigate the straggler effects: 

\begin{itemize}

\item \emph{Computation Load Allocations: }Coding techniques can be implemented together with computation load allocation strategies to reduce the computation latency in the distributed computing systems. It is important to take into account the variation in the computational capabilities of the processing nodes to allocate computation load. As such, different prediction methods such as an long short-term memory (LSTM) algorithm \cite{narra2019slack}, an Auto Regressive Integrated Moving Average (ARIMA) model and a Markov model \cite{yang2019timelythroughput} are used to estimate the processing speeds of the nodes. Generally, the objective of the load allocation strategies is to minimize computation latency. However, in some applications where strict deadlines are given, the load allocation strategies aim to maximize timely computation throughput \cite{yang2019timelythroughput}.


\item \emph{Approximate Coding: }For some applications, e.g., location-based recommendation systems, exact solutions are not necessary. The studies explore different coding approaches to obtain approximate solutions to the problems. The approximate coding methods relax the requirement for convergence and thus reduce the number of workers that are required to return their computed results. This can avoid stragglers to make an adverse effect to the system.

\item \emph{Exploitation of Stragglers: }The straggling nodes may have completed a fraction of the allocated computation tasks, of which is a waste to be ignored completely. In fact, the stragglers may not be persistent over the entire computation process \cite{ozfatura2019persistent} and thus their partial computed results can be useful to recover the final result. In order to maximize the resource utilization of the straggling nodes, the workers are allowed to sequentially process their allocated subtasks and transmit their partial computed results continuously \cite{ferdinand2018hierarchical}. However, this may come at the expense of higher communication load, which needs to be carefully managed. 

\end{itemize}

\subsection{Computation Load Allocation}

Apart from having different straggling rates that affect the completion of tasks, the processing nodes have different capabilities, e.g., storage capacities, computing resources and communication bandwidths. To better handle the straggling nodes in the distributed computing systems, an optimal load allocation strategy that takes into account these heterogeneities is necessary to minimize the overall job execution time. Given the computation time parameters, i.e., the straggling and shift parameters of each worker, the Heterogeneous Coded Matrix Multiplication (HCMM) algorithm \cite{reisizadeh2019} determines the allocation of computation load to each worker. The HCMM scheme exploits the benefits of both coding techniques and computation load allocation strategy to minimize the average computation time of the computation tasks. Given that it is difficult to derive closed-form expressions of expected computation time of the heterogeneous processing nodes, a two-step alternative problem formulation is proposed. In the first step, given a time period, the number of computed results by the workers is maximized by optimizing the load allocation. In the second step, given the derived load allocation in the first step, the time needed to ensure sufficient results are returned at a pre-defined probability is minimized. The simulation results show that when workers' computation time is assumed to follow a shifted exponential runtime distribution, HCMM reduces the average computation time by up to 71\%, 53\% and 39\% over uniform uncoded, load-balanced uncoded and uniform coded load allocation schemes, respectively. In practical experiments over Amazon EC2 clusters, the combination of HCMM and LT codes outperforms the uniform uncoded, load-balanced uncoded and uniform coded load allocation schemes by up to 61\%, 46\% and 36\%, respectively.

Although HCMM achieves asymptotically optimal computation time, the decoding complexity is high, which suggests the opportunity to further speed up the overall computation tasks. In practical distributed computing systems, some processing nodes have the same computational capabilities, in terms of the same distributions of computation time, and thus they can be grouped together. By exploiting the group structure and heterogeneities among different groups of processing nodes \cite{kim2019group}, \cite{kim2019optimal}, the implementation of a combination of group codes and an optimal load allocation strategy not only approaches the optimal computation time that is achieved by the MDS codes, but also has low decoding complexity. In addition, by varying the number of allocated rows of the matrix to the workers \cite{kim2019optimal}, the computation latency can be reduced by orders of magnitude over the MDS codes with fixed computation load allocation \cite{kim2019group} as the number of workers increases. The load allocation strategy proposed in \cite{kim2019optimal} focuses mainly on the design of an optimal MDS code. Other efficient coding algorithms such as LT codes \cite{mallick2019rateless} can also be explored in future studies.

In addition to the heterogeneous capabilities of the processing nodes, the amount of available resources of the processing nodes may vary over time. By always allocating computation tasks to the processing nodes with higher capabilities, delays may be incurred in completing the allocated tasks if the processing nodes start to work on the newly-allocated computation tasks in parallel. At the same time, the resources of the processing nodes with lower capabilities may be under-utilized. In order to maximize the resource utilization of the processing nodes, dynamic workload allocation algorithms which are adaptive to the time-varying capabilities of the processing nodes are proposed \cite{kesh2018dynamic}, \cite{narra2019slack}, \cite{ yang2019timelythroughput}. To provide robustness against the straggling nodes, the design of the load allocation algorithms often depends on the historical data of the processing nodes such as computation time, which can be used to predict the processing speeds by using an LSTM algorithm \cite{narra2019slack}, an ARIMA model or a Markov model \cite{yang2019timelythroughput}. The best performing LSTM model achieves 5\% lower prediction error than an ARIMA (1,0,0) model \cite{narra2019slack}.

In the study of \cite{kesh2018dynamic}, the authors propose Coded Cooperative Computation Protocol (C3P) which is a dynamic and adaptive coded cooperation framework that efficiently utilizes the available resources of the processing node while minimizing the overall job execution time. Specifically, the master node determines the coded packet transmission intervals based on the responsiveness of the processing nodes. For the processing nodes which are not able to complete the tasks within the given transmission interval, they suffer from longer waiting time for the next coded packets. In comparison to the HCMM scheme \cite{reisizadeh2019} which does not consider the dynamic resource heterogeneity in workers, the C3P framework achieves more than 30\% improvement in task completion delay.

The dynamic and adaptive load allocation algorithms are especially useful in providing timely services with deadline constraints which are common in many IoT applications. For such applications, instead of minimizing task completion delay, the objective of the load allocation algorithms is to maximize timely computation throughput, i.e., the average number of computation tasks that are successfully completed before the given deadline \cite{yang2019timelythroughput}. 

For some applications that may need timely but not necessarily optimal results, it is more important to recover the final result with the highest accuracy possible by the stipulated deadlines than to solve for an exact solution. The algorithm to solve for an approximate solution requires significantly shorter computation time than that of an algorithm that solves for an exact solution \cite{ferdinand2016anytime}. In the study of \cite{wang2019batchprocessing}, the batch-processing based coded computing (BPCC) algorithm is proposed. The workers first partition the allocated encoded matrix into several batches, i.e., submatrices, for computations. As soon as the workers complete the computation of each batch of submatrix with a given vector, they return the partial computation results which are used to generate the approximate solution. Based on the computation time parameters, i.e., the straggling and shift parameters of the workers, the BPCC algorithm is used to optimally allocate the computation load to each worker by considering two cases: (i) negligible batching overheads, and (ii) linear batching overheads. Hence, the allocation of computation load is optimized by jointly minimizing both the overall job execution time as well as the potential overhead of batch processing. In addition to reducing computation latency, the BPCC algorithm exploits the partial computation results of all processing nodes, including the straggling nodes, which contribute to approximate solutions of higher accuracy \cite{ferdinand2018hierarchical}. The simulation results show that BPCC algorithm with negligible batching overheads achieves up to 73\%, 56\% and 34\% reduction in average job execution time over uniform coded, load-balanced uncoded and HCMM \cite{reisizadeh2019} load allocation schemes respectively. The experimental results on Amazon EC2 clusters and Unmanned Aerial Vehicles (UAVs) based airborne computing platform also demonstrate similar results. 

\begin{table*}[!ht]
\caption{\small Approximate CDC schemes to mitigate the straggler effects.} 
\label{tab:approximate}
\centering
\begin{tabular}{|>{\centering\arraybackslash}m{2.0cm} |>{\centering\arraybackslash}m{1.5cm} |>{\centering\arraybackslash}m{3.3cm} |m{8cm}|}
\hline

\rowcolor{mygray}
\multicolumn{1}{|c|}{\textbf{Problems}}  & \textbf{Ref.} & \textbf{Coding Schemes} & \multicolumn{1}{c|}{\textbf{Key Ideas}}  \\ \hline

\multirow{5}{=}{Matrix-Vector}  & \cite{ferdinand2016anytime} & Anytime Coding & Computations can be stopped anytime and the approximate solution is derived from the processing nodes that have completed their tasks\\
\cline{2-4}
 &\cite{zhu2017sequential} & Coded Sequential Computation Scheme & A sequence of approximated problems are designed such that the time required to solve these problems is shorter than solving the original problem directly\\
\hline

\multirow{3}{=}{Matrix-Matrix}  &\cite{jahani2019codedsketch}& CodedSketch  & Use a combination of count-sketch technique and structured polynomial codes \\
\cline{2-4}
 &\cite{gupta2018oversketch} & OverSketch  & Divide the sketched matrices into blocks for computations\\
\hline

\multirow{8}{=}{Gradient Descent} &\cite{charles2017approximate} & Bernoulli Gradient Codes  & Use Bernoulli random variables as entries of the function assignment matrix\\
\cline{2-4}
 &\cite{charles2018gradient}& Stochastic Block Codes  & Interpolation between BGC and FRC \\
\cline{2-4}
 &\cite{bitar2020stochastic} & Stochastic Gradient Coding  & Distribute data based on pair-wise balanced scheme and provide a rigorous convergence analysis of the proposed coding scheme\\
\cline{2-4}
 &\cite{maity2019ldpc} & LDPC Codes  & Encode the second moment of the data points\\
\cline{2-4}
 &\cite{karakus2018redundancy},\cite{karakus2017encoded} & Encoded Optimization  & Encode both the labels and data such that the redundancy is introduced in the formulation of the optimization problems \\
\hline

\multirow{4}{=}{Non-Linear} & \cite{dutta2018polydot} & Generalized PolyDot & Generalization of PolyDot codes \cite{fahim2017optimal} and is used for the training of DNNs\\
\cline{2-4}
  & \cite{kosaian2018learning} & Learning-based Approach & Design neural network architectures to learn and train the encoding and decoding functions to approximate unavailable outputs\\
\hline

\end{tabular}
\end{table*}

\subsection{Approximate Coding}

For some applications, it is not necessary to obtain exact solutions. In this subsection, we review the coding approaches that are used to derive approximate solutions for the distributed computation tasks: (i) matrix multiplication, (ii) gradient descent, (iii) non-linear computations. Table~\ref{tab:approximate} summarizes the approximate coding schemes for various distributed computation tasks.

\subsubsection{Matrix multiplications} \label{subsubsec:matrix} The anytime coding \cite{ferdinand2016anytime} scheme derives an approximate solution by using the output results of the completed processing nodes at any given time. Based on singular value decomposition (SVD), the given computation task is decomposed into several subtasks of different priorities. More important subtasks are allocated more processing nodes for computations as they improve the accuracy of the approximation. To allow the users to receive useful information from time to time, approximate solutions can be transmitted to the users sequentially. This can be achieved by solving a sequence of approximated problems \cite{zhu2017sequential}, instead of solving the original problem directly. 

To further reduce the computation time of approximate matrix multiplications, sketching techniques \cite{woodruff2014sketching}, \cite{wang2015practical} can be used to remove redundancy in the structure of the matrices through dimensionality reduction. However, by using sketching techniques, the recovery threshold increases as the redundancy is removed. In contrast, coding techniques reduces recovering threshold by introducing redundancy. As such, a combination of both techniques that carefully manage the tradeoff between the recovery threshold and the amount of redundancy can be implemented to minimize computation latency. In particular, count-sketch technique \cite{cormode2008finding} is combined with structured codes to mitigate the straggler effects by preserving a certain amount of redundancy, thereby achieving the optimal recovery threshold and hence computation latency \cite{jahani2019codedsketch}, \cite{gupta2018oversketch}.


\subsubsection{Gradient descent} \label{subsubsec:gradient}To speed up the distributed gradient descent tasks, several approximate gradient coding schemes are proposed to approximately compute any sum of functions. Instead of constructing the gradient codes based on expander graphs \cite{raviv2017gradient} which are difficult to compute due to high complexity, a more efficient and simpler Bernoulli Gradient Code (BGC) is proposed by using sparse random graphs \cite{charles2017approximate} which introduce randomness into the entries of the function assignment matrix. Since the performance of the gradient codes depends on the efficiency of the decoding algorithms, the authors also present two decoding techniques to recover the approximate solution from the outputs of the non-straggling nodes. The simulation results show that the BGC schemes can handle adversaries with polynomial-time computations but at a cost of higher decoding error than the FRC schemes \cite{tandon2017gradient}. Besides, the optimal decoding algorithm always achieves a lower decoding error than that of the one-step decoding algorithm across various gradient coding schemes. A rigorous convergence analysis of the approximate gradient codes and the performance of BGC on different practical datasets such as the Amazon dataset, Covertype dataset and KC Housing dataset are presented in \cite{wang2019erasurehead}. Stochastic block code \cite{charles2018gradient} which is based on the stochastic block model from random graph theory, is an interpolation between FRC \cite{tandon2017gradient} and BGC \cite{charles2017approximate}. On one hand, the FRC schemes achieve small reconstruction errors under random straggler selection while on the other hand, the BGC schemes are robust against polynomial-time adversarial stragglers.

Other approximate gradient coding methods such as the Stochastic Gradient Coding (SGC) \cite{bitar2020stochastic} and LDPC codes \cite{maity2019ldpc} are used to obtain an estimate of the true gradient. Similar to the idea of encoding data batches in the PCR schemes \cite{li2018polynomially}, the data variables in the optimization formulation can be efficiently encoded to mitigate the straggler effects in more general large-scale optimization problems such as support vector machines, linear regressions and compressed sensing \cite{karakus2018redundancy}, \cite{karakus2017encoded}. Generally, in solving for approximate solutions to reduce computation latency of the distributed tasks, there is an inherent tradeoff between the recovery threshold and the approximation error where the recovery threshold can be reduced at the expense of larger approximation error \cite{chang2019random}.

\subsubsection{Non-linear computations}By leveraging on the enormous amounts of data generated, machine learning algorithms are useful in making predictions and allowing devices to respond intuitively to user demands without human interception. Different neural network architectures are developed to make accurate inference given the dataset. Since some of the layers of the neural networks such as the max-pooling functions and the activation layer are non-linear, the overall computation of the functions are non-linear. As a result, most of the prior works on linear computations which are discussed in Sections~\ref{subsec:codingtech}, \ref{subsubsec:matrix} and \ref{subsubsec:gradient}, are not applicable to the computation of the increasingly important non-linear neural networks, in which their performances are also limited by the straggling nodes. One of the few approaches that can be extended to the training of DNNs is the Generalized PolyDot codes \cite{dutta2018polydot} which are used to compute matrix-vector multiplications. The Generalized PolyDot codes are used to code the linear operations at each layer of the neural networks. This coding scheme allows for errors in the training of each layer. In other words, decoding can still be performed correctly given that the amount of errors does not exceed the maximum error tolerance level. The effectiveness of coding techniques in mitigating the straggler effects of different neural network architectures such as AlexNet \cite{alex2012imagenet} and Visual Geometry Group (VGG) \cite{simonyan2014deep} in an IoT system is illustrated in \cite{hadidi2019}. However, this unified coded DNN training strategy may not be relevant to the training of other neural networks which have large number of non-linear functions. As such, the authors in \cite{kosaian2018learning} propose a learning-based approach for designing codes. Based on the dilated CNN and multilayer perceptrons (MLP), neural network architectures and a training methodology are proposed to learn the encoding and decoding functions. The outputs of the decoding functions are used to approximate the unavailable outputs of any differentiable non-linear function. The simulation results show that the learning-based approach to designing the encoding and decoding functions can accurately reconstruct 95.71\%, 82.77\% and 60.74\% of the unavailable ResNet-18 classifier outputs on MNIST, Fashion-MNIST and CIFAR-10 datasets respectively.

\subsection{Exploitation of Stragglers}

To avoid delays caused by the straggling nodes in the network, most distributed computation schemes ignore the work completed by the straggling nodes by either increasing the workload of the non-straggling nodes or by obtaining less accurate approximate solutions. However, the amount of work that has been completed by the straggling nodes, especially the non-persistent stragglers, is non-negligible and can be better utilized.

In order to exploit the computational capacities of these non-persistent stragglers, multi-message communication (MMC) is used where the workers are allowed to send multiple messages to the master node at each iteration. This allows the workers to transmit their partial computed results whenever they complete a fraction of the allocated task, rather than completing the entire computation task before transmitting the computed result in a single communication round. The work in \cite{ozfatura2019persistent} considers the implementation of Lagrange Coded Computing (LCC) with MMC to minimize the average job execution time at the expense of higher communication load due to the increase in number of messages transmitted by the workers to the master node. Since the LCC scheme utilizes polynomial interpolation to recover the final result, the decoding complexity and the number of transmissions can be further reduced by increasing the number of polynomials used in decoding the computed results returned by the workers. The simulation results show that by exploiting the computing resources of the non-persistent stragglers via MMC, the average job execution time decreases as the computation load of each worker increases. However, since the communication load of the LCC-MMC scheme is constant as the computation load increases, it is suitable to be implemented only when computation time dominates the overall execution time of the distributed tasks. The total time needed to execute the distributed tasks includes the time needed for both computation and communication. Otherwise, if communication load is the cause of the bottleneck of the network, LCC without MMC should be used instead since the communication load can be reduced at the expense of higher computation load.

Given that MMC is allowed where the workers perform more than one round of communication for each iteration, the computation work done by the straggling nodes can be exploited by allowing sequential processing where the workers need to transmit the computation results of their completed subtask before working on the next subtask. To fully exploit the useful information provided by the straggling nodes, the hierarchical coded computation scheme is proposed in \cite{ferdinand2018hierarchical} to utilize the computations from all workers. Each worker first divides the allocated computation task into layers of sub-computations, which are processed sequentially, i.e., the result of a layer of sub-computation is transmitted to the master node before the computation of the next layer starts. Since the workers have different processing speeds and the sub-computations are performed sequentially, the finishing time for each layer is different. MDS codes are used to encode the layers so that the finishing time of each layer is approximately the same. The top layers which have lower probability of erasure are encoded with higher-rate MDS codes whereas the bottom layers are encoded with lower-rate MDS codes. The simulation results show that for the same amount of tasks to be completed, the proposed hierarchical coded computation scheme achieves $1.5$ factor improvement in expected finishing time as compared to the coded computation scheme proposed in \cite{lee2018speeding} which ignores the computations of the straggling nodes.


In the study of \cite{das2018c3les}, by computing the block products sequentially, the partial computation results from the straggling nodes are used to aggregate the final result. In order to preserve the sparsity of matrices for processing by the workers, instead of coding the entire matrix, the fraction of coded blocks can be specified to control the level to which coding is utilized in the solutions. There are two different approaches considered, depending on the placement of the coded blocks. When the coded blocks appear at the bottom of the workers, it is easier for the master to decode. When the coded blocks appear at the top of the nodes, it minimizes the computation load of the workers. As such, the placement can be calibrated based on the task requirements.


While the coded computation schemes achieve low communication load and reduce the average job execution time for each iteration, the uncoded computation schemes have their own benefits of having no decoding complexity and allowing partial gradient updates. In order for a system to benefit from the advantages of both schemes, coded partial gradient computation (CPGC) scheme is proposed in \cite{ozfatura2019partial}. In the CPGC scheme, uncoded submatrices are allocated for the first computation since there is a high probability for each worker to complete their first computation task. Subsequently, coded submatrices are allocated to handle the straggling nodes. The master node is able to update the gradient parameters by using the computation results from a subset of workers and by exploiting the partial computations of the straggling nodes. As a result, the average job execution time for each iteration is reduced. 

\begin{figure}
\centering
\includegraphics[width=\linewidth]{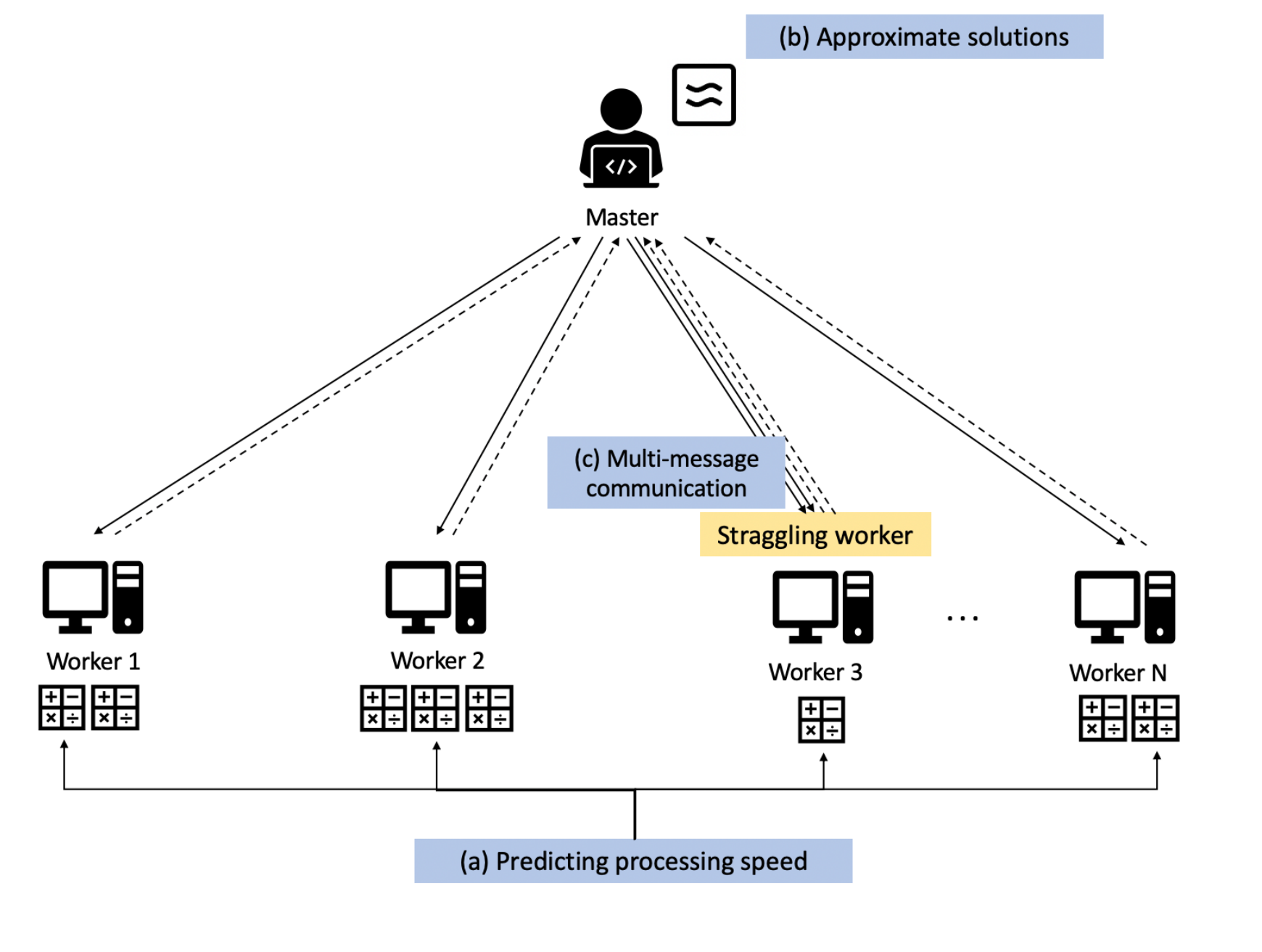}
\caption{\small Approaches to mitigate the straggler effects include (a) computation load allocation by predicting the speed of the processing nodes \cite{narra2019slack}, \cite{yang2019timelythroughput}, (b) approximate coding and (c) exploitation of stragglers by allowing multi-message communications \cite{ozfatura2019persistent}, \cite{ferdinand2018hierarchical}.}
\label{fig:mitigate}
\end{figure}

\subsection{Summary and Lessons Learned}

In this section, we have discussed three approaches (Fig.~\ref{fig:mitigate}) to mitigate the straggler effects. The lessons learned are as follows:

\begin{itemize}

\item The straggler effect is a key issue to be resolved in order to reduce computation latency, hence minimizing the overall job execution time. Due to various factors such as insufficient power, contention of shared resources, imbalance work allocation and network congestions \cite{dean2013tail, ananthanarayanan2010reining}, some processing nodes may run slower than the average or even be disconnected from the network. Since the computation tasks are only completed when all processing nodes complete their computations, the time needed to complete the tasks is determined by the slowest processing node. Coding techniques have shown their effectiveness in reducing computation latency by introducing redundancy \cite{lee2018speeding}. In this section, we have explored the use of coding techniques for different distributed computation tasks, e.g., matrix-vector multiplications, matrix-matrix multiplications, linear inverse problems, iterative algorithms, convolutions and non-linear problems. While most of the research focuses on the design of encoding techniques, the decoding complexity of the codes also affects the computation latency significantly. Apart from Reed-Solomon codes \cite{halbawi2018solomon} and LDPC codes \cite{maity2019ldpc}, more effective codes with low decoding complexity can be investigated in future studies. 

\item Considering heterogeneities in the capabilities of the processing nodes, effective computation load allocation strategies are implemented to allocate workload to the processing nodes. We have discussed the proposed computation load allocation algorithms under different constraints, e.g., strict deadlines and time-varying computing resources. 
In addition, different prediction methods such as the LSTM algorithm \cite{narra2019slack}, an ARIMA model and a Markov model \cite{yang2019timelythroughput} that predict the speeds of the processing nodes are explored. However, the stragglers may be non-persistent in nature and thus they may be useful when they are able to perform computations faster than the average rate. Hence, the load allocation based on the responsiveness of the processing nodes may be more useful in such situations.  

\item Instead of exact solutions, it is acceptable to derive approximate solutions for some applications, e.g., location-based recommender systems. Various coding techniques to derive approximate solutions are investigated. For example, in the studies of \cite{jahani2019codedsketch} and \cite{gupta2018oversketch}, sketching techniques are used with structured codes to minimize computation latency. However, there exists a tradeoff between the recovery threshold and the approximation error. For future works, an improvement to this tradeoff can be investigated.

\item Although the straggling nodes run slower than the average, the computations that are completed may still be useful. It is wasteful to not utilize the partial computed results of the straggling nodes. Besides, these partial computed results can help to improve the accuracy of the estimates. For example, in \cite{NIPS2017_inverse}, the stragglers are treated as soft errors instead of erasures to minimize the mean-squared error of the iterative linear inverse solvers under a deadline constraint by using approximate weights. Outputs from all computing nodes, including the straggling nodes are used to recover estimates that are as close as possible to the convergence values when the computation deadline is brought forward or when the number of computing nodes increases. Unfortunately, as the processing nodes are required to send their partial results once they complete, more communication rounds are performed. The high communication costs may be the bottleneck of the distributed computation tasks. Given the advantages of using partial results from the straggling nodes, optimization approaches to minimize the communication costs between the master node and the workers should be explored.

\item The current studies in this section have proposed effective coding schemes for implementation. However, they do not consider security in the design of the coding schemes. For example, the FRC scheme \cite{tandon2017gradient} achieves high accuracy in the presence of stragglers, but it is susceptible to attacks from adversarial stragglers, which turn more processing nodes into straggling nodes. Besides, other than the straggling workers, there may exist malicious or curious workers that may compromise the privacy and security of the system. Therefore, approaches to ensure secure coding are discussed in-depth in the next section. 

\end{itemize}

\section{Secure Coding for Distributed Computing}
\label{sec:security}

In distributed computing, the data owner, master node, and workers may not belong to the same entity. For example, the data owner may wish to perform a task on a massive dataset on which intensive computations have to be performed. The computations may be divided and distributed to multiple workers on third-party computing services. However, sensitive data, e.g., in healthcare services \cite{lim2020hierarchical}, may be involved. In this case, \textit{curious} workers may collude to obtain information about the raw data, whereas \textit{malicious} workers \cite{lim2020federated} may intentionally contribute erroneous inputs to introduce biases to the model. Besides, in some cases, the dataset does not belong to either the master node or the workers, and as such, the raw dataset has to be guarded against both parties.

\begin{figure}
\centering
\includegraphics[width=\linewidth]{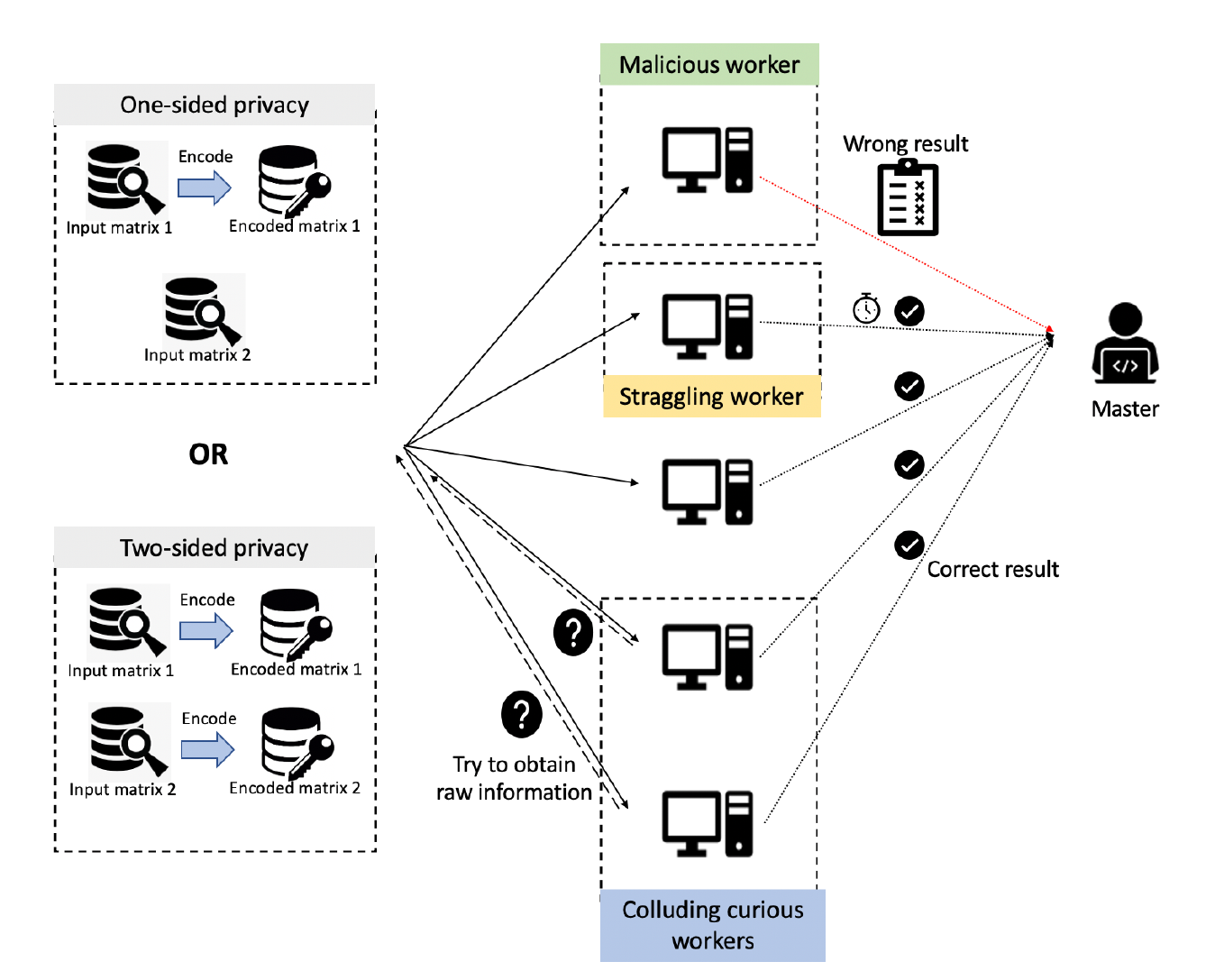}
\caption{\small Illustration of coding framework for the objectives of mitigating stragglers, colluding curious workers, and malicious workers.}
\label{fig:security}
\end{figure}

To ensure that privacy and security are preserved during the computing tasks, conventional methods such as homomorphic encryption \cite{gentry2009fully} have been proposed in which the data is first encrypted before being shared to workers. However, the encryption techniques are found to be costly in terms of computation and storage costs \cite{brakerski2014efficient}. Besides, the secure multi-party computation (MPC) approaches \cite{goldreich1998secure,huang2011faster,bogdanov2008sharemind} mainly focus on the correctness and privacy of the data \cite{nodehi2018limited}, whereas neglecting to reduce the complexity of computation at each workers, or the number of workers required to perform the task. Recently, coding techniques that have originally been introduced to mitigate straggling workers are increasingly utilized to provide information-theoretic privacy guarantees. Specifically, information-theoretic privacy considers the setup consisting honest but curious workers, in which collusions formed between $T$ of $N$ workers do not leak information about the training dataset. Coding schemes can be used to not only mitigate the stragglers but also the colluding curious workers and malicious workers as illustrated in Fig.~\ref{fig:security}. For ease of exposition, we classify the related studies into two main categories in this section:

\begin{itemize}

\item \textit{Secure Distributed Computing:} In this section, the studies aim to reduce the number of workers needed for information-theoretic privacy, i.e., where the colluding workers are unable to infer sensitive information from the dataset. For some studies, this objective is met while simultaneously preserving the efficiency of distributed computing, e.g., through providing resiliency against straggling workers \cite{yu2018lagrange}. 

\item \textit{Secure Distributed Matrix Multiplication (SDMM):} The studies in the aforementioned category mainly focus on generic operations, e.g., addition, subtraction, multiplication, or computation of polynomial functions, whereas the studies in this category focus specifically on matrix multiplication. One key difference between the two categories is that SDMM considers the specific scenario in which \textit{both} input matrices in the multiplication operation are private information, i.e., two-sided privacy \cite{chang2018capacity}, whereas the prior category mainly considers one-sided privacy, i.e., only one input matrix is private. In addition, a performance metric of interest in the SDMM literature is the download rate, i.e., the ratio of the size of desired result to the total amount of information downloaded by a user from the workers.

\end{itemize}

\subsection{Secure Distributed Computing}
\label{subsec:securedist}

In Section~\ref{subsec:codingtech}, we discuss that the polynomial codes proposed in \cite{yu2017polynomial} have the desirable property of an optimal recovery threshold that does not scale with the number of workers involved. In consideration of this useful property, the authors in \cite{nodehi2018limited} propose the polynomial sharing approach which combines the polynomial codes and the Ben-Or, Goldwasser, and Wigderson (BGW) scheme \cite{ben2019completeness}. The system model considered in this study is that the data originates from external sources, and thus has to be kept private against both the workers and master node. In contrast to the BGW approach which uses Shamir's scheme to  encode the dataset, the study of  \cite{nodehi2018limited} proposes to encoded the dataset using the polynomial coding scheme. The authors show that the polynomial sharing approach may be applied to perform several procedures, e.g., addition, multiplication, and the computation of polynomial functions, while reducing the number of workers required to complete the task as compared to conventional MPC approaches even when workers have capacity-limited communication links. 

Typically, in conventional polynomial coding schemes, the dataset on which computations are performed is divided into multiple sub-tasks, with one sub-task encoded and assigned to each worker. In this case, faster workers that complete their task will be idle while waiting for straggling workers. To further mitigate the straggler effects, the authors in \cite{kim2018private} leverage on computation redundancy to propose the private asynchronous polynomial coding scheme in which a computation task is divided into several relatively smaller sub-tasks for distribution to each worker. This results in two key advantages, in addition to retaining the privacy preservation properties of polynomial coding. Firstly, the smaller sub-tasks can be successfully completed by straggling workers with limited computing capacity. Secondly, workers of the fastest groups are assigned more tasks to continue working throughout the whole duration rather than wait for the stragglers, thus reducing the computation time. 

However, the studies \cite{nodehi2018limited,kim2018private} mainly utilize polynomial coding for privacy preservation which is restrictive in certain aspects, e.g., it only allows column-wise partitioning of the matrices \cite{yu2020fundamental}. As such, the entangled polynomial codes \cite{yu2020fundamental} are applied by \cite{nodehi2018entangled} as an extension to polynomial sharing, so as to  further reduce the restrictions during the data sharing phase, and hence, the number of workers required to perform the same computations while meeting privacy constraints.

While the studies of \cite{nodehi2018limited,nodehi2018entangled} consider the scenario in which honest-but-curious workers are involved, workers may randomly be malicious in nature. As an illustration, a group of workers may be involved to compute gradients towards training a machine learning model. However, the gradients may be intentionally misreported by the workers to introduce biases or inaccuracies to the model \cite{lim2020federated}. An existing approach is to perform median, rather than mean, based aggregation of the gradients to eliminate misreports that are usually outliers \cite{chen2017distributed}. However, the median based aggregation is computationally costly and faces convergence issues. As such, the study of \cite{chen2018draco} proposes DRACO, which is based on the coding of gradients and algorithmic redundancy, i.e., each worker evaluates redundant gradients, such that the accurate gradients may be derived even in the presence of adversarial nodes. The simulation results show that DRACO is more than 3 times faster in achieving 90\% test accuracy for gradient computations on MNIST dataset as compared to the geometric median method.

%

\begin{table}
\caption{\small Comparison of BGW, LCC, Harmonic coding schemes for Gradient-type computations\cite{ben2019completeness,yu2018lagrange, yu2019harmonic}.}
\label{table:lccharmonic}
\centering

\begin{tabular}{|>{\centering\arraybackslash}m{2.0cm}|c|c|c|}
\hline
\rowcolor{mygray}
 & \textbf{Sharmir} & \textbf{LCC} & \textbf{Harmonic} \\ \hline
 Min. number of workers & $K($deg $g+1)$ & $K$deg $g$+1 & $K($deg $ g-1) +2$ \\ \hline
 \end{tabular}

\end{table}

An improvement to the studies of \cite{nodehi2018limited,nodehi2018entangled,chen2018draco} is done in \cite{yu2018lagrange}, which proposes LCC to achieve an optimum tradeoff between resiliency against straggling workers, security against malicious workers, and information-theoretic privacy amid colluding workers. In LCC, the dataset of the master is encoded using the Lagrange polynomial to create computational redundancy. Then, the coded data is shared to the workers for computation on the encoded data, as if the coding did not take place. In comparison with the BGW MPC scheme \cite{ben2019completeness}, LCC requires more workers. However, the Lagrange polynomial based encoding leads to a reduction in the amount of randomness required to encode the data, which translates to lower storage and computation costs incurred by each worker. The LCC also outperforms the BGW based polynomial sharing \cite{nodehi2018limited} in terms of communication costs, given that the polynomial sharing scheme requires a communication round for each bilinear operations. In addition, LCC is less computationally costly than DRACO \cite{chen2018draco}, which does not utilize the algebraic structure of the encoded gradients. However, the Lagrange coding only works for computations involving arbitrary multivariate polynomial functions of the input dataset. As an extension, the study of \cite{so2019codedprivateml} proposes CodedPrivateML which adopts polynomial approximations to handle the non-linearities of the gradient computation when the sigmoid function is involved, such that logistic regression can be conducted on LCC-encoded data while providing information-theoretic privacy for the dataset. Given that the advantages of the LCC is preserved, the experiments conducted on Amazon EC2 clusters validate that the proposed scheme is close to 34 times faster than the BGW based MPC approaches.

In light of the growing popularity of machine learning, the study of \cite{yu2019harmonic} proposes Harmonic coding for tasks specific to gradient-type computations, e.g., for loss function minimization in distributed model training. Harmonic coding leverages on the structure of the intermediate partial gradients computed to enable the cancellation of redundant results, such that the encoding and decoding process is more efficient. As such, for the same level of privacy constraint, Harmonic coding improves on Shamir's secret sharing scheme \cite{shamir1979share} and LCC \cite{yu2018lagrange} in terms of requiring fewer workers to compute gradient-type functions. This result is further summarized in Table \ref{table:lccharmonic}, where we present a comparison of the minimum number of workers required for the discussed schemes. Note that $K$ refers to the number of partitions of the input dataset, $g$ refers to the fixed multivariate polynomial , and deg refers to the degree of $g$. Like LCC, Harmonic coding can also be applied universally to any gradient-type function computation. As such, the data encoding can be performed before the computing task is specified, thus further reducing the delay in computation.

\subsection{Secure Distributed Matrix Multiplication (SDMM)}

Matrix multiplication is a key operation in many popular machine learning algorithms \cite{bitar2017minimize}, e.g., principal component analysis \cite{abdi2010principal}, support vector machines \cite{hearst1998support}, and gradient-based computations. While the reviewed studies in Section~\ref{subsec:securedist} discuss coded computing for privacy preservation in general operations, the studies to be discussed consider tailored strategies for SDMM.

In \cite{bitar2020staircase} and \cite{bitar2017minimize}, the authors propose the use of staircase codes in place of linear secret sharing codes, e.g., Shamir's codes \cite{shamir1979share}. As an illustration, we consider that a master encodes its data $A$ with random matrix $R$ into three secret shares before transmitting a share to each of the workers to perform matrix multiplication. When linear secret sharing codes are used, the data and random matrix are not segmented but instead encoded and transmitted as a whole (Table~\ref{table:staircase}). As such, the master has to wait for two \textit{full} responses from any two of three workers before being able to decode and derive the desired results. In contrast, when the staircase code is used, the data and random matrices are  segmented into sub-shares before transmission to the workers. When sufficient sub-tasks have  been completed by the workers, the master can then instruct the workers to cease computation. Clearly, the staircase coding approach reduces the computation cost of workers and communication costs incurred by the master node. Accordingly, the staircase coding approach can outperform the classical secret sharing code in terms of mean waiting time. With 4 workers considered, experiments conducted on the Amazon EC2 clusters show a 59\% decrease in mean waiting time using staircase codes. 

However, \cite{bitar2020staircase} and \cite{bitar2017minimize} still consider the case of one-sided privacy, i.e., the approach is designed to keep only one of two input matrices involved in SDMM operations private. As such, several studies have shifted the focus towards applying coding techniques to the specific case in which two-sided privacy, i.e., in which both input matrices are private, is ensured. Among the first such study is that of \cite{chang2018capacity}, which applies the aligned secret sharing for two-sided privacy. Specifically, the input matrices are split into submatrices and encoded with random matrices. Then, the undesired terms are aligned  such that the server only recovers the desired results and saves on communication costs. This leads to an improved download rate, i.e., the ratio of size of desired result to the amount of information downloaded by a user, over conventional secret sharing schemes. 

Following \cite{chang2018capacity}, the study of \cite{kakar2018rateefficiency} proposes an inductive approach to find a close-to-optimal partition of the input matrices in consideration of two metrics namely the download rate and the minimum number of required workers. The proposed scheme improves on download rate, number of tolerable colluding servers, and computational complexity as compared to the study of \cite{chang2018capacity}. Inspired by \cite{yu2017polynomial}, the polynomial coding scheme is also extended to SDMM operations and specifically convolution tasks in \cite{yang2019secure}, while preserving two-sided privacy, download rate similar to that of \cite{kakar2018rateefficiency}, and further mitigating the straggler effects. For convolution tasks, the authors leverage on the inherent property in which the sums of convolutions of sub-vectors may be used to derive the convolution result. Then, the upper and lower bounds of the recovery threshold is derived to show that an order-optimal recovery threshold is achieved, i.e., it does not scale with number of workers. 

However, the key weakness of \cite{chang2018capacity} and \cite{kakar2018rateefficiency} as indicated in \cite{jia2019capacity} is that the proposed theoretical results do not clarify the effect of matrix dimensions on the download rate, i.e., the download rates are derived in the case whereby matrices are simply assumed to have large dimensions but without any specifications otherwise. However, the study of \cite{jia2019capacity} found that the results for \cite{chang2018capacity,kakar2018rateefficiency} may be violated in some cases for differing relative dimensions of the input matrices. Under this context, the model proposed in \cite{jia2019capacity} allows the matrix dimensions to be specified, and a new converse bound for the two-sided security SDMM is derived.

In general, the encoded results of matrix multiplication are sent to the master, where interpolation is performed to obtain the multiplication results, i.e., coefficients of a polynomial. In \cite{chang2018capacity,kakar2018rateefficiency}, the encoding of the private matrices is such that the coefficients are mainly non-zero. In contrast, the study of \cite{oliveira2020gasp} propose the Gap Additive Secure Polynomial (GASP) codes such that there are as many zero coefficients as possible. This allows the product to be interpolated and the desired results to be derived with fewer number of evaluations performed, which implies that fewer workers are required to perform the matrix multiplication. To assign the exponents for decodability while having as many zero coefficients as possible, the authors propose the degree table to solve the combinatorial problem. In \cite{rafael2019degree}, the authors further generalize the GASP codes to be applicable to different partitions of input matrices and security parameters. The GASP codes are shown to outperform the approaches in \cite{chang2018capacity,kakar2018rateefficiency} in terms of download rate.

\begin{table}[!]

\caption{\small Comparison of linear secret sharing codes and staircase codes in the distribution of computation tasks in a system with three workers \cite{bitar2020staircase, bitar2017minimize}.}
\label{table:staircase}
\centering
\scalebox{0.9}{
\begin{tabular}{ll ll l}
\hline
\rowcolor{mygray}
\multicolumn{1}{|l|}{} & \multicolumn{1}{l|}{$S_1$} & \multicolumn{1}{l|}{$S_2$} & \multicolumn{1}{l|}{$S_3$}  \\ \hline
\multicolumn{1}{|l|}{\begin{tabular}[c]{@{}l@{}}Linear secret\\ sharing code\end{tabular}} & \multicolumn{1}{l|}{$R$}  & \multicolumn{1}{l|}{$R+A$}   & \multicolumn{1}{l|}{$R+2A$}   \\ \hline
\multicolumn{1}{|l|}{Staircase code} & \multicolumn{1}{l|}{\begin{tabular}[c]{@{}l@{}}$A_1+2A_2+4R_1$\\ $R_1+R_2$\end{tabular}} & \multicolumn{1}{l|}{\begin{tabular}[c]{@{}l@{}}$A_1+2A_2+4R_1$\\ $R_1+2R_2$\end{tabular}}   & \multicolumn{1}{l|}{\begin{tabular}[c]{@{}l@{}}$A_1+2A_2+4R_1$\\ $R_1+3R_2$\end{tabular}}  \\ \hline
\end{tabular}
}
\end{table}

\subsection{Summary and Lessons Learned}

In this section, we have discussed studies that have adopted a coded computing approach towards ensuring secure distributed computing. Then, we discuss the specific applications of SDMM, which require two-sided privacy. The summary and lessons learned are as follows:

\begin{itemize}

\item Originally proposed to mitigate stragglers in distributed computing systems, coding approaches have also been extended to ensure information theoretic privacy \cite{nodehi2018limited} and for some studies, security against malicious workers \cite{chen2018draco,yu2018lagrange}. However, given that the coding techniques, e.g., polynomial codes, are designed with the intention to mitigate stragglers, the studies discussed in this section have focused on modifications to the approaches to simultaneously achieve both straggler mitigation and privacy. As was mentioned in this section, the coding approaches to ensure privacy preservation have outperformed conventional secret sharing schemes that mainly focused on data correctness and privacy. 

\item With the recent popularity of machine learning, several studies have also focused on tailoring the coding approaches to be utilized for various machine learning tasks, e.g., entangled polynomial coding for logistic regression \cite{nodehi2018entangled} and harmonic coding for gradient based computations \cite{yu2019harmonic}. In particular, given the importance of two-sided privacy in SDMM, we have also specifically discussed papers that focus on two-sided privacy separate from papers that only considered one-sided privacy. In general, papers that focus on taking advantage of the algebraic structure of the specific operations \cite{yu2018lagrange}, e.g., convolution tasks \cite{yang2019secure} or gradient computations \cite{yu2019harmonic}, for efficient encoding and decoding usually perform better.

\item In most of the studies that we have discussed, the focus tends to be on information-theoretic privacy. However, weak security is key and may be explored in future works. Specifically, a system is weakly secure if  attackers  are unable to learn the sensitive intermediate values without having received a certain number of coded packets \cite{chang2010achieving}. This may be relevant when there exist eavesdroppers with an access to communication links between the master and workers \cite{yang2019secure}. For example, in the study of \cite{zhao2018weakly}, a data shuffling design and redundancy reduction algorithm to assign computing tasks have been proposed to ensure weak security in the system. However, the focus has been on workers, rather than eavesdroppers that may not be involved in the computations.

\item Given that the straggler effect is a fundamental concern in distributed computing, most papers have considered workers with heterogeneous computing capabilities. However, as discussed in \cite{yang2019secure}, the issue of heterogeneous networks in other aspects have been under-explored in the aforementioned studies. For example, the workers may have different levels of reputation \cite{resnick2000reputation}. This enables the adoption of context-sensitive solutions, e.g., data security may be guarded against new workers or workers with low reputation, whereas it may not be required for trusted workers. With this, the computation complexity and duration may be reduced in trusted nodes.

\end{itemize}

\section{CDC Applications}
\label{sec:applications}

\begin{figure}
\centering
\includegraphics[width=\linewidth]{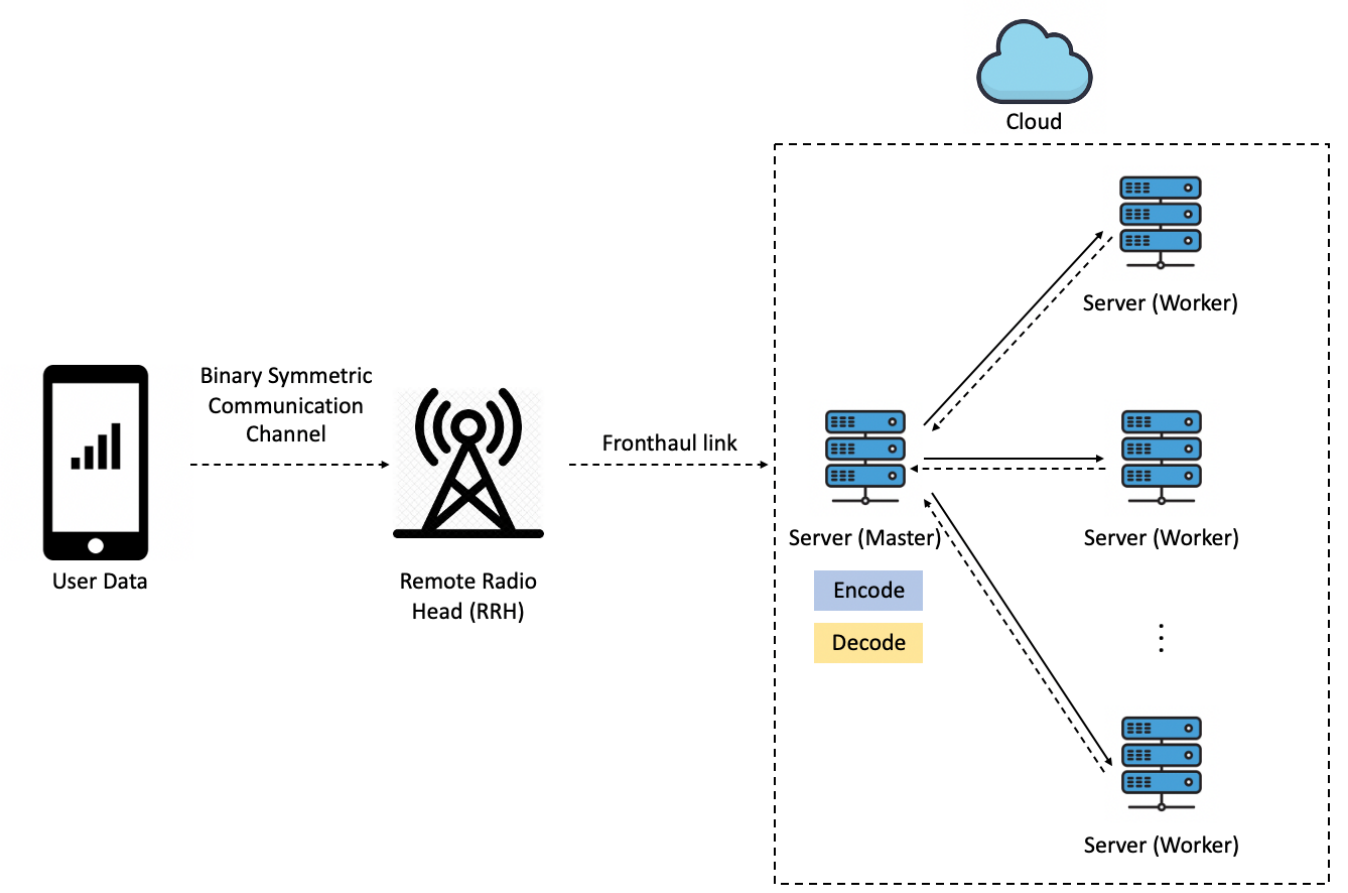}
\caption{\small Application of CDC to Network Function Virtualization (NFV) model for uplink channel decoding.}
\label{fig:nfv}
\end{figure}

\subsection{Network Function Virtualization (NFV)}

Network Function Virtualization serves as an enabling technology for optimizing the 5G as well as emerging 6G networks \cite{miljumbi2016network}, presenting a promising paradigm shift in the telecommunication service provisioning industry. By leveraging on virtualization technologies, NFV simplifies the management and operations of the networking services. In particular, NFV decouples the Network Functions (NFs) such as routing and baseband processing from the physical network equipment on which they operate. The NFs are mapped to the Virtual Network Functions (VNFs) that are supported by the Commercial Off-The-Shelf (COTS) physical resources which provide storage, networking and computational capabilities. As the software component in the network is decoupled from the hardware component by a virtualization layer, the VNFs can be easily implemented over the distributed network locations which have the required hardware resources. Due to the flexibility in the deployment of the VNFs, the NFV brings about significant reduction in operating and capital expenses. Moreover, the development of new networking services is faster and cheaper as the COTS resources can be instantiated easily to provide the required network connection services. Extensive literature has been carried out on the various aspects of NFV such as architectural designs \cite{miljumbi2016network, qi2014sdn}, resource allocation \cite{eramo2019proposal, sun2018reliable}, energy efficiency \cite{quzweeni2015energy}, performance improvement \cite{lingua2019acceleration} and security \cite{jang2015security, aljuhani2017virtualized}. More details can be found in \cite{miljumbi2016network, cherrared2019fault, anvith2019telecom}.

However, one of the limiting factors of the performance of NFV lies in the reliability of the COTS hardware resources \cite{miljumbi2016network, liu2016reliability}. Hardware failure due to several factors such as component malfunctioning, temporary unavailability and malicious attacks affects the implementation of NFV, hindering the provision of services. Apart from the fault-tolerant virtualization strategies that are based on the diversity approach which maps VNFs onto the various virtual machines (VMs) such that the probability of a disruptive failure is minimized \cite{cherrared2019fault}, coding approach can also be used to minimize computation latency in NFV. 

In the study of \cite{shuwaili2016fault}, the authors consider the highly complex uplink channel decoding of the Cloud Radio Access Network (C-RAN) architecture \cite{nikaein2015processing}, which is a key application of NFV. In this system, the users communicate with the cloud via Remote Radio Head (RRH). In order to ensure the reliability of channel decoding, the received data frames by the RRH are encoded by leveraging on their algebraic structures before being distributed to the different VMs as shown in Fig.~\ref{fig:nfv}. The simulation results show that the coding approach achieves lower probability of error for decoding at the cloud than that of the diversity-based approach. However, there are several assumptions that are made in this proposed scheme. Firstly, the binary symmetric communication channel is assumed between the users and the RRH. By considering other communication channels such as the additive Gaussian noise channels, different coding techniques may be applied. Secondly, this simple framework works well for a network with three processing nodes, but its performance for larger-size networks is not guaranteed. 

Considering the same issue of uplink channel decoding in the C-RAN architecture, the authors in \cite{aliasgari2019virtualized} propose a more generalized coded computation framework that works for any number of servers, random computing runtimes and random packet arrivals by adopting the coding approach proposed in \cite{shuwaili2016fault}. Given the randomness in the arrivals of data frames transmitted by the users, two queue management policies are considered: (i) per-frame decoding where one frame is decoded at any point of time, and (ii) continuous decoding where the servers start to decode the next packet of data frame upon completion of the first packet. There is an underlying tradeoff between the average decoding latency and the frame unavailability probability which is an indication of the reliability of the decoding process at the servers. The simulation results show that properly designed NFV codes are useful in achieving the desired tradeoffs by optimizing the minimum distance of the codes.

The idea of adopting coding approach in NFV is relatively new. Different coding techniques can be explored in the future. Besides, instead of addressing the uplink channel decoding, which is the most computationally-intensive baseband function \cite{nikaein2015processing}, other network functions such as routing and security functions can be considered.

\subsection{Edge Computing}

With the enhanced sensing capabilities of end devices, an overwhelming amount of data is produced at the edge of the network today. Traditional schemes of computation offloading to the cloud is thus unsustainable. Moreover, certain edge applications may involve end devices in remote areas that have limited connectivity. This necessitates a paradigm shift towards edge computing, in which computation is performed closer to the edge of the network where data is produced. However, resource-constrained devices may not be able to carry out complex computations individually, especially given the increasing size and complexity of state-of-the-art AI models \cite{frankle2018lottery}. As such, one of the enabling technologies of edge computing is cooperative computation in which the available resources of end devices and edge nodes, e.g., road side units in vehicular networking, can be pooled together to execute computation intensive tasks collaboratively \cite{lim2020incentive}. For ease of exposition, we refer to these participating end devices and edge nodes as workers in this section.

As the number of devices connected to the network increases, more information needs to be exchanged among the workers, resulting in high communication load. However, the communication bandwidth is fixed and thus the network is unable to handle the high communication load, causing a bottleneck as a result. Moreover, the heterogeneous nature of workers in the edge computing paradigm, e.g., in terms of computational and communication capabilities, can lead to the straggler effects.  In face of these challenges, coding techniques can be used.

In the study of \cite{li2017scalable}, the \emph{coded wireless distributed computing} (CWDC) framework is proposed. The system model consists of multiple devices, i.e., workers, involved in cooperative computation. As an illustration, a worker may have an input, e.g., an image, that has to be processed, e.g., for object recognition. Individually, a worker may not have the storage or computational capabilities to execute the task. Therefore, the inference model may be split and stored on each worker, whereas the cooperative computation of results can be implemented following the MapReduce framework as discussed in Section~\ref{subsec:mapreduce}. An access point, e.g., a base station or a Wi-Fi router can then be utilized to facilitate the exchange of intermediate results among workers. The proposed framework achieves communication loads that are independent of the size of the network and the storage size of the workers. Moreover, the CWDC framework can be generalized and applied to different types of applications.

In practical distributed computing systems, the workers may have heterogeneous computational, communication, and storage capabilities. Based on a similar system model proposed in \cite{li2017scalable} where the workers communicate with each other via an access point, the study in \cite{kiamari2017edge} considers devices with heterogeneous storage capacities over wireless networks. For uplink transmission, the allocation of files is based on the scheme proposed in \cite{kiamari2017heterogeneous} as previously discussed in Section~\ref{subsec:files} whereas for downlink transmission, data is encoded at the access point for the reduction of downlink communication load. However, the achievable scheme has only been validated in a small network that consists just three processing nodes. 

\begin{figure}
\centering
\includegraphics[width=\linewidth]{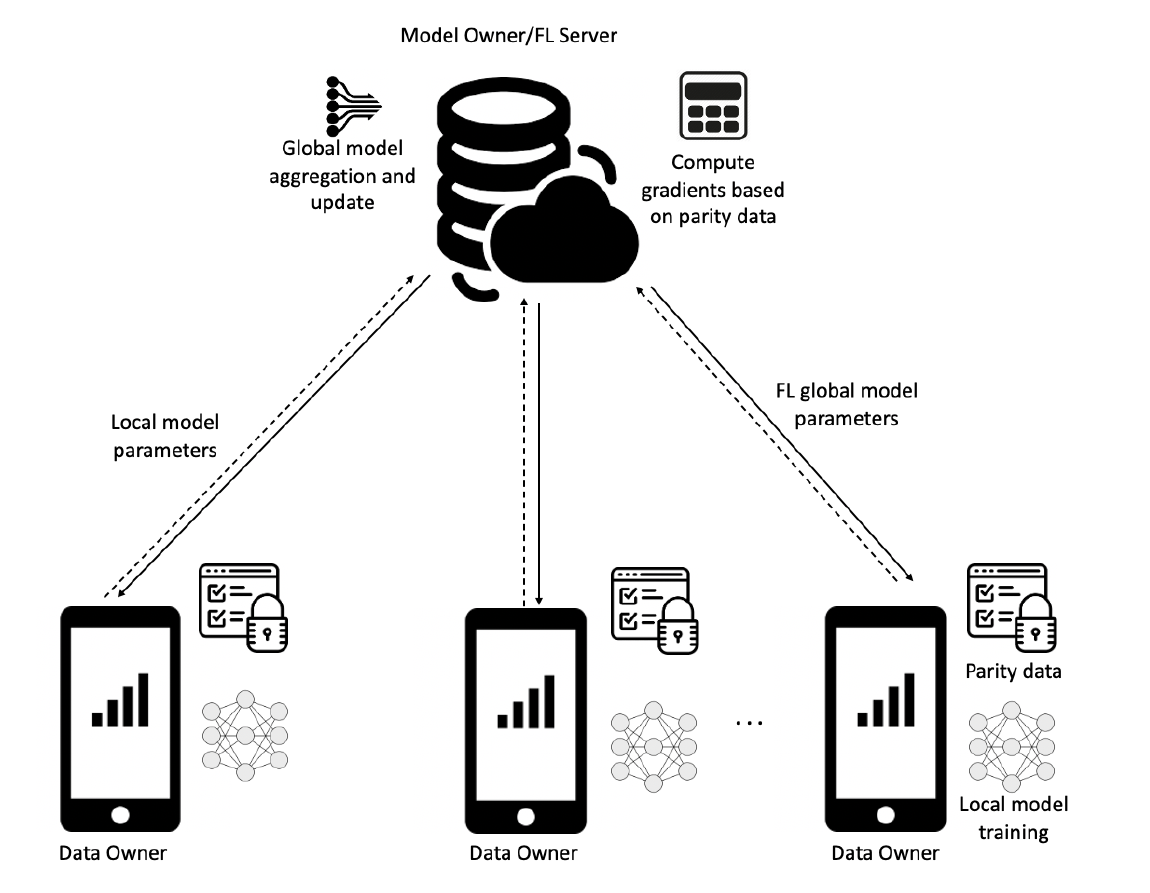}
\caption{\small Illustration of the Coded Federated Learning (CFL) scheme in a FL system that consists of multiple data owners and a FL model owner.}
\label{fig:flmodel}
\end{figure}

In light of the growing popularity of machine learning model training at the edge, the study of \cite{dhakal2019coded2} considers the distribution of gradient descent computations across workers in the network to train a linear regression model. The proposed heterogeneous coded gradient descent (HCGD) assigns each worker with an optimal load partition, through modelling the computation delay of devices with a shifted exponential distribution. In consideration of data privacy, the authors in \cite{dhakal2019coded} propose the Coded Federated Learning (CFL) approach for privacy-preserving linear regression. Federated Learning (FL) is a privacy preserving distributed machine learning paradigm proposed in \cite{mcmahan2017communication}, in which the sensitive data of data owners are kept locally, whereas only the model parameters are transmitted to the central server for aggregation to update the model. However, FL still suffers from the issues of straggling devices and communication inefficiency \cite{lim2020federated}. As such, in the proposed CFL scheme, each data owner first generates parity data from its local data for transmission to the central server. At the central server, the gradients are also computed from parity data simultaenously, such that only a subset of gradients from the data owners have to be received for completion of model update. Figure~\ref{fig:flmodel} illustrates the implementation of the CFL scheme. The CFL approach is also shown to converge almost four times faster than an uncoded approach. However, the communication and computation cost involved in generating and transmitting the parity data has not been well elaborated in the study. Clearly, a major weakness of aforementioned  studies of \cite{dhakal2019coded2,dhakal2019coded} is that they can only be applied to linear regression model training. The study of \cite{prakash2020coded} extends on the aforementioned works  with the  proposed CodedFedL designed to mitigate straggler mitigation in non-linear regression and classification tasks.

In some cases, the resource level of the devices may not be known by the network operator. To enable a dynamic and adaptive coded sub-task allocation for cooperative computation, an Automatic Repeat reQuest (ARQ) mechanism is proposed in the study of \cite{kesh2018dynamic}, in which devices are allocated with specific levels of packets for computation based on their responsiveness. Specifically, devices that are more responsive are assumed to have more available resources. These devices will thus be assigned with more sub-tasks for computation.

Instead of reducing communication load, it is also important to consider communication efficiency, i.e., achieved data rates, especially in wireless networks that have limited spectral resources or networks with mutual interference among users. In order to improve spectral efficiency, the co-channel communication model is proposed in the study of \cite{yang2019lowrank} which consists of two stages, i.e., the uplink multiple access stage and the downlink broadcasting stage. The communication model turns out to be equivalent to a multiple-input-multiple-output (MIMO) interference channel. Interference alignment \cite{zhao2016interference} has been an effective approach in handling mutual interference among users. The signals are precoded into the same subspaces at the unintended receivers and the desired signals are recovered at the intended receivers by using the decoding matrix. Linear coding scheme is adopted to establish the conditions for interference alignment. A low-rank optimization problem is formulated to minimize the number of channel used subject to the established interference alignment conditions. By solving this optimization problem, the achievable symmetric degree-of-freedom (DoF), which implies the extent to which interference is eliminated, can be maximized. In \cite{yang2019lowrank}, an efficient difference-of-convex-functions (DC) algorithm based on a DC representation for the rank function is proposed to solve the low-rank optimization problem. The performance of the DC algorithm is evaluated in two different scenarios by varying: (i) the number of files stored in the devices, and (ii) the number of antennas that are equipped by the devices. In the first scenario, as the number of files stored in the devices increases, the achievable DoF increases. Similarly in the second scenario, as the number of antennas equipped by the devices increases, the achievable DoF increases. Furthermore, the simulation results show that the DC approach achieves higher DoF than the existing benchmark algorithms, e.g., iterative reweighted least squares (IRLS) algorithm and the nuclear norm relaxation approach. However, the proposed scheme is based on a homogeneous network, i.e., the number of files stored is the same across all devices and all devices are equipped with the same number of antennas. As an extension, heterogeneous networks can be considered. 

Instead of communicating with each other via an access point, the devices can communicate directly with each other over wireless interference channels \cite{li2019wireless}. In particular, the transmission of data in the Shuffle phase operates over wireless interference channels. While the CDC scheme in \cite{li2018tradeoff} allows communications based on time-division multiple access (TDMA) scheme which allows each processing node to transmit one coded information packet at any time, the one-shot linear scheme adopted in \cite{li2019wireless} allows more than one processing node to transmit information simultaneously at any given time slot. The transmitted symbols are linear combinations of coded intermediate results from the processing nodes. The transmitted symbols are broadcasted to the processing nodes, following which the nodes can decode to recover the desired information. The study in \cite{li2019wireless} characterizes an improved computation-communication tradeoff as compared to the study in \cite{li2018tradeoff}. However, the proposed scheme operates under the assumption of perfect channel state information (CSI) where the CSI is available to all processing nodes. As such, the authors in \cite{ha2019wireless} propose superposition coding scheme which has better performance than that of the CDC scheme \cite{li2018tradeoff} and the one-shot linear scheme \cite{li2019wireless} under imperfect CSI condition. The study is extended to account for the presence of stragglers in the networks \cite{ha2019federated}.

In general, the aforementioned studies have considered conventional implementations of the CDC scheme in which each device has to complete the computation task first, before transmitting the intermediate results, e.g., to other devices via the access point. However, given the limited computational capabilities of devices, the latency involved can still be significant. As such, the studies in \cite{wang2019batchprocessing} and \cite{wang2019uav} consider the batch-processing based coded computing (BPCC) framework which allows each device to return only the partially completed results to a master node in batches, even before the task is fully completed. These partial results are then utilized to generate approximations to the full solution through the singular value decomposition (SVD) approach \cite{ferdinand2016anytime}. This is particularly useful in applications that require fast, but not necessarily optimized, results. The BPCC scheme has its effectiveness validated on the EC2 computing cluster, in which latency is proven to be reduced. Moreover, in consideration of the energy limitations of Unmanned Aerial Vehicles (UAV), the BPCC framework was proposed for UAV-based mobile edge computing to provide energy-efficient computation offloading support on-demand \cite{shyuan2020joint,lim2020towards}. 


\subsection{Summary and Lessons Learned}

In this section, we have discussed applications of CDC in NFV and edge computing. The lessons learned are as follows:

\begin{itemize}

\item The convergence of the recently popular edge computing and machine learning has given rise to edge intelligence, in which the computational capabilities of edge servers and devices have been increasingly leveraged to conduct machine learning model training closer to where the data is produced. While this leads to several benefits, e.g., lower training latency and enhanced privacy preservation, the problem of straggling devices is still a bottleneck. As such, CDC approaches have been increasingly applied in this context.

\item One major difference between distributed computing at the edge and at computing clusters is that the end devices and edge servers, i.e., workers, are not specifically dedicated to computations. For example, the workers may only share a fraction of their processing power \cite{dhakal2019coded2} at time slots in which they are idle. As such, the heterogeneity in computational capabilities of workers may also be greater in this regard. In face of these challenges, optimal load partition and allocation strategies have been explored. Moreover, for the future works, the dynamics of the system may be captured using Deep Reinforcement Learning based resource allocation \cite{mao2016resource}.

\item In edge  intelligence, there may be several data owners involved. Moreover, the data owners can be devices with computational constraints. As such, conventional techniques of data encoding before transmission to workers for computation may not be feasible. To meet this challenge, FL has been proposed in the work of \cite{mcmahan2017communication}, whereas CFL \cite{dhakal2019coded} and CodedFedL \cite{prakash2020coded} have been proposed to mitigate the straggler effects in FL. However, the proposed methods are highly restrictive. For example, the CFL can only be applied to linear regression problems, whereas both methods require costly computation and transmission of parity data. Moreover, they have not been implemented on typical end devices to validate the feasibility of the schemes in practical implementation. For the future works, studies on using CDC schemes in edge computing applications may adopt the approach of \cite{wang2019batchprocessing,wang2019uav}, in which the schemes are implemented under the context of practical hardware constraints.

\end{itemize}

\section{Challenges, Open Issues and Future Works}
\label{sec:challenges}

The utilization of CDC schemes to solve the implementation challenges of the distributed computing systems is a relatively new and recent approach. There are still challenges and open issues that have yet to be addressed and this provides opportunities for new research directions in the future. We present major challenges that need to be looked into for effective implementation of CDC schemes.

\begin{itemize}
 
\item \emph{Heterogeneous nodes: }As compared to traditional distributed computing clusters, heterogeneities among computing nodes are much more significant when the computing nodes are connected in the edge computing networks, e.g., smartphones and wearable devices. Many studies, e.g., \cite{kiamari2017heterogeneous}, \cite{woolsey2019coded},  \cite{xu2019heterogeneous}, \cite{reisizadeh2019} have considered heterogeneous systems where the processing nodes have different computational, communication and storage capabilities. For example, in \cite{xu2019heterogeneous}, a joint file and function allocation strategy is proposed to assign jobs to the processing nodes such that the communication load is minimized. In \cite{reisizadeh2019}, the computation load is allocated based on the capabilities of the processing nodes. However, other aspects of heterogeneity such as the reputation \cite{resnick2000reputation} and the willingness to participate of the processing nodes are not taken into account. New allocation strategies need to consider different aspects of heterogeneities of the processing devices so that coding techniques can be implemented effectively and securely. For example, the computation tasks can be allocated to workers with higher reputation which implies higher probability of the workers in completing their allocated tasks.

\item \emph{Encoding and decoding complexities: }The studies that we have discussed in Section~\ref{sec:comms} and Section~\ref{sec:stragglers} mainly minimize communication load in Shuffle phase and computation latency in the Map phase respectively. However, the complexities of encoding and decoding are often not evaluated. It is important to ensure low encoding and decoding complexities in order to minimize the overall job execution time. Otherwise, the speedup gain achieved by in specific phases, e.g., communication and computation phases may be offset by the high encoding and decoding complexities. For example, the UberShuffle algorithm \cite{chung2017ubershuffle} incurs high computational overhead under the fast broadcast environment, i.e., networks with large bandwidth, such that it is not feasible for implementation even though it achieves significant shuffling gain. Hence, to better assess the performance of the CDC schemes, the complexities of the encoding and decoding methods have to be evaluated.

\item \emph{Non-static computing nodes: }For the commonly used distributed computing models such as cluster computing, grid computing and cloud computing that are discussed in Section~\ref{sec:fundamentals}, the computing nodes are static, i.e., the nodes are located at fixed location. For example, the servers are located at specific data centers. Data required for computations is transmitted over the wireless communication channels to the servers. However, as the edge devices, e.g., IoT devices, wearable devices and vehicles, have greater communication and computational capabilities, new distributed computing models such as mobile edge computing \cite{mao2017mec}, \cite{abbas2018mec} and fog computing \cite{bonomi2012fog} have been developed recently. In \cite{li2017fog}, the basic CDC scheme is implemented in the context of fog computing. However, the edge devices are usually mobile. The data that is processed by the edge devices depends on the locations that they visit \cite{song2017benefit}, and hence the master node has no control of the data distribution to the workers. New coding approaches for edge and fog computing which involve moving workers can be proposed in future works. One proposed solution is to allocate Reduce functions based on the data stored at the processing nodes \cite{song2017benefit}.

\item \emph{Security concerns: }Coding techniques are able to mitigate the straggler effects while preserving privacy as shown in the studies of \cite{nodehi2018limited}, \cite{yu2018lagrange} and \cite{chen2018draco}. The proposed secure coding techniques are an extension to the coding techniques that are originally proposed to mitigate the straggler effects. As mentioned in \cite{yu2018lagrange}, the tradeoff between resiliency against straggling workers, security against malicious workers and information-theoretic privacy amid colluding workers needs to be carefully managed. In addition, there may exist eavesdroppers which tap on the less secure communication links between the master node and the workers. As such, more research effort can be directed towards developing weakly secure systems which prevents the eavesdroppers from retrieving sensitive information. 

\item \emph{Network architectures: }It is important to consider how the computing nodes are connected and communicate with each other for effective implementation of the CDC schemes. For example, in \cite{gupta2017locality} and  \cite{park2018hierarchical}, the authors introduce a hierarchical structure where the master communicates with multiple submaster and each submaster leads a group of computing nodes. From the studies that we have reviewed, the network architecture is only considered for the implementation of CDC schemes to reduce communication load. However, it is an important consideration when designing secure coding schemes. In practice, it may be safe for computing nodes within a group, e.g., from the same location, to share information freely with each other but not with computing nodes from another group, e.g., from different location. In addition, the communication channels are not perfect, i.e., they may not have perfect CSI \cite{ha2019wireless} or the transmitted information may have missing entries. Some networks may have limited spectral resources or suffer from mutual interference among users \cite{yang2019lowrank}. As such, future research can work towards designing coding schemes that can be implemented in practical distributed computing systems. Besides a need to design effective coding schemes, there is also a need to consider the design of low-cost, easily-implementable and scalable network architectures in which the coding schemes can be applied to.


\item \emph{Different computation frameworks: }Currently, most of the studies are based on the MapReduce computing model. Specifically, Coded MapReduce is proposed in \cite{li2015mapreduce} by implementing coding techniques in the MapReduce framework. However, there are limitations to MapReduce model that hinder its wide adoption for all types of distributed computation tasks as explained in Section~\ref{subsec:mapreduce}. In fact, there are other computing models such as Spark, Dryad and CIEL which support iterative algorithms, in which the feasibility of implementing coding techniques has not been explored. As such, the importance of these computing models motivates future directions such as the design of coding schemes that are specific to these computing models in order to solve any distributed computation tasks, e.g., convolution, Fourier transform and non-linear computations. 

\item \emph{Coding for both communication reduction and stragglers mitigation: }As we have discussed previously in Section~\ref{sec:comms} and Section~\ref{sec:stragglers}, coding techniques are used to either reduce communication load or mitigate the straggler effects. However, coding techniques cannot be applied to solve these implementation challenges simultaneously. As characterized in \cite{li2016unified}, there is a tradeoff between communication load and computation latency. However, in an ideal situation, both communication load and computation latency need to be minimized. Thus, it is important to carefully manage the tradeoff to achieve optimal performance of the distributed computing systems. For future works on CDC schemes, there is a need to improve the latency-communication tradeoff curve so that the time taken to execute the allocated tasks can be significantly reduced. In addition, instead of the two-dimensional tradeoff, the three-dimensional tradeoff between computation, communication and storage cost \cite{yan2018storage}, which is much more challenging to manage, should be considered.

\item \emph{CDC applications: }Given the advantages, CDC schemes are implemented in distributed computing applications such as the NFV and edge computing. Apart from the UAVs, CDC schemes can be extended to edge computing applications in other areas, e.g., vehicular networks, healthcare systems and industrial operations. For the studies that we have reviewed in CDC applications, the main focus lies in the implementation of coding techniques in various applications, without considering privacy and security. Given the importance of secure coding as discussed in Section~\ref{sec:security}, secure coding techniques need to be considered in the implementation of CDC applications. Besides, application-specific issues need to be addressed. For example, in vehicular networks where the vehicles are constantly moving, the CDC schemes need to be robust to vehicles which do not have consistent access to the wireless communication channels.

\end{itemize}

The idea of using coding techniques to overcome the challenges in distributed computing systems is relatively new. For effective implementation in practical distributed computing systems, various aspects such as the heterogeneities of the computing nodes and the network architectures are worth the in-depth studies. Some promising research directions presented in this survey serve as useful guidelines and valuable references for future research in CDC. 

\section{Conclusion}
\label{sec:conclusion}

In this paper, we provided a tutorial of CDC schemes and a comprehensive survey on the two main lines of CDC works. We first motivated the need for CDC schemes. The current performance of distributed computing systems can be improved using coding schemes. Then, we described the fundamentals and principles of CDC schemes. We also reviewed CDC works which aim to minimize communication costs, mitigate straggler effects as well as enhance privacy and security. In addition, we discussed the implementation of CDC schemes in practical distributed computing applications. Finally, we highlighted the challenges and discussed promising research directions.

\bibliographystyle{ieeetr}
\bibliography{cdc}

\begin{thebibliography}{100}

\bibitem{cristea2010large}
V.~Cristea, C.~Dobre, C.~Stratan, F.~Pop, and A.~Costan, {\em {Large-Scale
  Distributed Computing and Applications: Models and Trends: Models and
  Trends}}.
\newblock IGI Global, 2010.

\bibitem{kumar2005distributed}
V.~P. Kumar, V.~K. Prasanna, S.~Iyengar, P.~Spirakis, and M.~Welsh, {\em
  {Distributed Computing in Sensor Systems}}.
\newblock Springer Science \& Business Media, 2005.

\bibitem{corbett2013spanner}
J.~C. Corbett, J.~Dean, M.~Epstein, A.~Fikes, C.~Frost, J.~J. Furman,
  S.~Ghemawat, A.~Gubarev, C.~Heiser, P.~Hochschild, {\em et~al.}, ``{Spanner:
  Google's Globally Distributed Database},'' {\em ACM Transactions on Computer
  Systems (TOCS)}, vol.~31, no.~3, pp.~1--22, 2013.

\bibitem{gonzaga2009ann}
J.~Gonzaga, L.~A.~C. Meleiro, C.~Kiang, and R.~Maciel~Filho, ``{ANN-based
  Soft-sensor for Real-time Process Monitoring and Control of an Industrial
  Polymerization Process},'' {\em Computers \& chemical engineering}, vol.~33,
  no.~1, pp.~43--49, 2009.

\bibitem{de2015structured}
A.~E. De~Giusti, ``{Structured Parallel Programming: Patterns for Efficient
  Computation},'' {\em Journal of Computer Science and Technology}, vol.~15,
  no.~01, pp.~43--44, 2015.

\bibitem{dean2008mapreduce}
J.~Dean and S.~Ghemawat, ``{MapReduce: Simplified Data Processing on Large
  Clusters},'' {\em Commun. ACM}, vol.~51, p.~107–113, Jan. 2008.

\bibitem{ahmad2012tarazu}
F.~Ahmad, S.~T. Chakradhar, A.~Raghunathan, and T.~N. Vijaykumar, ``{Tarazu:
  Optimizing MapReduce on Heterogeneous Clusters},'' {\em SIGARCH Comput.
  Archit. News}, vol.~40, p.~61–74, Mar. 2012.

\bibitem{guo2017ishuffle}
Y.~{Guo}, J.~{Rao}, D.~{Cheng}, and X.~{Zhou}, ``{iShuffle: Improving Hadoop
  Performance with Shuffle-on-Write},'' {\em IEEE Transactions on Parallel and
  Distributed Systems}, vol.~28, no.~6, pp.~1649--1662, 2017.

\bibitem{chowdhury2011managing}
M.~Chowdhury, M.~Zaharia, J.~Ma, M.~I. Jordan, and I.~Stoica, ``{Managing Data
  Transfers in Computer Clusters with Orchestra},'' in {\em Proceedings of the
  ACM SIGCOMM 2011 Conference}, SIGCOMM ’11, (New York, NY, USA),
  p.~98–109, Association for Computing Machinery, 2011.

\bibitem{zhang2013performance}
Z.~{Zhang}, L.~{Cherkasova}, and B.~T. {Loo}, ``{Performance Modeling of
  MapReduce Jobs in Heterogeneous Cloud Environments},'' in {\em 2013 IEEE
  Sixth International Conference on Cloud Computing}, pp.~839--846, 2013.

\bibitem{he2016deep}
K.~He, X.~Zhang, S.~Ren, and J.~Sun, ``{Deep Residual Learning for Image
  Recognition},'' in {\em The IEEE Conference on Computer Vision and Pattern
  Recognition (CVPR)}, June 2016.

\bibitem{alex2012imagenet}
A.~Krizhevsky, I.~Sutskever, and G.~E. Hinton, ``{ImageNet Classification with
  Deep Convolutional Neural Networks},'' in {\em Advances in Neural Information
  Processing Systems 25} (F.~Pereira, C.~J.~C. Burges, L.~Bottou, and K.~Q.
  Weinberger, eds.), pp.~1097--1105, Curran Associates, Inc., 2012.

\bibitem{attia2017combating}
M.~A. Attia and R.~Tandon, ``{Combating Computational Heterogeneity in
  Large-Scale Distributed Computing via Work Exchange},'' {\em arXiv preprint
  arXiv:1711.08452}, 2017.

\bibitem{wang2015replication}
D.~Wang, G.~Joshi, and G.~Wornell, ``{Using Straggler Replication to Reduce
  Latency in Large-Scale Parallel Computing},'' {\em SIGMETRICS Perform. Eval.
  Rev.}, vol.~43, p.~7–11, Nov. 2015.

\bibitem{maity2019ldpc}
R.~K. {Maity}, A.~{Singh Rawa}, and A.~{Mazumdar}, ``{Robust Gradient Descent
  via Moment Encoding and LDPC Codes},'' in {\em 2019 IEEE International
  Symposium on Information Theory (ISIT)}, pp.~2734--2738, 2019.

\bibitem{maddah2014fundamental}
M.~A. Maddah-Ali and U.~Niesen, ``{Fundamental Limits of Caching},'' {\em IEEE
  Transactions on Information Theory}, vol.~60, no.~5, pp.~2856--2867, 2014.

\bibitem{li2018tradeoff}
S.~{Li}, M.~A. {Maddah-Ali}, Q.~{Yu}, and A.~S. {Avestimehr}, ``{A Fundamental
  Tradeoff Between Computation and Communication in Distributed Computing},''
  {\em IEEE Transactions on Information Theory}, vol.~64, no.~1, pp.~109--128,
  2018.

\bibitem{li2015mapreduce}
S.~{Li}, M.~A. {Maddah-Ali}, and A.~S. {Avestimehr}, ``{Coded MapReduce},'' in
  {\em 2015 53rd Annual Allerton Conference on Communication, Control, and
  Computing (Allerton)}, pp.~964--971, 2015.

\bibitem{lee2018speeding}
K.~{Lee}, M.~{Lam}, R.~{Pedarsani}, D.~{Papailiopoulos}, and K.~{Ramchandran},
  ``{Speeding Up Distributed Machine Learning Using Codes},'' {\em IEEE
  Transactions on Information Theory}, vol.~64, no.~3, pp.~1514--1529, 2018.

\bibitem{nodehi2018limited}
H.~A. {Nodehi} and M.~A. {Maddah-Ali}, ``{Limited-Sharing Multi-Party
  Computation for Massive Matrix Operations},'' in {\em 2018 IEEE International
  Symposium on Information Theory (ISIT)}, pp.~1231--1235, 2018.

\bibitem{yu2018lagrange}
Q.~Yu, S.~Li, N.~Raviv, S.~M.~M. Kalan, M.~Soltanolkotabi, and S.~Avestimehr,
  ``{Lagrange Coded Computing: Optimal Design for Resiliency, Security and
  Privacy},'' {\em arXiv preprint arXiv:1806.00939}, 2018.

\bibitem{shi2016edge}
W.~{Shi}, J.~{Cao}, Q.~{Zhang}, Y.~{Li}, and L.~{Xu}, ``{Edge Computing: Vision
  and Challenges},'' {\em IEEE Internet of Things Journal}, vol.~3, no.~5,
  pp.~637--646, 2016.

\bibitem{lim2020incentive}
W.~Y.~B. Lim, J.~S. Ng, Z.~Xiong, D.~Niyato, C.~Leung, C.~Miao, and Q.~Yang,
  ``{Incentive Mechanism Design for Resource Sharing in Collaborative Edge
  Learning},'' {\em arXiv preprint arXiv:2006.00511}, 2020.

\bibitem{mao2017mec}
Y.~{Mao}, C.~{You}, J.~{Zhang}, K.~{Huang}, and K.~B. {Letaief}, ``{A Survey on
  Mobile Edge Computing: The Communication Perspective},'' {\em IEEE
  Communications Surveys Tutorials}, vol.~19, no.~4, pp.~2322--2358, 2017.

\bibitem{krauter2002taxonomy}
K.~Krauter, R.~Buyya, and M.~Maheswaran, ``{A Taxonomy and Survey of Grid
  Resource Management Systems for Distributed Computing},'' {\em Software:
  Practice and Experience}, vol.~32, no.~2, pp.~135--164, 2002.

\bibitem{hussain2013survey}
H.~Hussain, S.~U.~R. Malik, A.~Hameed, S.~U. Khan, G.~Bickler, N.~Min-Allah,
  M.~B. Qureshi, L.~Zhang, W.~Yongji, N.~Ghani, J.~Kolodziej, A.~Y. Zomaya,
  C.-Z. Xu, P.~Balaji, A.~Vishnu, F.~Pinel, J.~E. Pecero, D.~Kliazovich,
  P.~Bouvry, H.~Li, L.~Wang, D.~Chen, and A.~Rayes, ``{A Survey on Resource
  Allocation in High Performance Distributed Computing Systems},'' {\em
  Parallel Computing}, vol.~39, no.~11, pp.~709 -- 736, 2013.

\bibitem{datla2012wireless}
D.~Datla, X.~Chen, T.~Tsou, S.~Raghunandan, S.~S. Hasan, J.~H. Reed, C.~B.
  Dietrich, T.~Bose, B.~Fette, and J.-H. Kim, ``{Wireless Distributed
  Computing: A Survey of Research Challenges},'' {\em IEEE Communications
  Magazine}, vol.~50, no.~1, pp.~144--152, 2012.

\bibitem{ahuja2006survey}
S.~P. Ahuja and J.~R. Myers, ``{A Survey on Wireless Grid Computing},'' {\em
  The Journal of Supercomputing}, vol.~37, no.~1, pp.~3--21, 2006.

\bibitem{valentini2013overview}
G.~L. Valentini, W.~Lassonde, S.~U. Khan, N.~Min-Allah, S.~A. Madani, J.~Li,
  L.~Zhang, L.~Wang, N.~Ghani, J.~Kolodziej, H.~Li, A.~Y. Zomaya, C.-Z. Xu,
  P.~Balaji, A.~Vishnu, F.~Pinel, J.~E. Pecero, D.~Kliazovich, and P.~Bouvry,
  ``{An Overview of Energy Efficiency Techniques in Cluster Computing
  Systems},'' {\em Cluster Computing}, vol.~16, no.~1, pp.~3--15, 2013.

\bibitem{sadashiv2011comparison}
N.~{Sadashiv} and S.~M.~D. {Kumar}, ``{Cluster, Grid and Cloud Computing: A
  Detailed Comparison},'' in {\em 2011 6th International Conference on Computer
  Science Education (ICCSE)}, pp.~477--482, 2011.

\bibitem{idrissi2013cloud}
H.~{Kamal Idrissi}, A.~{Kartit}, and M.~{El Marraki}, ``{A Taxonomy and Survey
  of Cloud Computing},'' in {\em 2013 National Security Days (JNS3)}, pp.~1--5,
  2013.

\bibitem{xu2004distributed}
Y.~Xu and H.~Qi, ``{Distributed Computing Paradigms for Collaborative Signal
  and Information Processing in Sensor Networks},'' {\em Journal of Parallel
  and Distributed Computing}, vol.~64, no.~8, pp.~945 -- 959, 2004.

\bibitem{khan2017handbook}
S.~U. Khan, A.~Y. Zomaya, and A.~Abbas, {\em {Handbook of Large-Scale
  Distributed Computing in Smart Healthcare}}.
\newblock Springer, 2017.

\bibitem{tang107incorporating}
B.~{Tang}, Z.~{Chen}, G.~{Hefferman}, S.~{Pei}, T.~{Wei}, H.~{He}, and
  Q.~{Yang}, ``{Incorporating Intelligence in Fog Computing for Big Data
  Analysis in Smart Cities},'' {\em IEEE Transactions on Industrial
  Informatics}, vol.~13, no.~5, pp.~2140--2150, 2017.

\bibitem{raghavan2002dpac}
N.~R.~S. {Raghavan} and T.~{Waghmare}, ``{DPAC: An Object-oriented Distributed
  and Parallel Computing Framework for Manufacturing Applications},'' {\em IEEE
  Transactions on Robotics and Automation}, vol.~18, no.~4, pp.~431--443, 2002.

\bibitem{juan2013vehicle}
A.~A. Juan, J.~Faulin, J.~Jorba, J.~Caceres, and J.~M. Marqu{\`e}s, ``{Using
  Parallel \& Distributed Computing for Real-time Solving of Vehicle Routing
  Problems with Stochastic Demands},'' {\em Annals of Operations Research},
  vol.~207, no.~1, pp.~43--65, 2013.

\bibitem{ahmad1995task}
I.~{Ahmad}, M.~K. {Dhodhi}, and A.~{Ghafoor}, ``{Task Assignment in Distributed
  Computing Systems},'' in {\em Proceedings International Phoenix Conference on
  Computers and Communications}, pp.~49--53, 1995.

\bibitem{kafil1998optimal}
M.~{Kafil} and I.~{Ahmad}, ``{Optimal Task Assignment in Heterogeneous
  Distributed Computing Systems},'' {\em IEEE Concurrency}, vol.~6, no.~3,
  pp.~42--50, 1998.

\bibitem{woo1997task}
{Sung-Ho Woo}, {Sung-Bong Yang}, {Shin-Dug Kim}, and {Tack-Don Han}, ``{Task
  Scheduling in Distributed Computing Systems with a Genetic Algorithm},'' in
  {\em Proceedings High Performance Computing on the Information Superhighway.
  HPC Asia '97}, pp.~301--305, 1997.

\bibitem{lopes2016taxonomy}
R.~V. {Lopes} and D.~{Menascé}, ``{A Taxonomy of Job Scheduling on Distributed
  Computing Systems},'' {\em IEEE Transactions on Parallel and Distributed
  Systems}, vol.~27, no.~12, pp.~3412--3428, 2016.

\bibitem{ranjan2008incentive}
R.~Ranjan, A.~Harwood, and R.~Buyya, ``{A Case for Cooperative and
  Incentive-based Federation of Distributed Clusters},'' {\em Future Generation
  Computer Systems}, vol.~24, no.~4, pp.~280 -- 295, 2008.

\bibitem{duan2012incentive}
L.~{Duan}, T.~{Kubo}, K.~{Sugiyama}, J.~{Huang}, T.~{Hasegawa}, and
  J.~{Walrand}, ``{Incentive Mechanisms for Smartphone Collaboration in Data
  Acquisition and Distributed Computing},'' in {\em 2012 Proceedings IEEE
  INFOCOM}, pp.~1701--1709, 2012.

\bibitem{xiao2007security}
Y.~Xiao, {\em {Security in Distributed, Grid, Mobile, and Pervasive
  Computing}}.
\newblock CRC Press, 2007.

\bibitem{pllana2007performance}
S.~Pllana, I.~Brandic, and S.~Benkner, ``{Performance Modeling and Prediction
  of Parallel and Distributed Computing Systems: A Survey of the State of the
  Art},'' in {\em First International Conference on Complex, Intelligent and
  Software Intensive Systems (CISIS'07)}, pp.~279--284, IEEE, 2007.

\bibitem{jiang2010performance}
D.~Jiang, B.~C. Ooi, L.~Shi, and S.~Wu, ``{The Performance of MapReduce: An
  in-Depth Study},'' {\em Proc. VLDB Endow.}, vol.~3, p.~472–483, Sept. 2010.

\bibitem{zaharia2010spark}
M.~Zaharia, M.~Chowdhury, M.~J. Franklin, S.~Shenker, I.~Stoica, {\em et~al.},
  ``{Spark: Cluster Computing with Working Sets.},'' {\em HotCloud}, vol.~10,
  no.~10-10, p.~95, 2010.

\bibitem{isard2007dryad}
M.~Isard, M.~Budiu, Y.~Yu, A.~Birrell, and D.~Fetterly, ``{Dryad: Distributed
  Data-Parallel Programs from Sequential Building Blocks},'' in {\em
  Proceedings of the 2nd ACM SIGOPS/EuroSys European Conference on Computer
  Systems 2007}, EuroSys ’07, (New York, NY, USA), p.~59–72, Association
  for Computing Machinery, 2007.

\bibitem{murray2011ciel}
D.~G. Murray, M.~Schwarzkopf, C.~Smowton, S.~Smith, A.~Madhavapeddy, and
  S.~Hand, ``{CIEL: A Universal Execution Engine for Distributed Data-flow
  Computing},'' in {\em Proc. 8th ACM/USENIX Symposium on Networked Systems
  Design and Implementation}, pp.~113--126, 2011.

\bibitem{joshi2017redundancy}
G.~Joshi, E.~Soljanin, and G.~Wornell, ``{Efficient Redundancy Techniques for
  Latency Reduction in Cloud Systems},'' {\em ACM Trans. Model. Perform. Eval.
  Comput. Syst.}, vol.~2, Apr. 2017.

\bibitem{weets2015limitations}
J.~{Weets}, M.~K. {Kakhani}, and A.~{Kumar}, ``{Limitations and Challenges of
  HDFS and MapReduce},'' in {\em 2015 International Conference on Green
  Computing and Internet of Things (ICGCIoT)}, pp.~545--549, 2015.

\bibitem{hadoop_terasort}
``{Hadoop Terasort},'' Aug. 13 2020.
\newblock
  \url{https://hadoop.apache.org/docs/current/api/org/apache/hadoop/examples/terasort/package-summary.html}.

\bibitem{retch2013parallel}
B.~Recht and C.~R{\'e}, ``{Parallel Stochastic Gradient Algorithms for
  Large-Scale Matrix Completion},'' {\em Mathematical Programming Computation},
  vol.~5, no.~2, pp.~201--226, 2013.

\bibitem{bottou2012stochastic}
L.~Bottou, ``{Stochastic Gradient Descent Tricks},'' in {\em Neural networks:
  Tricks of the trade}, pp.~421--436, Springer, 2012.

\bibitem{ioffe2015batch}
S.~Ioffe and C.~Szegedy, ``{Batch Normalization: Accelerating Deep Network
  Training by Reducing Internal Covariate Shift},'' {\em arXiv preprint
  arXiv:1502.03167}, 2015.

\bibitem{ahmad2013marco}
F.~Ahmad, S.~Lee, M.~Thottethodi, and T.~Vijaykumar, ``Mapreduce with
  communication overlap (marco),'' {\em Journal of Parallel and Distributed
  Computing}, vol.~73, no.~5, pp.~608 -- 620, 2013.

\bibitem{nicolae2017leveraging}
B.~{Nicolae}, C.~H.~A. {Costa}, C.~{Misale}, K.~{Katrinis}, and Y.~{Park},
  ``{Leveraging Adaptive I/O to Optimize Collective Data Shuffling Patterns for
  Big Data Analytics},'' {\em IEEE Transactions on Parallel and Distributed
  Systems}, vol.~28, no.~6, pp.~1663--1674, 2017.

\bibitem{yu2015virtual}
W.~{Yu}, Y.~{Wang}, X.~{Que}, and C.~{Xu}, ``{Virtual Shuffling for Efficient
  Data Movement in MapReduce},'' {\em IEEE Transactions on Computers}, vol.~64,
  no.~2, pp.~556--568, 2015.

\bibitem{isard2009quincy}
M.~Isard, V.~Prabhakaran, J.~Currey, U.~Wieder, K.~Talwar, and A.~Goldberg,
  ``{Quincy: Fair Scheduling for Distributed Computing Clusters},'' in {\em
  Proceedings of the ACM SIGOPS 22nd Symposium on Operating Systems
  Principles}, SOSP ’09, (New York, NY, USA), p.~261–276, Association for
  Computing Machinery, 2009.

\bibitem{suresh2014scheduling}
S.~Suresh and N.~Gopalan, ``{An Optimal Task Selection Scheme for Hadoop
  Scheduling},'' {\em IERI Procedia}, vol.~10, pp.~70 -- 75, 2014.
\newblock International Conference on Future Information Engineering (FIE
  2014).

\bibitem{xie2013research}
J.~Xie, F.~Meng, H.~Wang, H.~Pan, J.~Cheng, and X.~Qin, ``Research on
  scheduling scheme for hadoop clusters,'' {\em Procedia Computer Science},
  vol.~18, pp.~2468 -- 2471, 2013.
\newblock 2013 International Conference on Computational Science.

\bibitem{hadoop_2020}
``{Hadoop: Fair Scheduler},'' Jul. 6 2020.
\newblock
  \url{https://hadoop.apache.org/docs/current/hadoop-yarn/hadoop-yarn-site/FairScheduler.html}.

\bibitem{zaharia2010delay}
M.~Zaharia, D.~Borthakur, J.~Sen~Sarma, K.~Elmeleegy, S.~Shenker, and
  I.~Stoica, ``{Delay Scheduling: A Simple Technique for Achieving Locality and
  Fairness in Cluster Scheduling},'' in {\em Proceedings of the 5th European
  Conference on Computer Systems}, EuroSys ’10, (New York, NY, USA),
  p.~265–278, Association for Computing Machinery, 2010.

\bibitem{rao2012survey}
B.~T. Rao and L.~S.~S. Reddy, ``{Survey on Improved Scheduling in Hadoop
  MapReduce in Cloud Environments},'' {\em arXiv preprint arXiv:1207.0780},
  2012.

\bibitem{dean2013tail}
J.~Dean and L.~A. Barroso, ``{The Tail at Scale},'' {\em Commun. ACM}, vol.~56,
  p.~74–80, Feb. 2013.

\bibitem{ananthanarayanan2010reining}
G.~Ananthanarayanan, S.~Kandula, A.~G. Greenberg, I.~Stoica, Y.~Lu, B.~Saha,
  and E.~Harris, ``{Reining in the Outliers in Map-Reduce Clusters using
  Mantri.},'' in {\em Osdi}, p.~24, 2010.

\bibitem{blumofe1999stealing}
R.~D. Blumofe and C.~E. Leiserson, ``{Scheduling Multithreaded Computations by
  Work Stealing},'' {\em J. ACM}, vol.~46, p.~720–748, Sept. 1999.

\bibitem{gardner2015exact}
K.~Gardner, S.~Zbarsky, S.~Doroudi, M.~Harchol-Balter, and E.~Hyytia,
  ``{Reducing Latency via Redundant Requests: Exact Analysis},'' in {\em
  Proceedings of the 2015 ACM SIGMETRICS International Conference on
  Measurement and Modeling of Computer Systems}, SIGMETRICS ’15, (New York,
  NY, USA), p.~347–360, Association for Computing Machinery, 2015.

\bibitem{shah2016redundant}
N.~B. {Shah}, K.~{Lee}, and K.~{Ramchandran}, ``{When Do Redundant Requests
  Reduce Latency?},'' {\em IEEE Transactions on Communications}, vol.~64,
  no.~2, pp.~715--722, 2016.

\bibitem{aktas2018relaunch}
M.~F. Aktas, P.~Peng, and E.~Soljanin, ``{Straggler Mitigation by Delayed
  Relaunch of Tasks},'' {\em SIGMETRICS Perform. Eval. Rev.}, vol.~45,
  p.~224–231, Mar. 2018.

\bibitem{aktas2017clones}
M.~F. Aktas, P.~Peng, and E.~Soljanin, ``{Effective Straggler Mitigation: Which
  Clones Should Attack and When?},'' {\em SIGMETRICS Perform. Eval. Rev.},
  vol.~45, p.~12–14, Oct. 2017.

\bibitem{yan2018storage}
Q.~{Yan}, S.~{Yang}, and M.~{Wigger}, ``{Storage, Computation, and
  Communication: A Fundamental Tradeoff in Distributed Computing},'' in {\em
  2018 IEEE Information Theory Workshop (ITW)}, pp.~1--5, 2018.

\bibitem{li2017terasort}
S.~{Li}, S.~{Supittayapornpong}, M.~A. {Maddah-Ali}, and S.~{Avestimehr},
  ``{Coded TeraSort},'' in {\em 2017 IEEE International Parallel and
  Distributed Processing Symposium Workshops (IPDPSW)}, pp.~389--398, 2017.

\bibitem{ezzeldin2017alternative}
Y.~H. {Ezzeldin}, M.~{Karmoose}, and C.~{Fragouli}, ``{Communication vs
  Distributed Computation: An Alternative Trade-off Curve},'' in {\em 2017 IEEE
  Information Theory Workshop (ITW)}, pp.~279--283, 2017.

\bibitem{mallick2019rateless}
A.~Mallick, M.~Chaudhari, U.~Sheth, G.~Palanikumar, and G.~Joshi, ``{Rateless
  Codes for Near-Perfect Load Balancing in Distributed Matrix-Vector
  Multiplication},'' {\em Proc. ACM Meas. Anal. Comput. Syst.}, vol.~3, Dec.
  2019.

\bibitem{dutta2016shortdot}
S.~Dutta, V.~Cadambe, and P.~Grover, ``{Short-Dot: Computing Large Linear
  Transforms Distributedly Using Coded Short Dot Products},'' in {\em Advances
  in Neural Information Processing Systems 29} (D.~D. Lee, M.~Sugiyama, U.~V.
  Luxburg, I.~Guyon, and R.~Garnett, eds.), pp.~2100--2108, Curran Associates,
  Inc., 2016.

\bibitem{wang2018fundamental}
S.~Wang, J.~Liu, N.~Shroff, and P.~Yang, ``{Fundamental Limits of Coded Linear
  Transform},'' {\em arXiv preprint arXiv:1804.09791}, 2018.

\bibitem{lee2017high}
K.~{Lee}, C.~{Suh}, and K.~{Ramchandran}, ``{High-dimensional Coded Matrix
  Multiplication},'' in {\em 2017 IEEE International Symposium on Information
  Theory (ISIT)}, pp.~2418--2422, 2017.

\bibitem{yu2017polynomial}
Q.~Yu, M.~Maddah-Ali, and S.~Avestimehr, ``{Polynomial Codes: an Optimal Design
  for High-Dimensional Coded Matrix Multiplication},'' in {\em Advances in
  Neural Information Processing Systems 30} (I.~Guyon, U.~V. Luxburg,
  S.~Bengio, H.~Wallach, R.~Fergus, S.~Vishwanathan, and R.~Garnett, eds.),
  pp.~4403--4413, Curran Associates, Inc., 2017.

\bibitem{fahim2017optimal}
M.~{Fahim}, H.~{Jeong}, F.~{Haddadpour}, S.~{Dutta}, V.~{Cadambe}, and
  P.~{Grover}, ``{On the Optimal Recovery Threshold of Coded Matrix
  Multiplication},'' in {\em 2017 55th Annual Allerton Conference on
  Communication, Control, and Computing (Allerton)}, pp.~1264--1270, 2017.

\bibitem{wang2018coded}
S.~Wang, J.~Liu, and N.~Shroff, ``{Coded Sparse Matrix Multiplication},'' {\em
  arXiv preprint arXiv:1802.03430}, 2018.

\bibitem{tandon2017gradient}
R.~Tandon, Q.~Lei, A.~G. Dimakis, and N.~Karampatziakis, ``{Gradient Coding:
  Avoiding Stragglers in Distributed Learning},'' in {\em Proceedings of the
  34th International Conference on Machine Learning} (D.~Precup and Y.~W. Teh,
  eds.), vol.~70 of {\em Proceedings of Machine Learning Research},
  (International Convention Centre, Sydney, Australia), pp.~3368--3376, PMLR,
  06--11 Aug 2017.

\bibitem{raviv2017gradient}
N.~Raviv, I.~Tamo, R.~Tandon, and A.~G. Dimakis, ``{Gradient Coding from Cyclic
  MDS Codes and Expander Graphs},'' {\em arXiv preprint arXiv:1707.03858},
  2017.

\bibitem{halbawi2018solomon}
W.~{Halbawi}, N.~{Azizan}, F.~{Salehi}, and B.~{Hassibi}, ``{Improving
  Distributed Gradient Descent Using Reed-Solomon Codes},'' in {\em 2018 IEEE
  International Symposium on Information Theory (ISIT)}, pp.~2027--2031, 2018.

\bibitem{li2018near}
S.~{Li}, S.~M. {Mousavi Kalan}, A.~S. {Avestimehr}, and M.~{Soltanolkotabi},
  ``{Near-Optimal Straggler Mitigation for Distributed Gradient Methods},'' in
  {\em 2018 IEEE International Parallel and Distributed Processing Symposium
  Workshops (IPDPSW)}, pp.~857--866, 2018.

\bibitem{li2018polynomially}
S.~Li, S.~M.~M. Kalan, Q.~Yu, M.~Soltanolkotabi, and A.~S. Avestimehr,
  ``{Polynomially Coded Regression: Optimal Straggler Mitigation via Data
  Encoding},'' {\em arXiv preprint arXiv:1805.09934}, 2018.

\bibitem{dutta2017convolution}
S.~{Dutta}, V.~{Cadambe}, and P.~{Grover}, ``{Coded Convolution for Parallel
  and Distributed Computing within a Deadline},'' in {\em 2017 IEEE
  International Symposium on Information Theory (ISIT)}, pp.~2403--2407, 2017.

\bibitem{yu2017fourier}
Q.~{Yu}, M.~A. {Maddah-Ali}, and A.~S. {Avestimehr}, ``{Coded Fourier
  Transform},'' in {\em 2017 55th Annual Allerton Conference on Communication,
  Control, and Computing (Allerton)}, pp.~494--501, 2017.

\bibitem{blaum1999lowest}
M.~{Blaum} and R.~M. {Roth}, ``{On Lowest Density MDS Codes},'' {\em IEEE
  Transactions on Information Theory}, vol.~45, no.~1, pp.~46--59, 1999.

\bibitem{balakrishnama1998linear}
S.~Balakrishnama and A.~Ganapathiraju, ``{Linear Discriminant Analysis - A
  Brief Tutorial},'' in {\em Institute for Signal and information Processing},
  vol.~18, pp.~1--8, 1998.

\bibitem{abdi2010principal}
H.~Abdi and L.~J. Williams, ``{Principal Component Analysis},'' {\em Wiley
  interdisciplinary reviews: computational statistics}, vol.~2, no.~4,
  pp.~433--459, 2010.

\bibitem{severinson2019block}
A.~{Severinson}, A.~{Graell i Amat}, and E.~{Rosnes}, ``{Block-Diagonal and LT
  Codes for Distributed Computing With Straggling Servers},'' {\em IEEE
  Transactions on Communications}, vol.~67, no.~3, pp.~1739--1753, 2019.

\bibitem{park2019irregular}
H.~{Park} and J.~{Moon}, ``{Irregular Product Coded Computation for
  High-Dimensional Matrix Multiplication},'' in {\em 2019 IEEE International
  Symposium on Information Theory (ISIT)}, pp.~1782--1786, 2019.

\bibitem{baharav2018proofing}
T.~{Baharav}, K.~{Lee}, O.~{Ocal}, and K.~{Ramchandran}, ``{Straggler-Proofing
  Massive-Scale Distributed Matrix Multiplication with D-Dimensional Product
  Codes},'' in {\em 2018 IEEE International Symposium on Information Theory
  (ISIT)}, pp.~1993--1997, 2018.

\bibitem{yu2020fundamental}
Q.~{Yu}, M.~A. {Maddah-Ali}, and A.~S. {Avestimehr}, ``{Straggler Mitigation in
  Distributed Matrix Multiplication: Fundamental Limits and Optimal Coding},''
  {\em IEEE Transactions on Information Theory}, vol.~66, no.~3,
  pp.~1920--1933, 2020.

\bibitem{fahim2019numerically}
M.~{Fahim} and V.~R. {Cadambe}, ``{Numerically Stable Polynomially Coded
  Computing},'' in {\em 2019 IEEE International Symposium on Information Theory
  (ISIT)}, pp.~3017--3021, 2019.

\bibitem{dutta2018polydot}
S.~{Dutta}, Z.~{Bai}, H.~{Jeong}, T.~M. {Low}, and P.~{Grover}, ``{A Unified
  Coded Deep Neural Network Training Strategy based on Generalized PolyDot
  codes},'' in {\em 2018 IEEE International Symposium on Information Theory
  (ISIT)}, pp.~1585--1589, 2018.

\bibitem{suh2017sparse}
G.~{Suh}, K.~{Lee}, and C.~{Suh}, ``{Matrix Sparsification for Coded Matrix
  Multiplication},'' in {\em 2017 55th Annual Allerton Conference on
  Communication, Control, and Computing (Allerton)}, pp.~1271--1278, 2017.

\bibitem{ye2018communicationcomputation}
M.~Ye and E.~Abbe, ``{Communication-Computation Efficient Gradient Coding},''
  {\em arXiv preprint arXiv:1802.03475}, 2018.

\bibitem{li2016unified}
S.~{Li}, M.~A. {Maddah-Ali}, and A.~S. {Avestimehr}, ``{A Unified Coding
  Framework for Distributed Computing with Straggling Servers},'' in {\em 2016
  IEEE Globecom Workshops (GC Wkshps)}, pp.~1--6, 2016.

\bibitem{zhang2019improved}
J.~{Zhang} and O.~{Simeone}, ``{Improved Latency-communication Trade-off for
  Map-shuffle-reduce Systems with Stragglers},'' in {\em ICASSP 2019 - 2019
  IEEE International Conference on Acoustics, Speech and Signal Processing
  (ICASSP)}, pp.~8172--8176, 2019.

\bibitem{parrinello2018}
E.~{Parrinello}, E.~{Lampiris}, and P.~{Elia}, ``{Coded Distributed Computing
  with Node Cooperation Substantially Increases Speedup Factors},'' in {\em
  2018 IEEE International Symposium on Information Theory (ISIT)},
  pp.~1291--1295, 2018.

\bibitem{konstantinidis2018leveraging}
K.~{Konstantinidis} and A.~{Ramamoorthy}, ``{Leveraging Coding Techniques for
  Speeding up Distributed Computing},'' in {\em 2018 IEEE Global Communications
  Conference (GLOBECOM)}, pp.~1--6, 2018.

\bibitem{prakash2018graph}
S.~{Prakash}, A.~{Reisizadeh}, R.~{Pedarsani}, and S.~{Avestimehr}, ``{Coded
  Computing for Distributed Graph Analytics},'' in {\em 2018 IEEE International
  Symposium on Information Theory (ISIT)}, pp.~1221--1225, 2018.

\bibitem{srini2018random}
S.~R. {Srinivasavaradhan}, L.~{Song}, and C.~{Fragouli}, ``{Distributed
  Computing Trade-offs with Random Connectivity},'' in {\em 2018 IEEE
  International Symposium on Information Theory (ISIT)}, pp.~1281--1285, 2018.

\bibitem{li2016multistage}
S.~{Li}, M.~A. {Maddah-Ali}, and A.~S. {Avestimehr}, ``{Coded Distributed
  Computing: Straggling Servers and Multistage Dataflows},'' in {\em 2016 54th
  Annual Allerton Conference on Communication, Control, and Computing
  (Allerton)}, pp.~164--171, 2016.

\bibitem{woolsey2019cascaded}
N.~{Woolsey}, R.~{Chen}, and M.~{Ji}, ``{Cascaded Coded Distributed Computing
  on Heterogeneous Networks},'' in {\em 2019 IEEE International Symposium on
  Information Theory (ISIT)}, pp.~2644--2648, 2019.

\bibitem{kiamari2017heterogeneous}
M.~Kiamari, C.~Wang, and A.~S. Avestimehr, ``{On Heterogeneous Coded
  Distributed Computing},'' {\em arXiv preprint arXiv:1709.00196}, 2017.

\bibitem{attia2016theoretic}
M.~A. {Attia} and R.~{Tandon}, ``{Information Theoretic Limits of Data
  Shuffling for Distributed Learning},'' in {\em 2016 IEEE Global
  Communications Conference (GLOBECOM)}, pp.~1--6, 2016.

\bibitem{gupta2017locality}
S.~{Gupta} and V.~{Lalitha}, ``{Locality-aware Hybrid Coded MapReduce for
  Server-Rack Architecture},'' in {\em 2017 IEEE Information Theory Workshop
  (ITW)}, pp.~459--463, 2017.

\bibitem{stinson2007combinatorial}
D.~Stinson, {\em {Combinatorial Designs: Constructions and Analysis}}.
\newblock Springer Science \& Business Media, 2007.

\bibitem{konstantinidis2019resolvable}
K.~Konstantinidis and A.~Ramamoorthy, ``{Resolvable Designs for Speeding up
  Distributed Computing},'' {\em arXiv preprint arXiv:1908.05666}, 2019.

\bibitem{li2018compressed}
S.~{Li}, M.~A. {Maddah-Ali}, and A.~S. {Avestimehr}, ``{Compressed Coded
  Distributed Computing},'' in {\em 2018 IEEE International Symposium on
  Information Theory (ISIT)}, pp.~2032--2036, 2018.

\bibitem{woolsey2018combinatorial}
N.~{Woolsey}, R.~{Chen}, and M.~{Ji}, ``{A New Combinatorial Design of Coded
  Distributed Computing},'' in {\em 2018 IEEE International Symposium on
  Information Theory (ISIT)}, pp.~726--730, 2018.

\bibitem{jiang2020coded}
J.~Jiang and L.~Qu, ``{Coded Distributed Computing Schemes with Smaller Numbers
  of Input Files and Output Functions},'' {\em arXiv preprint
  arXiv:2001.04194}, 2020.

\bibitem{yan2018pda}
Q.~{Yan}, X.~{Tang}, and Q.~{Chen}, ``{Placement Delivery Array and Its
  Applications},'' in {\em 2018 IEEE Information Theory Workshop (ITW)},
  pp.~1--5, 2018.

\bibitem{ramkumar2019pda}
V.~{Ramkumar} and P.~V. {Kumar}, ``{Coded MapReduce Schemes Based on Placement
  Delivery Array},'' in {\em 2019 IEEE International Symposium on Information
  Theory (ISIT)}, pp.~3087--3091, 2019.

\bibitem{alistarh2017qsgd}
D.~Alistarh, D.~Grubic, J.~Li, R.~Tomioka, and M.~Vojnovic, ``{QSGD:
  Communication-Efficient SGD via Gradient Quantization and Encoding},'' in
  {\em Advances in Neural Information Processing Systems 30} (I.~Guyon, U.~V.
  Luxburg, S.~Bengio, H.~Wallach, R.~Fergus, S.~Vishwanathan, and R.~Garnett,
  eds.), pp.~1709--1720, Curran Associates, Inc., 2017.

\bibitem{song2020pliable}
L.~{Song}, C.~{Fragouli}, and T.~{Zhao}, ``{A Pliable Index Coding Approach to
  Data Shuffling},'' {\em IEEE Transactions on Information Theory}, vol.~66,
  no.~3, pp.~1333--1353, 2020.

\bibitem{brahma2015pliable}
S.~{Brahma} and C.~{Fragouli}, ``{Pliable Index Coding},'' {\em IEEE
  Transactions on Information Theory}, vol.~61, no.~11, pp.~6192--6203, 2015.

\bibitem{charles2017approximate}
Z.~Charles, D.~Papailiopoulos, and J.~Ellenberg, ``{Approximate Gradient Coding
  via Sparse Random Graphs},'' {\em arXiv preprint arXiv:1711.06771}, 2017.

\bibitem{haddadpour2018cross}
F.~{Haddadpour}, Y.~{Yang}, V.~{Cadambe}, and P.~{Grover}, ``{Cross-Iteration
  Coded Computing},'' in {\em 2018 56th Annual Allerton Conference on
  Communication, Control, and Computing (Allerton)}, pp.~196--203, 2018.

\bibitem{haddadpour2018stragglerresilient}
F.~Haddadpour, Y.~Yang, M.~Chaudhari, V.~R. Cadambe, and P.~Grover,
  ``{Straggler-Resilient and Communication-Efficient Distributed Iterative
  Linear Solver},'' {\em arXiv preprint arXiv:1806.06140}, 2018.

\bibitem{zhang2013statistical}
Y.~Zhang, J.~C. Duchi, and M.~J. Wainwright, ``{Communication-Efficient
  Algorithms for Statistical Optimization},'' {\em J. Mach. Learn. Res.},
  vol.~14, p.~3321–3363, Jan. 2013.

\bibitem{martin2010parallelized}
M.~Zinkevich, M.~Weimer, L.~Li, and A.~J. Smola, ``{Parallelized Stochastic
  Gradient Descent},'' in {\em Advances in Neural Information Processing
  Systems 23} (J.~D. Lafferty, C.~K.~I. Williams, J.~Shawe-Taylor, R.~S. Zemel,
  and A.~Culotta, eds.), pp.~2595--2603, Curran Associates, Inc., 2010.

\bibitem{wan2020topological}
K.~Wan, M.~Ji, and G.~Caire, ``{Topological Coded Distributed Computing},''
  {\em arXiv preprint arXiv:2004.04421}, 2020.

\bibitem{xia2017data}
W.~{Xia}, P.~{Zhao}, Y.~{Wen}, and H.~{Xie}, ``{A Survey on Data Center
  Networking (DCN): Infrastructure and Operations},'' {\em IEEE Communications
  Surveys Tutorials}, vol.~19, no.~1, pp.~640--656, 2017.

\bibitem{fares2008scalable}
M.~Al-Fares, A.~Loukissas, and A.~Vahdat, ``{A Scalable, Commodity Data Center
  Network Architecture},'' in {\em Proceedings of the ACM SIGCOMM 2008
  Conference on Data Communication}, SIGCOMM ’08, (New York, NY, USA),
  p.~63–74, Association for Computing Machinery, 2008.

\bibitem{chung2017ubershuffle}
J.~Chung, K.~Lee, R.~Pedarsani, D.~Papailiopoulos, and K.~Ramchandran,
  ``{Ubershuffle: Communication-efficient Data Shuffling for SGD via Coding
  Theory},'' {\em Advances in NIPS}, 2017.

\bibitem{woolsey2019coded}
N.~Woolsey, R.-R. Chen, and M.~Ji, ``{Coded Distributed Computing with
  Heterogeneous Function Assignments},'' {\em arXiv preprint arXiv:1902.10738},
  2019.

\bibitem{song2017benefit}
L.~{Song}, S.~R. {Srinivasavaradhan}, and C.~{Fragouli}, ``{The Benefit of
  Being Flexible in Distributed Computation},'' in {\em 2017 IEEE Information
  Theory Workshop (ITW)}, pp.~289--293, 2017.

\bibitem{xu2019heterogeneous}
F.~Xu and M.~Tao, ``{Heterogeneous Coded Distributed Computing: Joint Design of
  File Allocation and Function Assignment},'' {\em arXiv preprint
  arXiv:1908.06715}, 2019.

\bibitem{yu2017allocate}
Q.~{Yu}, S.~{Li}, M.~A. {Maddah-Ali}, and A.~S. {Avestimehr}, ``{How to
  Optimally Allocate Resources for Coded Distributed Computing?},'' in {\em
  2017 IEEE International Conference on Communications (ICC)}, pp.~1--7, 2017.

\bibitem{zhao2019load}
M.~Zhao, W.~Wang, Y.~Wang, and Z.~Zhang, ``{Load Scheduling for Distributed
  Edge Computing: A Communication-Computation Tradeoff},'' {\em Peer-to-Peer
  Networking and Applications}, vol.~12, no.~5, pp.~1418--1432, 2019.

\bibitem{lee2017multicore}
K.~{Lee}, R.~{Pedarsani}, D.~{Papailiopoulos}, and K.~{Ramchandran}, ``{Coded
  Computation for Multicore Setups},'' in {\em 2017 IEEE International
  Symposium on Information Theory (ISIT)}, pp.~2413--2417, 2017.

\bibitem{zaharia2008improving}
M.~Zaharia, A.~Konwinski, A.~D. Joseph, R.~H. Katz, and I.~Stoica, ``{Improving
  MapReduce Performance in Heterogeneous Environments.},'' in {\em Osdi}, p.~7,
  2008.

\bibitem{narra2019slack}
K.~G. Narra, Z.~Lin, M.~Kiamari, S.~Avestimehr, and M.~Annavaram, ``{Slack
  Squeeze Coded Computing for Adaptive Straggler Mitigation},'' in {\em
  Proceedings of the International Conference for High Performance Computing,
  Networking, Storage and Analysis}, SC ’19, (New York, NY, USA), Association
  for Computing Machinery, 2019.

\bibitem{yang2019timelythroughput}
C.-S. Yang, R.~Pedarsani, and A.~S. Avestimehr, ``{Timely-Throughput Optimal
  Coded Computing over Cloud Networks},'' {\em arXiv preprint
  arXiv:1904.05522}, 2019.

\bibitem{ozfatura2019persistent}
E.~{Ozfatura}, D.~{Gündüz}, and S.~{Ulukus}, ``{Speeding Up Distributed
  Gradient Descent by Utilizing Non-persistent Stragglers},'' in {\em 2019 IEEE
  International Symposium on Information Theory (ISIT)}, pp.~2729--2733, 2019.

\bibitem{ferdinand2018hierarchical}
N.~{Ferdinand} and S.~C. {Draper}, ``{Hierarchical Coded Computation},'' in
  {\em 2018 IEEE International Symposium on Information Theory (ISIT)},
  pp.~1620--1624, 2018.

\bibitem{reisizadeh2019}
A.~{Reisizadeh}, S.~{Prakash}, R.~{Pedarsani}, and A.~S. {Avestimehr}, ``{Coded
  Computation Over Heterogeneous Clusters},'' {\em IEEE Transactions on
  Information Theory}, vol.~65, no.~7, pp.~4227--4242, 2019.

\bibitem{kim2019group}
M.~{Kim}, J.~{Sohn}, and J.~{Moon}, ``{Coded Matrix Multiplication on a
  Group-Based Model},'' in {\em 2019 IEEE International Symposium on
  Information Theory (ISIT)}, pp.~722--726, 2019.

\bibitem{kim2019optimal}
D.~Kim, H.~Park, and J.~Choi, ``{Optimal Load Allocation for Coded Distributed
  Computation in Heterogeneous Clusters},'' {\em arXiv preprint
  arXiv:1904.09496}, 2019.

\bibitem{kesh2018dynamic}
Y.~{Keshtkarjahromi}, Y.~{Xing}, and H.~{Seferoglu}, ``{Dynamic
  Heterogeneity-Aware Coded Cooperative Computation at the Edge},'' in {\em
  2018 IEEE 26th International Conference on Network Protocols (ICNP)},
  pp.~23--33, 2018.

\bibitem{ferdinand2016anytime}
N.~S. {Ferdinand} and S.~C. {Draper}, ``{Anytime Coding for Distributed
  Computation},'' in {\em 2016 54th Annual Allerton Conference on
  Communication, Control, and Computing (Allerton)}, pp.~954--960, 2016.

\bibitem{wang2019batchprocessing}
B.~Wang, J.~Xie, K.~Lu, Y.~Wan, and S.~Fu, ``{On Batch-Processing Based Coded
  Computing for Heterogeneous Distributed Computing Systems},'' {\em arXiv
  preprint arXiv:1912.12559}, 2019.

\bibitem{zhu2017sequential}
J.~{Zhu}, Y.~{Pu}, V.~{Gupta}, C.~{Tomlin}, and K.~{Ramchandran}, ``{A
  Sequential Approximation Framework for Coded Distributed Optimization},'' in
  {\em 2017 55th Annual Allerton Conference on Communication, Control, and
  Computing (Allerton)}, pp.~1240--1247, 2017.

\bibitem{jahani2019codedsketch}
T.~{Jahani-Nezhad} and M.~A. {Maddah-Ali}, ``{CodedSketch: Coded Distributed
  Computation of Approximated Matrix Multiplication},'' in {\em 2019 IEEE
  International Symposium on Information Theory (ISIT)}, pp.~2489--2493, 2019.

\bibitem{gupta2018oversketch}
V.~{Gupta}, S.~{Wang}, T.~{Courtade}, and K.~{Ramchandran}, ``{OverSketch:
  Approximate Matrix Multiplication for the Cloud},'' in {\em 2018 IEEE
  International Conference on Big Data (Big Data)}, pp.~298--304, 2018.

\bibitem{charles2018gradient}
Z.~Charles and D.~Papailiopoulos, ``{Gradient Coding via the Stochastic Block
  Model},'' {\em arXiv preprint arXiv:1805.10378}, 2018.

\bibitem{bitar2020stochastic}
R.~{Bitar}, M.~{Wootters}, and S.~{El Rouayheb}, ``{Stochastic Gradient Coding
  for Straggler Mitigation in Distributed Learning},'' {\em IEEE Journal on
  Selected Areas in Information Theory}, pp.~1--1, 2020.

\bibitem{karakus2018redundancy}
C.~Karakus, Y.~Sun, S.~Diggavi, and W.~Yin, ``{Redundancy Techniques for
  Straggler Mitigation in Distributed Optimization and Learning},'' {\em arXiv
  preprint arXiv:1803.05397}, 2018.

\bibitem{karakus2017encoded}
C.~{Karakus}, Y.~{Sun}, and S.~{Diggavi}, ``{Encoded Distributed
  Optimization},'' in {\em 2017 IEEE International Symposium on Information
  Theory (ISIT)}, pp.~2890--2894, 2017.

\bibitem{kosaian2018learning}
J.~Kosaian, K.~V. Rashmi, and S.~Venkataraman, ``{Learning a Code: Machine
  Learning for Approximate Non-Linear Coded Computation},'' {\em arXiv preprint
  arXiv:1806.01259}, 2018.

\bibitem{woodruff2014sketching}
D.~P. Woodruff, ``{Sketching as a Tool for Numerical Linear Algebra},'' {\em
  arXiv preprint arXiv:1411.4357}, 2014.

\bibitem{wang2015practical}
S.~Wang, ``{A Practical Guide to Randomized Matrix Computations with MATLAB
  Implementations},'' {\em arXiv preprint arXiv:1505.07570}, 2015.

\bibitem{cormode2008finding}
G.~Cormode and M.~Hadjieleftheriou, ``{Finding Frequent Items in Data
  Streams},'' {\em Proc. VLDB Endow.}, vol.~1, p.~1530–1541, Aug. 2008.

\bibitem{wang2019erasurehead}
H.~Wang, Z.~Charles, and D.~Papailiopoulos, ``{ErasureHead: Distributed
  Gradient Descent without Delays Using Approximate Gradient Coding},'' {\em
  arXiv preprint arXiv:1901.09671}, 2019.

\bibitem{chang2019random}
W.~{Chang} and R.~{Tandon}, ``{Random Sampling for Distributed Coded Matrix
  Multiplication},'' in {\em ICASSP 2019 - 2019 IEEE International Conference
  on Acoustics, Speech and Signal Processing (ICASSP)}, pp.~8187--8191, 2019.

\bibitem{simonyan2014deep}
K.~Simonyan and A.~Zisserman, ``{Very Deep Convolutional Networks for
  Large-Scale Image Recognition},'' {\em arXiv preprint arXiv:1409.1556}, 2014.

\bibitem{hadidi2019}
R.~Hadidi, J.~Cao, M.~S. Ryoo, and H.~Kim, ``{Robustly Executing DNNs in IoT
  Systems Using Coded Distributed Computing},'' in {\em Proceedings of the 56th
  Annual Design Automation Conference 2019}, DAC ’19, (New York, NY, USA),
  Association for Computing Machinery, 2019.

\bibitem{das2018c3les}
A.~B. {Das}, L.~{Tang}, and A.~{Ramamoorthy}, ``{C3LES: Codes for Coded
  Computation that Leverage Stragglers},'' in {\em 2018 IEEE Information Theory
  Workshop (ITW)}, pp.~1--5, 2018.

\bibitem{ozfatura2019partial}
E.~{Ozfatura}, S.~{Ulukus}, and D.~{Gündüz}, ``{Distributed Gradient Descent
  with Coded Partial Gradient Computations},'' in {\em ICASSP 2019 - 2019 IEEE
  International Conference on Acoustics, Speech and Signal Processing
  (ICASSP)}, pp.~3492--3496, 2019.

\bibitem{NIPS2017_inverse}
Y.~Yang, P.~Grover, and S.~Kar, ``{Coded Distributed Computing for Inverse
  Problems},'' in {\em Advances in Neural Information Processing Systems 30}
  (I.~Guyon, U.~V. Luxburg, S.~Bengio, H.~Wallach, R.~Fergus, S.~Vishwanathan,
  and R.~Garnett, eds.), pp.~709--719, Curran Associates, Inc., 2017.

\bibitem{lim2020hierarchical}
W.~Y.~B. Lim, Z.~Xiong, C.~Miao, D.~Niyato, Q.~Yang, C.~Leung, and H.~V. Poor,
  ``{Hierarchical Incentive Mechanism Design for Federated Machine Learning in
  Mobile Networks},'' {\em IEEE Internet of Things Journal}, 2020.

\bibitem{lim2020federated}
W.~Y.~B. Lim, N.~C. Luong, D.~T. Hoang, Y.~Jiao, Y.-C. Liang, Q.~Yang,
  D.~Niyato, and C.~Miao, ``{Federated Learning in Mobile Edge Networks: A
  Comprehensive Survey},'' {\em IEEE Communications Surveys \& Tutorials},
  2020.

\bibitem{gentry2009fully}
C.~Gentry, ``Fully homomorphic encryption using ideal lattices,'' in {\em
  Proceedings of the forty-first annual ACM symposium on Theory of computing},
  pp.~169--178, 2009.

\bibitem{brakerski2014efficient}
Z.~Brakerski and V.~Vaikuntanathan, ``{Efficient Fully Homomorphic Encryption
  from (Standard) LWE},'' {\em SIAM Journal on Computing}, vol.~43, no.~2,
  pp.~831--871, 2014.

\bibitem{goldreich1998secure}
O.~Goldreich, ``{Secure Multi-Party Computation},'' {\em Manuscript.
  Preliminary version}, vol.~78, 1998.

\bibitem{huang2011faster}
Y.~Huang, D.~Evans, J.~Katz, and L.~Malka, ``{Faster Secure Two-Party
  Computation Using Garbled Circuits},'' in {\em USENIX Security Symposium},
  vol.~201, pp.~331--335, 2011.

\bibitem{bogdanov2008sharemind}
D.~Bogdanov, S.~Laur, and J.~Willemson, ``{Sharemind: A Framework for Fast
  Privacy-Preserving Computations},'' in {\em European Symposium on Research in
  Computer Security}, pp.~192--206, Springer, 2008.

\bibitem{chang2018capacity}
W.~{Chang} and R.~{Tandon}, ``{On the Capacity of Secure Distributed Matrix
  Multiplication},'' in {\em 2018 IEEE Global Communications Conference
  (GLOBECOM)}, pp.~1--6, 2018.

\bibitem{ben2019completeness}
M.~Ben-Or, S.~Goldwasser, and A.~Wigderson, {\em {Completeness Theorems for
  Non-Cryptographic Fault-Tolerant Distributed Computation}}, p.~351–371.
\newblock New York, NY, USA: Association for Computing Machinery, 2019.

\bibitem{kim2018private}
M.~Kim, H.~Yang, and J.~Lee, ``{Private Coded Computation for Machine
  Learning},'' {\em arXiv preprint arXiv:1807.01170}, 2018.

\bibitem{nodehi2018entangled}
H.~A. {Nodehi}, S.~R.~H. {Najarkolaei}, and M.~A. {Maddah-Ali}, ``{Entangled
  Polynomial Coding in Limited-Sharing Multi-Party Computation},'' in {\em 2018
  IEEE Information Theory Workshop (ITW)}, pp.~1--5, 2018.

\bibitem{chen2017distributed}
Y.~Chen, L.~Su, and J.~Xu, ``{Distributed Statistical Machine Learning in
  Adversarial Settings: Byzantine Gradient Descent},'' {\em Proceedings of the
  ACM on Measurement and Analysis of Computing Systems}, vol.~1, no.~2,
  pp.~1--25, 2017.

\bibitem{chen2018draco}
L.~Chen, H.~Wang, Z.~Charles, and D.~Papailiopoulos, ``{DRACO:
  Byzantine-resilient Distributed Training via Redundant Gradients},'' {\em
  arXiv preprint arXiv:1803.09877}, 2018.

\bibitem{yu2019harmonic}
Q.~{Yu} and A.~S. {Avestimehr}, ``{Harmonic Coding: An Optimal Linear Code for
  Privacy-Preserving Gradient-Type Computation},'' in {\em 2019 IEEE
  International Symposium on Information Theory (ISIT)}, pp.~1102--1106, 2019.

\bibitem{so2019codedprivateml}
J.~So, B.~Guler, A.~S. Avestimehr, and P.~Mohassel, ``{CodedPrivateML: A Fast
  and Privacy-Preserving Framework for Distributed Machine Learning},'' {\em
  arXiv preprint arXiv:1902.00641}, 2019.

\bibitem{shamir1979share}
A.~Shamir, ``{How to Share a Secret},'' {\em Communications of the ACM},
  vol.~22, no.~11, pp.~612--613, 1979.

\bibitem{bitar2017minimize}
R.~{Bitar}, P.~{Parag}, and S.~{El Rouayheb}, ``{Minimizing Latency for Secure
  Distributed Computing},'' in {\em 2017 IEEE International Symposium on
  Information Theory (ISIT)}, pp.~2900--2904, 2017.

\bibitem{hearst1998support}
M.~A. Hearst, S.~T. Dumais, E.~Osuna, J.~Platt, and B.~Scholkopf, ``{Support
  Vector Machines},'' {\em IEEE Intelligent Systems and their applications},
  vol.~13, no.~4, pp.~18--28, 1998.

\bibitem{bitar2020staircase}
R.~{Bitar}, P.~{Parag}, and S.~{El Rouayheb}, ``{Minimizing Latency for Secure
  Coded Computing Using Secret Sharing via Staircase Codes},'' {\em IEEE
  Transactions on Communications}, pp.~1--1, 2020.

\bibitem{kakar2018rateefficiency}
J.~Kakar, S.~Ebadifar, and A.~Sezgin, ``{Rate-Efficiency and
  Straggler-Robustness through Partition in Distributed Two-Sided Secure Matrix
  Computation},'' {\em arXiv preprint arXiv:1810.13006}, 2018.

\bibitem{yang2019secure}
H.~{Yang} and J.~{Lee}, ``{Secure Distributed Computing With Straggling Servers
  Using Polynomial Codes},'' {\em IEEE Transactions on Information Forensics
  and Security}, vol.~14, no.~1, pp.~141--150, 2019.

\bibitem{jia2019capacity}
Z.~Jia and S.~A. Jafar, ``{On the Capacity of Secure Distributed Matrix
  Multiplication},'' {\em arXiv preprint arXiv:1908.06957}, 2019.

\bibitem{oliveira2020gasp}
R.~G.~L. {D’Oliveira}, S.~{El Rouayheb}, and D.~{Karpuk}, ``{GASP Codes for
  Secure Distributed Matrix Multiplication},'' {\em IEEE Transactions on
  Information Theory}, pp.~1--1, 2020.

\bibitem{rafael2019degree}
G.~L. {Rafael D’Oliveira}, S.~E. {Rouayheb}, D.~{Heinlein}, and D.~{Karpuk},
  ``{Degree Tables for Secure Distributed Matrix Multiplication},'' in {\em
  2019 IEEE Information Theory Workshop (ITW)}, pp.~1--5, 2019.

\bibitem{chang2010achieving}
X.~Chang, J.~Wang, J.~Wang, V.~Lee, K.~Lu, and Y.~Yang, ``{On Achieving Maximum
  Secure Throughput Using Network Coding Against Wiretap Attack},'' in {\em
  2010 IEEE 30th International Conference on Distributed Computing Systems},
  pp.~526--535, IEEE, 2010.

\bibitem{zhao2018weakly}
R.~{Zhao}, J.~{Wang}, K.~{Lu}, J.~{Wang}, X.~{Wang}, J.~{Zhou}, and C.~{Cao},
  ``{Weakly Secure Coded Distributed Computing},'' in {\em 2018 IEEE
  SmartWorld, Ubiquitous Intelligence Computing, Advanced Trusted Computing,
  Scalable Computing Communications, Cloud Big Data Computing, Internet of
  People and Smart City Innovation
  (SmartWorld/SCALCOM/UIC/ATC/CBDCom/IOP/SCI)}, pp.~603--610, 2018.

\bibitem{resnick2000reputation}
P.~Resnick, K.~Kuwabara, R.~Zeckhauser, and E.~Friedman, ``{Reputation
  Systems},'' {\em Communications of the ACM}, vol.~43, no.~12, pp.~45--48,
  2000.

\bibitem{miljumbi2016network}
R.~{Mijumbi}, J.~{Serrat}, J.~{Gorricho}, N.~{Bouten}, F.~{De Turck}, and
  R.~{Boutaba}, ``{Network Function Virtualization: State-of-the-Art and
  Research Challenges},'' {\em IEEE Communications Surveys Tutorials}, vol.~18,
  no.~1, pp.~236--262, 2016.

\bibitem{qi2014sdn}
Q.~{Qi}, W.~{Wang}, X.~{Gong}, and X.~{Que}, ``{A SDN-based Network
  Virtualization Architecture with Autonomie Management},'' in {\em 2014 IEEE
  Globecom Workshops (GC Wkshps)}, pp.~178--182, 2014.

\bibitem{eramo2019proposal}
V.~{Eramo}, T.~{Catena}, and F.~G. {Lavacca}, ``{Proposal and Investigation of
  an Optical Reconfiguration Cost Aware Policy for Resource Allocation in
  Network Function Virtualization Infrastructures},'' in {\em 2019 21st
  International Conference on Transparent Optical Networks (ICTON)}, pp.~1--5,
  2019.

\bibitem{sun2018reliable}
J.~{Sun}, G.~{Zhu}, G.~{Sun}, D.~{Liao}, Y.~{Li}, A.~K. {Sangaiah},
  M.~{Ramachandran}, and V.~{Chang}, ``{A Reliability-Aware Approach for
  Resource Efficient Virtual Network Function Deployment},'' {\em IEEE Access},
  vol.~6, pp.~18238--18250, 2018.

\bibitem{quzweeni2015energy}
A.~{Al-Quzweeni}, T.~E.~H. {El-Gorashi}, L.~{Nonde}, and J.~M.~H. {Elmirghani},
  ``{Energy Efficient Network Function Virtualization in 5G networks},'' in
  {\em 2015 17th International Conference on Transparent Optical Networks
  (ICTON)}, pp.~1--4, 2015.

\bibitem{lingua2019acceleration}
L.~{Linguaglossa}, S.~{Lange}, S.~{Pontarelli}, G.~{Rétvári}, D.~{Rossi},
  T.~{Zinner}, R.~{Bifulco}, M.~{Jarschel}, and G.~{Bianchi}, ``{Survey of
  Performance Acceleration Techniques for Network Function Virtualization},''
  {\em Proceedings of the IEEE}, vol.~107, no.~4, pp.~746--764, 2019.

\bibitem{jang2015security}
H.~{Jang}, J.~{Jeong}, H.~{Kim}, and J.~{Park}, ``{A Survey on Interfaces to
  Network Security Functions in Network Virtualization},'' in {\em 2015 IEEE
  29th International Conference on Advanced Information Networking and
  Applications Workshops}, pp.~160--163, 2015.

\bibitem{aljuhani2017virtualized}
A.~{Aljuhani} and T.~{Alharbi}, ``{Virtualized Network Functions Security
  Attacks and Vulnerabilities},'' in {\em 2017 IEEE 7th Annual Computing and
  Communication Workshop and Conference (CCWC)}, pp.~1--4, 2017.

\bibitem{cherrared2019fault}
S.~{Cherrared}, S.~{Imadali}, E.~{Fabre}, G.~{Gössler}, and I.~G.~B. {Yahia},
  ``{A Survey of Fault Management in Network Virtualization Environments:
  Challenges and Solutions},'' {\em IEEE Transactions on Network and Service
  Management}, vol.~16, no.~4, pp.~1537--1551, 2019.

\bibitem{anvith2019telecom}
P.~v.~{Anvith}, N.~{Gunavathi}, B.~{Malarkodi}, and B.~{Rebekka}, ``{A Survey
  on Network Functions Virtualization for Telecom Paradigm},'' in {\em 2019
  TEQIP III Sponsored International Conference on Microwave Integrated
  Circuits, Photonics and Wireless Networks (IMICPW)}, pp.~302--306, 2019.

\bibitem{liu2016reliability}
J.~{Liu}, Z.~{Jiang}, N.~{Kato}, O.~{Akashi}, and A.~{Takahara}, ``{Reliability
  Evaluation for NFV Deployment of Future Mobile Broadband Networks},'' {\em
  IEEE Wireless Communications}, vol.~23, no.~3, pp.~90--96, 2016.

\bibitem{shuwaili2016fault}
A.~{Al-Shuwaili}, O.~{Simeone}, J.~{Kliewer}, and P.~{Popovski}, ``{Coded
  Network Function Virtualization: Fault Tolerance via In-Network Coding},''
  {\em IEEE Wireless Communications Letters}, vol.~5, no.~6, pp.~644--647,
  2016.

\bibitem{nikaein2015processing}
N.~Nikaein, ``{Processing Radio Access Network Functions in the Cloud: Critical
  Issues and Modeling},'' in {\em Proceedings of the 6th International Workshop
  on Mobile Cloud Computing and Services}, MCS ’15, (New York, NY, USA),
  p.~36–43, Association for Computing Machinery, 2015.

\bibitem{aliasgari2019virtualized}
M.~{Aliasgari}, J.~{Kliewer}, and O.~{Simeone}, ``{Coded Computation Against
  Processing Delays for Virtualized Cloud-Based Channel Decoding},'' {\em IEEE
  Transactions on Communications}, vol.~67, no.~1, pp.~28--38, 2019.

\bibitem{frankle2018lottery}
J.~Frankle and M.~Carbin, ``{The Lottery Ticket Hypothesis: Finding Sparse,
  Trainable Neural Networks},'' {\em arXiv preprint arXiv:1803.03635}, 2018.

\bibitem{li2017scalable}
S.~{Li}, Q.~{Yu}, M.~A. {Maddah-Ali}, and A.~S. {Avestimehr}, ``{A Scalable
  Framework for Wireless Distributed Computing},'' {\em IEEE/ACM Transactions
  on Networking}, vol.~25, no.~5, pp.~2643--2654, 2017.

\bibitem{kiamari2017edge}
M.~{Kiamari}, C.~{Wang}, and A.~S. {Avestimehr}, ``{Coding for Edge-facilitated
  Wireless Distributed Computing with Heterogeneous Users},'' in {\em 2017 51st
  Asilomar Conference on Signals, Systems, and Computers}, pp.~536--540, 2017.

\bibitem{dhakal2019coded2}
S.~Dhakal, S.~Prakash, Y.~Yona, S.~Talwar, and N.~Himayat, ``{Coded Computing
  for Distributed Machine Learning in Wireless Edge Network},'' in {\em 2019
  IEEE 90th Vehicular Technology Conference (VTC2019-Fall)}, pp.~1--6, IEEE,
  2019.

\bibitem{dhakal2019coded}
S.~Dhakal, S.~Prakash, Y.~Yona, S.~Talwar, and N.~Himayat, ``{Coded Federated
  Learning},'' in {\em 2019 IEEE Globecom Workshops (GC Wkshps)}, pp.~1--6,
  IEEE, 2019.

\bibitem{mcmahan2017communication}
B.~McMahan, E.~Moore, D.~Ramage, S.~Hampson, and B.~A. y~Arcas,
  ``{Communication-efficient Learning of Deep Networks from Decentralized
  Data},'' in {\em Artificial Intelligence and Statistics}, pp.~1273--1282,
  2017.

\bibitem{prakash2020coded}
S.~Prakash, S.~Dhakal, M.~Akdeniz, A.~S. Avestimehr, and N.~Himayat, ``{Coded
  Computing for Federated Learning at the Edge},'' {\em arXiv preprint
  arXiv:2007.03273}, 2020.

\bibitem{yang2019lowrank}
K.~{Yang}, Y.~{Shi}, and Z.~{Ding}, ``{Data Shuffling in Wireless Distributed
  Computing via Low-Rank Optimization},'' {\em IEEE Transactions on Signal
  Processing}, vol.~67, no.~12, pp.~3087--3099, 2019.

\bibitem{zhao2016interference}
N.~{Zhao}, F.~R. {Yu}, M.~{Jin}, Q.~{Yan}, and V.~C.~M. {Leung},
  ``{Interference Alignment and Its Applications: A Survey, Research Issues,
  and Challenges},'' {\em IEEE Communications Surveys Tutorials}, vol.~18,
  no.~3, pp.~1779--1803, 2016.

\bibitem{li2019wireless}
F.~{Li}, J.~{Chen}, and Z.~{Wang}, ``{Wireless MapReduce Distributed
  Computing},'' {\em IEEE Transactions on Information Theory}, vol.~65, no.~10,
  pp.~6101--6114, 2019.

\bibitem{ha2019wireless}
S.~{Ha}, J.~{Zhang}, O.~{Simeone}, and J.~{Kang}, ``{Wireless Map-Reduce
  Distributed Computing with Full-Duplex Radios and Imperfect CSI},'' in {\em
  2019 IEEE 20th International Workshop on Signal Processing Advances in
  Wireless Communications (SPAWC)}, pp.~1--5, 2019.

\bibitem{ha2019federated}
S.~{Ha}, J.~{Zhang}, O.~{Simeone}, and J.~{Kang}, ``{Coded Federated Computing
  in Wireless Networks with Straggling Devices and Imperfect CSI},'' in {\em
  2019 IEEE International Symposium on Information Theory (ISIT)},
  pp.~2649--2653, 2019.

\bibitem{wang2019uav}
B.~{Wang}, J.~{Xie}, K.~{Lu}, Y.~{Wan}, and S.~{Fu}, ``{Coding for
  Heterogeneous UAV-Based Networked Airborne Computing},'' in {\em 2019 IEEE
  Globecom Workshops (GC Wkshps)}, pp.~1--6, 2019.

\bibitem{shyuan2020joint}
J.~S. Ng, W.~Y.~B. Lim, H.-N. Dai, Z.~Xiong, J.~Huang, D.~Niyato, X.-S. Hua,
  C.~Leung, and C.~Miao, ``{Joint Auction-Coalition Formation Framework for
  Communication-Efficient Federated Learning in UAV-Enabled Internet of
  Vehicles},'' {\em arXiv preprint arXiv:2007.06378}, 2020.

\bibitem{lim2020towards}
W.~Y.~B. Lim, J.~Huang, Z.~Xiong, J.~Kang, D.~Niyato, X.-S. Hua, C.~Leung, and
  C.~Miao, ``{Towards Federated Learning in UAV-Enabled Internet of Vehicles: A
  Multi-Dimensional Contract-Matching Approach},'' {\em arXiv preprint
  arXiv:2004.03877}, 2020.

\bibitem{mao2016resource}
H.~Mao, M.~Alizadeh, I.~Menache, and S.~Kandula, ``{Resource Management with
  Deep Reinforcement Learning},'' in {\em Proceedings of the 15th ACM Workshop
  on Hot Topics in Networks}, pp.~50--56, 2016.

\bibitem{abbas2018mec}
N.~{Abbas}, Y.~{Zhang}, A.~{Taherkordi}, and T.~{Skeie}, ``{Mobile Edge
  Computing: A Survey},'' {\em IEEE Internet of Things Journal}, vol.~5, no.~1,
  pp.~450--465, 2018.

\bibitem{bonomi2012fog}
F.~Bonomi, R.~Milito, J.~Zhu, and S.~Addepalli, ``{Fog Computing and Its Role
  in the Internet of Things},'' in {\em Proceedings of the First Edition of the
  MCC Workshop on Mobile Cloud Computing}, MCC ’12, (New York, NY, USA),
  p.~13–16, Association for Computing Machinery, 2012.

\bibitem{li2017fog}
S.~{Li}, M.~A. {Maddah-Ali}, and A.~S. {Avestimehr}, ``{Coding for Distributed
  Fog Computing},'' {\em IEEE Communications Magazine}, vol.~55, no.~4,
  pp.~34--40, 2017.

\bibitem{park2018hierarchical}
H.~{Park}, K.~{Lee}, J.~{Sohn}, C.~{Suh}, and J.~{Moon}, ``{Hierarchical Coding
  for Distributed Computing},'' in {\em 2018 IEEE International Symposium on
  Information Theory (ISIT)}, pp.~1630--1634, 2018.

\end{thebibliography}

\end{document}